\documentstyle[12pt,epsfig]{article}

\def \sect #1 {\setcounter{equation} 0\section{#1}}
\def \be  {\begin{equation}}
\def \ee  {\end{equation}}
\def \ba  {\begin{eqnarray}}
\def \ea  {\end{eqnarray}}
\def \baa {\begin{eqnarray*}}
\def \eaa {\end{eqnarray*}}
\def \bb  {\begin{thebibliography}}
\def \eb  {\end{thebibliography}}

\def \lab #1 {\label{#1}}
\newcommand\re[1]{(\ref{#1})}

\def \fracs #1#2 {\mbox{\small $\frac{#1}{#2}$}}

\def \bin #1#2 {{\left({#1}\atop{#2}\right)}}
\def \as {\relax\ifmmode\alpha_s\else{$\alpha_s${ }}\fi}

\def \al #1 {\frac {\as({#1})}{\pi} }
\def \ds #1 {\ooalign{$\hfil/\hfil$\crcr$#1$}}
\def \MS {\overline{\rm MS}}

\def \d {{\rm d}}

\def\np#1#2#3  {{Nucl. Phys.~{\bf #1} (19#3) #2}}
\def\nc#1#2#3  {{Nuovo. Cim.~{\bf #1} (19#3) #2}}
\def\pl#1#2#3  {{Phys. Lett.~{\bf #1} (19#3) #2}}
\def\pr#1#2#3  {{Phys. Rev.~{\bf #1} (19#3) #2}}
\def\prl#1#2#3  {{Phys. Rev. Lett.~{\bf #1} (19#3) #2}}
\def\prep#1#2#3 {{Phys. Rep.~{\bf #1} (19#3) #2}}
\def\zp#1#2#3  {{Z. Phys.~{\bf #1} (19#3) #2}}
\def\epj#1#2#3  {{Eur. Phys. J.~{\bf #1} (19#3) #2}}
\def\rmp#1#2#3  {{Rev. Mod. Phys.~{\bf #1} (19#3) #2}}
\def\JETP#1#2#3 {{Sov.\ Phys.\ JETP~{\bf #1} (19#3) #2}}
\def\sj#1#2#3 {{Sov.\ J.\ Nucl.\ Phys.~{\bf #1} (19#3) #2}}
\def\hepph  #1 {{\tt hep-ph/#1}}

\begin{document}

\begin{flushright}
NIKHEF/00-15\\
YITP-SB-00-03\\
BNL-HET-00/36 \\
RBRC-143 \\
\today\\
\end{flushright}

\vspace*{18mm}

\begin{center}
{\LARGE \bf Recoil and Threshold
Corrections

\bigskip

in Short-distance Cross Sections}
\par\vspace*{20mm}\par
{\large Eric Laenen$^a$,
George Sterman$^{b,c}$ and Werner Vogelsang$^{b,d}$}

\bigskip

{\em $^a$NIKHEF Theory Group, Kruislaan 409\\ 1098 SJ Amsterdam, The
Netherlands}

\bigskip

{\em $^b$C.N.\ Yang Institute for Theoretical Physics,
SUNY Stony Brook\\
Stony Brook, New York 11794 -- 3840, U.S.A.}

\bigskip

{\em $^c$Physics Department, Brookhaven National Laboratory,\\
Upton, NY 11973, U.S.A.}

\bigskip

{\em $^d$RIKEN-BNL Research Center, Brookhaven National Laboratory,\\
Upton, NY 11973, U.S.A.}
\end{center}
\vspace*{15mm}

\begin{abstract}
We identify and resum corrections
associated with the kinematic recoil 
of the hard scattering against soft-gluon emission
in single-particle inclusive
cross sections. The method avoids double counting and conserves
the flow of partonic energy.  It reproduces
threshold resummation for high-$p_T$
single-particle cross sections, when recoil is
neglected, and $Q_T$-resummation at 
low $Q_T$, when higher-order threshold
logarithms are suppressed. We exhibit explicit resummed cross sections, 
accurate to next-to-leading logarithm, for electroweak annihilation and
prompt photon inclusive cross sections.
\end{abstract}

\section{Introduction}

A large class of hard-scattering cross sections in QCD are factorized 
into convolutions of parton distributions and
fragmentation functions with hard-scattering
functions \cite{cssrv}.
Important and representative
examples are Higgs production
and Drell-Yan cross sections, at measured invariant
mass $Q$ and transverse momentum
$Q_T$.  We shall refer to these reactions
collectively as electroweak annihilation.
At fixed, large $Q_T$, electroweak annihilation cross sections
are written in collinear-factorized form as
\be
{d\sigma_{AB\to V}\over dQ^2dQ_T^2}
=
\sum_{ab}
\int dx_a\; \phi_{a/A}(x_a,\mu)\,
\int dx_b\; \phi_{b/B}(x_b,\mu)\;
\hat \sigma_{ab\to V}\left(Q^2/\hat
s,Q_T^2/Q^2,Q^2/\mu^2,\mu,\alpha_s(\mu)\right)\, ,
\label{dyqtcofact}
\ee
in terms of evolved, nonperturbative distributions
(densities)  $\phi_{a/A}(x,\mu)$
of parton $a$ in hadron $A$, and hard-scattering
functions $\hat\sigma_{ab\to V}
\left(Q^2/\hat s,Q_T^2/Q^2,Q^2/\mu^2,\mu,\alpha_s(\mu)\right)$, computed as
power series in $\as$.  Here $\hat s\equiv x_ax_bS$ is
the partonic invariant mass squared, while $\mu$ is a factorization scale,
which for the time being we equate to the renormalization
scale.  General single-particle inclusive (1PI) cross sections for 
photons and light hadrons
at high $p_T$ take a similar form, including a fragmentation function.

Many hard-scattering functions have been computed
to next-to-leading order (NLO) in
$\alpha_s(\mu)$. Analytic calculations of still higher-order
contributions~\cite{VNrev} to $\hat\sigma$ are as yet too complex
to carry out, except for fully inclusive processes,
such as the Drell-Yan production of lepton pairs at
measured invariant mass~\cite{VN}.  Nevertheless, general
arguments show that the functions $\hat\sigma$ are
infrared safe to all orders \cite{cssfact}.

Starting at NLO, the computation of $\hat\sigma$ involves
cancellations between soft gluon emission
and virtual corrections.
These cancellations produce
plus distributions and delta functions,
which require integration
against smooth functions, such as parton densities.
The finite integrals, in turn,
are potential sources of numerically large corrections at each order
in perturbation
theory.  Because of their connection to soft-gluon emission,
however, such corrections can sometimes be resummed
to all orders in perturbation theory.

For example, in Eq.\ (\ref{dyqtcofact}),
$\hat\sigma_{ab\to V}$ includes distributions that are singular
at partonic threshold, $\hat{s}=Q^2$,
where partons $a$ and $b$ have just enough
invariant mass to produce the observed final state.
Defining $z=Q^2/\hat s$, we find at $n$th order
singularities as strong as $\as^n[(1-z)^{-1}\ln^{2n-1}(1-z)]_+$.
Threshold resummations, which organize these distributions, have been 
developed for
a large class of cross sections \cite{dyresumgs,dyresumct,kidrv}.
Although these singularities
are manifest in $\hat\sigma_{ab\to V}$, they
do not generally result in large logarithms in the physical cross section,
because they are smoothed by the integrals over $x_a$ and $x_b$
in Eq.\ (\ref{dyqtcofact}).
Thus, threshold
resummation is not a summation of kinematic logarithms in the 
physical cross section.
It is rather an attempt to quantify the effect on the physical cross 
section of a
well-defined set of  corrections in $\hat\sigma$ to all orders.

Threshold singularities are not the only singular
distributions encountered in the computation of $\hat\sigma$.
In addition, the perturbative cross section
is singular up to $\as^n[(1/Q_T^2)\, \ln^{2n-1}(Q_T^2/Q^2)]_+$ in
$\hat\sigma$, Eq.\ (\ref{dyqtcofact}),
when the  transverse momentum, $Q_T$, of the electroweak boson
is small compared to its mass, $Q$ \cite{ktresumold,cscss}.  
At each order, $Q_T$ is balanced by soft gluons,  and
singularities in the differential
cross section at $Q_T=0$ reflect collinear
divergences in $\hat{\sigma}$ not eliminated by factorization.
These divergences, resummed or not, cancel in the
$Q_T$-integrated cross section, even before
the integrals over the partonic fractions $x_a$ and $x_b$, although
the remainder is still singular at partonic threshold.

Much of the recent interest in soft-gluon recoil
effects has centered on the normalization and $p_T$-dependence of
single-particle inclusive cross sections \cite{owensrv}, particularly
direct-photon production at
fixed-target energies
\cite{dgamexp,kteff1,kteff2,mrstkt,kmrkt,afgkpw,LiLai}.
The  formalism of $Q_T$ resummation
for Eq.\ (\ref{dyqtcofact}) is not immediately applicable to inclusive
high-$p_T$ cross sections, because in this case most of the transverse
momentum of the observed particle is recoil against other
high-$p_T$ particles, while only a  small portion is from
soft radiation. A rough-and-ready approach to soft-gluon radiation is 
to introduce intrinsic transverse momentum for the partons in factorized 
expressions like Eq.\ (\ref{dyqtcofact}), typically in the form of an  
energy-dependent Gaussian smearing of standard parton densities, 
which enhances the cross section. This method, however, certainly involves
double counting, and does not respect the conservation of partonic energy.
Some time ago,
Li and Lai explored the possibility that nonperturbative $k_T$
smearing in high-$p_T$ cross sections has the same origin as
in the low-$Q_T$ Drell-Yan cross sections described by the
$Q_T$-resummation formalism \cite{LiLai}.  More recently,  Li
\cite{Liunified} has shown how
threshold and transverse momentum resummation may be derived from
the same parton distribution, defined in transverse momentum
space, as in Ref.\ \cite{cscss}.

When the conservation of energy is taken into account, however, it
is no longer obvious whether the inclusion of recoil
effects will lead to an enhancement or a suppression, because
the extra radiation involves a number of  
competing effects.   On the one hand, a substantial $k_T$ from
initial-state radiation allows a softer $2\to 2$ subprocess
at the hard scattering, which clearly acts toward enhancement.
On the other hand, the extra energy of the initial-state
radiation drives the physical parton distributions to
larger $x$, which may more than make up for enhancements
in the hard scattering if the distributions are decreasing
with $x$.  At the same time, larger $x$ 
is associated with larger threshold enhancements
in general.   The only way to estimate the
influence of recoil effects on cross sections is to
develop a self-consistent resummation formalism.  

In this paper, we shall
take up, and we hope clarify, this general viewpoint.
Our reasoning is based on a generalization of threshold resummation
which, as we have seen, controls singular distributions at $z=1$.
For electroweak annihilation and single-particle inclusive
cross sections, such contributions are always associated with an underlying
$2\to2$ hard scattering \cite{dyresumgs}.  We use the $2\to 2$ subprocess to 
define the relevant transverse momentum $Q_T$, whose singularities
we resum.  The recoil we discuss below is always 
the recoil of a $2\to 2$ subprocess.
Thus, just as for threshold resummation, we reorganize
a well-defined set of higher order corrections in hard scattering
functions, always working at leading power in the hard scale, $Q$,
within collinear factorization.  
We do not exclude the possibility of nonperturbative
effects, however.  Indeed, we will observe that nonperturbative 
corrections arise quite naturally from our resummed expressions.  A summary
of our results, applied to prompt photon cross sections, was described
in Ref.\ \cite{LSV}.

Let us offer a few additional comments on nonperturbative
effects in these cross sections.  Nonperturbative
effects play a crucial role in the phenomenological
description of electroweak annihilation cross sections
at low $Q_T$, even for a large 
final state mass scale $Q$.  This is the case, even
after the resummation of logarithms of $Q_T/Q$ that
can give a well-defined perturbative prediction for small $Q_T$.
Incorporating nonperturbative
effects, of course, requires the introduction of new
parameters~\cite{ktresumold,yuancollab,evst}.  
In each of these cases, the form of nonperturbative corrections is
suggested by perturbation theory~\cite{powerCS,irrold,nphh}.
In contrast, nonperturbative effects (beyond fragmentation
functions) have not been incorporated in prompt photon
and other single-particle inclusive cross sections, where
there is no need for them at fixed order in perturbation
theory.  For threshold-resummed cross sections the situation
is somewhat more subtle, but ``minimal" formulations of
threshold resummation allow for a class of purely
perturbative predictions~\cite{scalereduce,thrphen},
with no new parameters.  Of course, the existence of such a
formalism does not by itself
preclude the importance of nonperturbative effects.
In this paper, we develop a perturbative formalism that
links both sorts of cross sections, and which is
consistent with known results that have suggested
nonperturbative corrections at measured $Q_T$ in
electroweak annihilation.  Part of our goal is
to open the door, not only to further perturbative
analysis, but also to the study of similarities and
differences in the roles of nonperturbative corrections
in these cases.      

We choose to work in the formalism of collinear factorization
because  we do not wish to
introduce a new set of phenomenological parton distributions, depending on
transverse as well as longitudinal degrees of freedom, except where
absolutely necessary~\footnote{This may well be the case for vector
boson production at low $Q_T$ \cite{yuancollab}. Nevertheless, we
feel that it is important to explore fully the simpler formalism.}.
The resulting combination of threshold and transverse momentum resummations
is at least as technically challenging as NLO factorization,
let alone $Q_T$-resummation,
and the new formalism will require some time to understand and develop as
a practical tool.  We therefore do not attempt to draw immediate
phenomenological
conclusions in this paper.
Instead, we shall concentrate on the formal development,
and (especially in Appendix A) the theoretical
underpinning of these ideas.  We have attempted
to be as explicit as possible in our arguments
and in specifying the functions whose momentum-dependence
controls the set of higher-order corrections that
we study.   This has resulted in a paper
of substantial, although we hope not excessive, length.

We begin in Sec.\ 2 with a treatment of electroweak annihilation
processes, such as Drell-Yan
and Higgs production, whose singular behavior
at vanishing transverse momentum has been
studied intensively over the years \cite{ktresumold,cscss},
and which is in many ways the archetype for resummation.  We show
how to introduce threshold resummation consistently
at measured $Q_T$ for these processes.  
Our approach to resummation is through a ``refactorization"
of partonic cross sections near threshold \cite{kidrv,CLS}.
In this discussion, we shall review the refactorizations
at the basis of $Q_T$ and threshold resummations, and
define a set of new functions which 
control singular behavior in $1-z$ and $Q_T$
jointly.  These will serve 
as building blocks both for electroweak annihilation
cross sections, in Sec.\ 3, and for single-particle
inclusive processes, in Sec.\ 4. 
 
Resummation at threshold and in transverse momentum
is most often formulated in Mellin ($N$) moment
space for the former, and impact parameter ($b$)
space for the latter.  Resummed logarithms of these
parameters exponentiate in the relevant limits,
so that the resummed cross sections are inverse
transforms. In Sec.\ 3 we resum logarithms of $b$ and $N$
in the electroweak annihilation
cross sections.  We begin by deriving 
a relation for the hadronic $d\sigma/dQ^2dQ_T^2$ 
in terms of parton distributions, eikonal cross sections for partons,
and universal anomalous dimensions.  
We observe that this
jointly-resummed cross section determines the pattern of
power corrections in $Q$ and $b$ that are
implied by the behavior of the strong coupling in perturbation theory.
In particular, we find that in QCD such power corrections
appear only at even powers of the invariant
mass $Q$ and impact parameter $b$.  

Sec.\ 4 deals first with prompt photon production, and then with general
high-$p_T$ single-hadron or photon inclusive cross sections.  
For the former, we derive
the joint resummation applied in Ref.\ \cite{LSV}, and for the latter
we discuss the additional resummation associated with fragmentation.

Explicit NLL expressions
for jointly-resummed exponents in electroweak annihilation
and prompt photon production are given in Sec.\ 5,
along with a few comments on the source of enhancement at NLL.
Following our conclusions, we include two appendices. The first
gives the necessary arguments for factorization
and refactorization, and the second gives
explicit one-loop results for some of the functions
that play an important role in the refactorizations
of hard-scattering cross sections.

\section{Refactorization for Electroweak Annihilation}

As above, $Q$ denotes
the mass of an electroweak final state, such
as a vector boson, a Drell-Yan pair or a Higgs boson.
The cross section  $d\sigma/dQ^2dQ_T^2$, at measured $Q^2$ and $Q_T$,
given in factorized form in Eq.\ (\ref{dyqtcofact}),
is singular at $Q_T=0$, order-by-order in perturbation theory.
There are a number of phenomenological applications of 
$Q_T$-resummation for these singularities 
\cite{yuancollab,evst}.  We know of  no simultaneous 
application of threshold resummation, however, although
the cross sections are  singular as well at partonic
threshold, $\hat s=Q$.

As pointed out above, distributions that are singular at threshold are
smoothed in the physical cross section by integration with the parton 
densities.
Nevertheless, there is a good deal to be learned by resumming singular
behavior from the limit $\hat s\to Q^2$, even at fixed, measured $Q_T$
\cite{DDKid}.
To derive a cross section
resummed both in threshold and $Q_T$ variables,
we study partonic cross sections  $a+b\to V+X$
near threshold, where $a$ and $b$ are partons, and where $V$ denotes
the heavy electroweak final state $V={\rm W,Z,H}$, etc.  We begin by
formulating the problem in
a standard form, through collinear factorization.

In this section, which is rather technical in parts, we
lay the groundwork for our derivation of jointly resummed
cross sections.  We have chosen to present our new formalism
in the context of a review of existing resummations, 
for the purpose of motivation, and  also to bring together 
a set of results and methods that are somewhat scattered
in the literature.
It  may be helpful, therefore, to outline the 
contents and aims of the subsections that follow.

We begin (subsection 2.1) by relating the hard-scattering
functions that we will resum to partonic cross
sections.  In subsection 2.2, we review existing refactorizations
for partonic cross sections, which have been used to derive
resummations for electroweak annihilation at low transverse momentum 
\cite{cscss} and at partonic threshold \cite{dyresumgs,dyresumct}.  
We then go on to present a novel refactorization that combines the two 
(Eq.\ (\ref{qtthfact})), and observe how refactorization provides a natural 
formulation of the effects of recoil.
The all-orders justifications for all of these refactorizations
are presented in Appendix A.   Eq.\ (\ref{qtthfact})
involves new perturbative 
functions, denoted ${\cal R}_{i/j}$.  The field-theoretic content
of these functions is the subject of subsection 2.3, which
begins with a review of the analogous definitions for 
light cone parton distributions \cite{matdefs}, as well as the fixed-energy
distributions introduced in Ref.\ \cite{dyresumgs}.  
Each of the refactorization theorems in subsection 2.2 also includes
a function that describes coherent radiation,
which summarizes the interference between emission by incoming
and outgoing hard partons.  An analysis of coherent
radiation is especially important for processes in
which colored particles emerge from the hard scattering \cite{kidrv},
the simplest of which is prompt photon production.  
This interference may be treated in eikonal approximation.
The analysis of the eikonal
approximation to soft gluon radiation is the subject of subsection
2.4, in which various eikonal analogs of the densities in 2.3
are introduced.  Finally (subsection 2.5), we review the use of Mellin and
Fourier transforms to isolate hard-scattering functions.
The new results derived in this section are applied in Sec.\ 3
to electroweak annihilation, and in Sec.\ 4 to single-particle
inclusive cross sections.

\subsection{The hard-scattering function}

Although the hard-scattering function  $\hat\sigma_{ab\to V}$ in
Eq.\ (\ref{dyqtcofact}) is singular at $Q_T=0$, these singularities 
may be determined at the same time as threshold
singularities at $Q^2/\hat s=1$.   To be specific, we shall
derive an expression for $\hat\sigma_{ab\to V}$
in terms of its moments with respect to $z\equiv Q^2/\hat s$,
\be
\hat\sigma_{ab\to V}\left(N,Q_T^2/Q^2,Q^2/\mu^2,\mu,\alpha_s(\mu)\right)
=
\int_0^1 dz\; z^{N-1}\;
\hat \sigma_{ab\to V}\left(z,Q_T^2/Q^2,Q^2/\mu^2,\mu,\alpha_s(\mu)\right)\, ,
\label{tildesigdef}
\ee
where for economy of notation, we denote the transform of
$\hat\sigma_{ab\to V}$ with respect
to $z$ ($Q_T$) by its argument $N$ ($b$).  The hat refers to
its role as a hard-scattering function, from which collinear divergences
are subtracted.
The inverse of the Mellin moment (\ref{tildesigdef}) is, as usual,
\be
\hat \sigma_{ab\to V}\left(z,Q_T^2/Q^2,Q^2/\mu^2,\mu,\alpha_s(\mu)\right)
=
\int_{C-i\infty}^{C+i\infty} {dN\over 2\pi i}\; z^{-N}\;
\hat \sigma_{ab\to V}\left(N,Q_T^2/Q^2,Q^2/\mu^2,\mu,\alpha_s(\mu)\right)\, .
\label{Ninverse}
\ee

We are able to construct $\hat \sigma_{ab\to V}(N)$ because it
also emerges from the factorization of {\it partonic} cross sections with
respect to $\tau\equiv Q^2/S$.
Up to corrections due to parton mixing, which we may neglect to
leading power in
the moment variable, we have, from Eq.\ (\ref{dyqtcofact}),
\be
\hat \sigma_{ab\to V}\left(N,Q_T^2/Q^2,Q^2/\mu^2,\mu,\alpha_s(\mu)\right)
= {1\over \tilde\phi_{a/a}(N+1,\mu)\;  \tilde\phi_{b/b}(N+1,\mu)}\;
\int_0^1 d\tau\; \tau^{N-1} {d\sigma_{ab\to V}\over dQ^2dQ_T^2}
\, .
\label{normalmoment}
\ee
The moments, $\tilde\phi_{i/i}(N+1,\mu)$, 
of the parton-in-parton distributions 
cancel collinear singularities
in the moments of the partonic cross section, and
the right-hand side of this expression
is infrared safe, order-by-order in perturbation theory when
$Q_T\ne 0$.
Our goal now is to determine the singular structure of  $\hat\sigma_{ab\to V}$
at both $z=1$ and at $Q_T=0$.
To control these singularities, we follow
Refs.\
\cite{dyresumgs,qqbaresumKS,qqbaresumBCMN,1pIresumLOS,1pIresumCMN,jetresum},
and
refactorize the (collinear-regularized)
partonic cross section $d\sigma_{ab\to V} /dQ^2 dQ_T^2$ in this limit.
The discussion of the following subsection applies entirely
to these, purely partonic, cross sections.

\subsection{Refactorization and recoil in the partonic cross section}

To motivate the refactorization appropriate to
joint resummation in $Q_T$ and $1-z$, it may be
useful to review the relevant features of the separate
resummation formalisms for  transverse momentum
and threshold.   We will
continue to work in the context of perturbation theory
as in Eq.\ (\ref{normalmoment}), because our
aim is always to analyze higher orders in partonic
hard-scattering functions.   Each of the refactorizations
given below involves the introduction
of new parton distributions, variously at measured
transverse momentum and/or energy fraction.
The new functions are not to be interpreted as physically-accessible
distributions.  Rather,
they are perturbative constructs useful for the analysis
of the hard-scattering functions of Eq.\ (\ref{normalmoment}).

In the formalism of Ref.\ \cite{dyqtfact}, the measured-$Q_T$ 
cross section is written as a  convolution of
(parton-in-parton) distributions ${\cal P}_{i/j}(x,{\bf k})$,
at fixed parton transverse momentum ${\bf k}$,
and light-cone momentum  fraction $x$,
along
with an additional, eikonal function $U_{cd}({\bf q})$
that describes coherent
soft-gluon emission at fixed transverse  momentum,
\ba
{d\sigma_{ab\to V}\over dQ^2 d^2{\bf Q}_T}
&=&  \sum_{cd}\ \sigma_{cd\to V}^{({\rm 0)}}(Q^2) \;
h_{cd}^{\rm (kt)}(\alpha_s(Q))\;
\int dx_a d^2{\bf k}_a\,{\cal P}_{c/a}(x_a,{\bf k}_a,Q)
\int dx_b d^2{\bf k}_b\, {\cal P}_{d/b}(x_b,{\bf k}_b,Q)
\nonumber\\
&\ & \times
\int  d^2{\bf q}\ U_{cd}({\bf q}/\mu,\alpha_s(\mu))\ \delta(Q^2-x_ax_bS)\
\delta^2 \left({\bf Q}_T+ {\bf k}_a+{\bf k}_b+{\bf q}\right) + Y_{\rm kt}\, ,
\label{cssfact}
\ea
where $\sigma^{(0)}$ is the Born total cross section for the process,
for example, $\sigma_{a\bar a\to \gamma^*}^{(0)}=4\pi \alpha^2/3N_C \, Q^2$, 
with $N_C$ the number of colors.
The remainder, $Y_{\rm kt}$, does not diverge as a power at $Q_T=0$.
Note that because the functions ${\cal P}_{i/j}$ are defined at
measured ${\bf k}$, no factorization scale is necessary,
although the distributions still depend on the overall
momentum scale $Q$.  The variable $\mu$ in the arguments of
$U$ is therefore a renormalization scale.  The additive convolution
in this expression implies that the cross section
breaks up into a product under a Fourier transform
to impact parameter ($b$) space \cite{ktresumold}.
The function $h_{cd}^{\rm (kt)}(\alpha_s(Q))=1 + {\cal O}(\alpha_s(Q))$
absorbs hard-gluon corrections that appear in coefficients
of $\delta^2({\bf Q}_T)$.  The combination $\sigma^{(0)}\; h^{\rm (kt)}$
is a truly short-distance function, dominated by lines off-shell
by ${\cal O}(Q^2)$~\cite{ddf}.  
In contrast, the full hard-scattering function
in Eq.\ (\ref{normalmoment}) in general contains lines that
are off-shell only by ${\cal O}(Q_T^2)$.  This  
hierarchy of perturbative scales is characteristic of
resummation. We shall use the term ``short-distance'' to refer
specifically to functions that depend only on the largest
scale, $Q$ in this case. 

The refactorized cross section for threshold resummation,
with integrated ${\bf Q}_T$, 
has many of the same features.  Now, however, the
parton-in-parton distributions $\psi_{i/j}(x)$ are defined at measured
fraction of the energy of parton $j$ (in the center-of-mass
frame for the hard scattering),
rather than light-cone fraction, as is the new eikonal
function $U_{cd}(w_s)$, with total energy $w_sQ$ for
soft gluon radiation into the final state.
Working to leading power in $1-Q^2/S$ leads to important
simplifications.  First,
the nondiagonal parton-in-parton distributions, $\psi_{c/a}(x,Q)$
begin at order $\alpha_s$ with the emission of a soft
fermion (not a pair) into the final state, which results in a
suppression of order $1-x$ in the distribution, and of
$1-Q^2/S$ in the cross section \cite{dyresumgs}.
To leading power, therefore,
we may neglect parton mixing, just as at leading power in
the moments, Eq.\ (\ref{normalmoment}).     The refactorized
expression is \cite{dyresumgs}
\ba
{d\sigma_{ab\to V}\over dQ^2 }
&=&  {1\over S}\; \sigma_{ab\to V}^{({\rm 0)}}(Q^2) \; h_{ab}^{\rm
(th)}(\alpha_s(Q))\
\int dx_a\,\psi_{a/a}(x_a,Q)
\int dx_b\, \psi_{b/b}(x_b,Q)
\nonumber\\
&\ & \hspace{-10mm} \times
\int  dw_s\ U_{ab}(w_sQ/\mu,\alpha_s(\mu))\ \delta\left(1-{Q^2/S} -
(1-x_a)-(1-x_b) - w_s \right)
     + Y_{\rm th}\, ,
\label{thfact}
\ea
where $Y_{\rm th}$ is nonleading by a power of $1-Q^2/S$.
Even though transverse momenta in $\psi$ are integrated, the
phase space for radiation is finite for fixed
parton energy, and $\mu$ again
denotes the renormalization scale.
An explicit  definition of $\psi_{c/a}$ as a matrix element will
be given below.  The remainder, $Y_{\rm th}$,
does not diverge as a power of $1-Q^2/S$ at threshold.  The 
short-distance function $h_{ab}^{\rm (kt)}=1+{\cal O}(\alpha_s)$ 
organizes infrared safe coefficients of $\delta(1-z)$ in this case.

It is most natural to analyze the cross section near threshold,
Eq.\ (\ref{thfact}), in terms of a Laplace transform, $\int d\tau
\exp[-N(1-\tau)]$,
with $\tau=Q^2/S$.
For $N$ large, we can readily relate this Laplace transform to the Mellin
moments in Eq.\ (\ref{normalmoment}).
This follows because generally,
\be
{\rm e}^{-N(1-\xi)}\sim \xi^N \, ,
\label{LapMel}
\ee
with corrections that are suppressed by a power of $N$, and because 
in Eq.\ (\ref{thfact}),
\be
1 - (1-x_a)-(1-x_b) - w_s
\sim
x_ax_b(1-w_s) + {\cal O}\left(\left[1-Q^2/S\right]^2 \right) \, .
\label{threshapprox}
\ee
The Laplace moments of Eq.\ (\ref{thfact}) are therefore equivalent
to its Mellin moments to leading power in $N$, and hence in $1-Q^2/S$.

The close correspondence
between the factorizations at low $Q_T$ and near threshold makes it
rather natural
to combine the two.  We therefore propose a convolution at
fixed transverse  momentum {\em and}  energy fraction:
\ba
{ d \sigma_{ab\to V} \over dQ^2 d^2{\bf Q}_T}
&=&
{1\over S}\; \sigma_{ab\to V}^{({\rm 0)}}(Q^2) \; h_{ab}^{\rm
(j)}(\alpha_s(Q))\;
\int dx_a d^2{\bf k}_a\, {\cal R}_{a/a}(x_a,{\bf k}_a,Q)\;
\int dx_b d^2{\bf k}_b\, {\cal R}_{b/b}(x_b,{\bf k}_b,Q)
\nonumber \\
&\ & \hspace{5mm} \times
\int dw_s d^2{\bf k}_s\, U_{ab}(w_s,Q,{\bf k}_s)\;
\delta(1-Q^2/S-(1-x_a)-(1-x_b)-w_s)
\nonumber \\
&\ & \hspace{5mm} \times
\; \delta^2 \left({\bf Q}_T+ {\bf k}_a+{\bf k}_b+{\bf k}_s\right)
+Y_{\rm j}\, .
\label{qtthfact}
\ea
The short-distance function $h_{ab}^{\rm (j)}(\as)$ is again an
infrared-safe
series in the running coupling, which begins with unity at
zeroth order, and which absorbs, in this case, the coefficients
of $\delta(1-z)\, \delta^2({\bf Q}_T)$ at one loop and beyond.
The remainder $Y_{\rm j}$ is free of power singularities at ${\bf Q}_T=0$  
at leading power in $1-Q^2/S$. As in threshold resummation, only 
flavor-diagonal hard scatterings contribute at ${\cal O}[1/(1-Q^2/S)]$. 
It is important to note that in terms that are {\em not} singular in $Q_T$, 
this leading power emerges only after integration over $Q_T$. This is
because at fixed energy $(1-z) Q$, the phase space in $Q_T$ behaves
as: $\int_0^{Q^2 (1-z)^2} dQ_T^2 = (1-z)^2 Q^2$.   

\begin{figure}[t]
\begin{center}
\hspace*{-7mm}
\epsfig{file=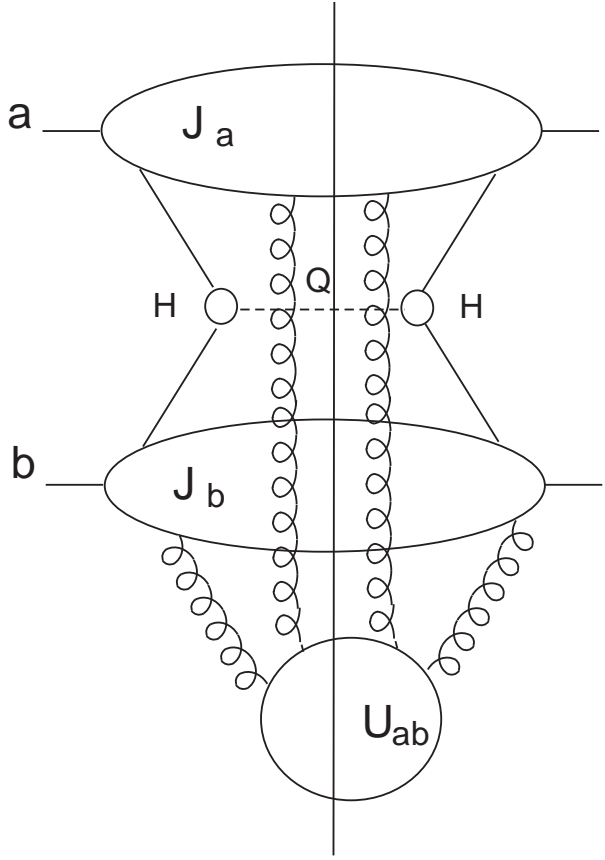,width=8cm}
\end{center}
\caption{Leading region for electroweak annihilation
in cut diagram form.  The vertical line represents
the final state that includes an electroweak boson,
labelled $Q$. The subdiagrams $J_a$, $J_b$, $U_{ab}$ 
and $H$ incorporate, respectively, on-shell lines with momenta
collinear to $p_a$, lines with on-shell momenta parallel to
$p_b$, lines with soft momenta, and lines off-shell by order 
$Q$.}
\label{EWAleadingfig}
\end{figure}

Eq.\ (\ref{qtthfact}),
and indeed each of the refactorizations discussed above, may
be represented as in Fig.\ \ref{EWAleadingfig}.  In the terminology
of Ref.\ \cite{cssfact} and Appendix~A below, Figure~\ref{EWAleadingfig} 
represents the general ``leading regions" in momentum space for this 
cross section. The subdiagrams $J_{a,b}$ include lines collinear to the 
incoming partons, $H$ lines off-shell by order $Q$, and $U$ soft radiation.

The refactorizations of Eqs.\ (\ref{cssfact}) and (\ref{qtthfact})
themselves define the concept of recoil that we will
use in this paper.  The short-distance function 
$\sigma^{(0)}\; h^{\rm (kt)}$ is
computed with on-shell external momenta, collinear to the incoming
lines.   All unintegrated transverse momentum dependence is contained in
the generalized parton densities ${\cal P}$ in (\ref{cssfact}) and ${\cal R}$
in Eq.\ (\ref{qtthfact}).   The dependence of highly off-shell lines on the 
transverse momenta ${\bf k}_a$ and ${\bf k}_b$ of initial-state 
partons is to be absorbed into
higher orders of the short-distance function, by the usual methods of
collinear factorization.  On the other hand, in both transverse 
momentum and joint resummation,
we retain the kinematic linkage of the partonic transverse momentum with
the electroweak final state.  This is what we shall mean by including 
recoil effects.

\subsection{Matrix elements}

The refactorization theorems above, and the resummations derived from them,
involve a number of new functions. We now give explicit definitions
for the various parton distributions, $\cal P$, ${\cal R}$ and $\psi$, 
when the incoming partons are quarks, as well as for the
eikonal functions $U$.  Gluonic distributions can be defined similarly,
following Ref.\ \cite{matdefs}.

The parton densities ${\cal R}_{a/a}$ and the eikonal functions 
$U_{f\bar f}$, defined at fixed energy and transverse momentum are, 
like Eq.\ (\ref{qtthfact}) itself, straightforward variations of functions 
identified for the $Q_T$ and threshold
resummed cases.   The prototype for these expressions is the
partonic light-cone distribution, written as \cite{matdefs}
\be
\phi_{f/f}(x,\mu,\epsilon)
=
\frac{1}{4N_C} \int
{d\lambda\over 2\pi}\; {\rm e}^{-i\lambda xp^+ }\;
\langle f(p)| \bar q_f\big(\lambda u) \gamma\cdot u\; q_f(0) |f(p)\rangle\, ,
\label{phidef}
\ee
where
$\mu$ is the scale at which the product of quark fields, which are connected
by a lightlike separation, $\lambda u^\mu$, $u^2=0$, is renormalized.
An average over colors and spins is included in the definition.
In this expression, we have suppressed an ordered exponential, 
$\Phi_u^{(q)}\left(\lambda,0;0 \right)$, which we shall also refer to 
as a nonabelian phase line, of the gauge field along the light 
cone vector between the quark fields, in the notation
\be
\Phi_{\beta}^{(f)}(\lambda_2,\lambda_1;X)
=
P\exp\left(-ig\int_{\lambda_1}^{\lambda_2}d{\eta}\;
{\beta}{\cdot} A^{(f)} ({\eta}{\beta}+X)\right)\, .
\label{Phidef}
\ee
Here the gauge field is a matrix in the representation
of parton $f$.
In momentum space, these operators correspond to eikonal
lines. Equivalently, we may define the matrix element (\ref{phidef})
in $u\cdot A=0$ gauge. The perturbative $\overline{\rm MS}$
distribution, computed as a power series in $\alpha_s(\mu)$, is
independent of the momentum $p^\mu$; it is a ``pure counterterm",
that is, a series of poles in $\epsilon=2-D/2$ in $D$ dimensions,
a dependence which we exhibit among its
arguments.  

We may regard perturbative $\MS$ distributions as
defined by their evolution equations, which in moment space are 
\be
\mu^2 {d\over d \mu^2}\, \tilde \phi_{f/f}(N,\mu,\epsilon)
=
\gamma_{ff}\left( N,\as(\mu)\right)\;
\tilde \phi_{ff}(N,\mu,\epsilon)+{\cal O}(1/N)\, ,
\label{evolmsbar}
\ee
with $\gamma_{ff}\left( N,\as,\epsilon\right)
= \int_0^1 dx x^{N-1} P_{ff}(x,\as)$
the moments of the splitting function for flavor $f$.
As usual, up to corrections of order $1/N$, we may
neglect flavor mixing.
     A very useful explicit form for the $\MS$ distributions
is found by solving this equation, with the boundary
condition $\tilde\phi(N,\mu=0,\epsilon)=1$,
\be
\tilde \phi_{f/f}(N,\mu,\epsilon)
=
\exp\left[ \int_0^{\mu^2}{d\mu'{}^2\over \mu'{}^2}\; \gamma_{ff}
\left( N,\as(\mu')\right)  \right]\, .
\label{msbarexp}
\ee
This expression is meaningful for the collinear-regularized distribution,
defined for $D>4$, or equivalently $\epsilon<0$, because of the 
$\epsilon$-dependence
of the strong coupling.  The $\mu'$ integral is regulated by
reexpressing $\as(\mu')$ in terms of the strong coupling evaluated at the
fixed scale $\mu$: $\as(\mu)=(\mu'/\mu)^{2\epsilon} \as(\mu') + \dots$.
The $\MS$ prescription then consists of reinterpreting the upper limit
as $\mu_F^2\ln(4\pi{\rm e}^{-\gamma}_E)$,
with $\gamma_E$ the Euler constant.  We shall not generally exhibit this
modification below, nor indicate explicitly the $\epsilon$-dependence
of the coupling.

To leading power in $N$, $\gamma_{ff}$ is found
from moments of the expansion
     \be
P_{ff}(z)
=
A_f(\as)\, \left[{1\over 1-z}\right]_+
+B_f(\as)\delta(1-z) + {\cal O}\left([1-z]^0\right)\, .
\label{splitAB}
\ee
The plus distribution appears only as a
power series in $\as$ times $[1/(1-z)]_+$.
It is worth noting here that the form of Eq.\ (\ref{splitAB}), with
no explicit powers of $\ln(1-z)$ in the plus distribution, is required 
by collinear factorization, and is not an additional assumption \cite{GK89}.  
We shall see this result emerge below in Section~3. A similar observation
was made recently by Albino and Ball~\cite{alba}.

For NLL expansions, we will need the anomalous dimensions $A_a(\as)$  to
two loops.  For flavor $a$, they are given by the familiar expansion,
$A_a(\as)=\sum_n\, (\as/\pi)^n\; A_a^{(n)}$, with\footnote{The
function $A_a$ 
{\protect{\cite{KoTr}}} is proportional to $\Gamma_{\rm cusp}$ in Refs.\
{\protect{\cite{KR,KorSt95}}}.}
\ba
A_a^{(1)} &=& C_a \nonumber \\
A_a^{(2)} &=& \frac{1}{2} C_a K \equiv \frac{1}{2} C_a \left[ C_A \left(
\frac{67}{18} - \frac{\pi^2}{6} \right) -\frac{10}{9} T_R N_f \right]\, ,
\label{explicitA}
\ea
where $C_q=C_F$, $C_g=C_A$. To lowest order, which is the accuracy 
necessary for NLL, we have $B_a = (\as/\pi) B_a^{(1)}$, 
where $B^{(1)}_q$ and $B^{(1)}_g$ are given by
\be
B^{(1)}_q = \frac{3}{4} C_F \; , \quad
B^{(1)}_g= {\beta_0\over 4} \, ,
\label{explicitB}
\ee
with $\beta_0=11N_C/3-2n_f/3$, the lowest-order coefficient of the
QCD beta function.

Matrix element representations of the functions ${\cal P}_{f/f}$
are similar \cite{matdefs},
\be
{\cal P}_{f/f}(x,{\bf k},p\cdot n,\epsilon)
=
\frac{1}{4N_C} \int
{d\lambda\over 2\pi}\; \; {d^2{\bf b}\over (2\pi)^2}\; {\rm
e}^{-i\lambda xp\cdot u+i{\bf b}\cdot {\bf
k}}\;
\langle f(p)| \bar q_f\big(0^+,\lambda,{\bf b}\big) \gamma\cdot u\;
q_f(0) |f(p)\rangle\, .
\label{Pdef}
\ee
This matrix element is defined in an axial, $n\cdot A=0$ gauge, which
is how it acquires $p\cdot n$ dependence.   It also
requires collinear regularization in perturbation theory.

The densities $\psi_{f/f}$ are the distributions of quarks of fixed
energy $p_0=Q/2x\sim Q/2$,
in the center-of-mass of the produced pair,
while ${\cal R}_{f/f}(x,{\bf k},p^0)$ are distributions in energy
{\em and} transverse
momentum ${\bf k}$.  The external line is an on-shell quark of four-momentum
$p^\mu=(Q/x\sqrt{2}) v^\mu$,
with $v^2=0$.
The inclusive energy distribution $\psi_{f/f}$ is then
given by
\be
\psi_{f/f}(x,2p_0,\epsilon)
=
\frac{1}{2N_C}\frac{p_0}{2 p\cdot u} \int
{d\lambda\over 2\pi} \;  {\rm e}^{-i\lambda xp_0}\;\langle
f(p)| \bar q_f (\lambda \hat n) \; \gamma\cdot u \; \;  q_f(0)
|f(p)\rangle\, ,
\label{psidef}
\ee
where $q_f(x)$ is the field for flavor $f$,
$u^\mu$ is the light-cone unit vector opposite to $v^\mu$, $u\cdot
v=1$, and $\hat n^\mu$ is the unit vector in the time direction,
$\hat n^\mu=(1,{\vec 0})$.
Following Ref.\ \cite{dyresumgs}, we evaluate the matrix element \re{psidef} in
$\hat n\cdot A=A^0=0$ gauge in the center-of-mass frame,
which turns out to be convenient for calculational purposes.  Were
$\psi_{f/f}$ defined in a spacelike axial gauge, it would differ
only by finite corrections. The operator product separated by a timelike 
distance requires no new renormalization.
Correspondingly, the functions ${\cal R}_{q/q}$ 
may be defined as matrix elements by
\be
{\cal R}_{f/f}(x,{\bf k},2p_0,\epsilon)
=
\frac{1}{2N_C}\frac{p_0}{2 p\cdot u} \int
{d\lambda\over 2\pi}\; {d^2{\bf b} \over (2\pi)^2}\; {\rm
e}^{-i\lambda xp_0 +i{\bf b}\cdot {\bf
k}}\;
\langle f(p)| \bar q_f\big(\lambda \hat n+{\bf b}\big) \gamma\cdot u
q_f(0) |f(p)\rangle\, ,
\label{Rdef}
\ee
again evaluated in $A^0=0$ gauge.  In these expressions,
and in the remainder of this section, we suppress
dependence on the renormalization scale, which we take equal
to $Q=2p_0$.  We will return to the choice of renormalization
scale later.

\subsection{Eikonal functions and factorizations}
\label{sec:eikon-funct-fact}

Near partonic threshold, all radiation is soft,
compared to the hard scattering function.
It is thus natural to study the
eikonal approximation for the
cross section and for the factorizations
that characterize the dynamics.
The discussion below follows Refs.\ \cite{jetresum,jccsud}.

The eikonal cross section is built from ordered
exponentials, $\Phi_{\beta}^{(f)}(0,-\infty;X)$,
of the gauge field in the group representation
of the incoming partons, extending from minus
infinity to the point of annihilation, in the notation
of (\ref{Phidef}).  We introduce a product that
represents the annihilating combination of two nonabelian phase operators:
\be
{\cal W}^{(c\bar d)}(X)
=
\Phi^{(\bar d)}_{\beta'}\left(0,-\infty;X\right)\;
\Phi^{(c)}_{\beta}\left(0,-\infty;X\right)\, ,
\label{Wdef}
\ee
where for quarks, $c$ and $\bar d$ may carry different
flavors, as in the case of $u+\bar d\to {\rm W}^+$. From 
the operators ${\cal W}^{(c\bar d)}$,  we define an eikonal cross section
at fixed energy and transverse momentum,
which represents the QCD radiation generated by the
annihilation of the two incoming color sources,
neglecting recoil,
\ba
\sigma^{\rm (eik)}_{c \bar d}(\xi, Q,{\bf k},\epsilon)
&=&
Q\, \int
{d\lambda \over 2\pi} \; {d^2{\bf b}\over (2\pi)^2}\ {\rm
e}^{-i\lambda \xi Q/2+i{\bf b}\cdot {\bf k}}\;\nonumber\\
&\ &  \hspace{-20mm} \times\; {1\over d(c)}\, {\rm Tr}\
\langle 0|\,  \bar {\rm T}\, \left[{\cal W}^{(c\bar d)}(0)^\dagger]\;
    {\rm T}\, \left[ {\cal W}^{(c\bar d)}(\lambda \hat n+{\bf
b}\right)\right]\, |0\rangle\, .
\label{sigeikdef}
\ea
As above $\hat n^\mu\equiv\delta_{\mu 0}$, so that
     $\lambda \hat n+{\bf b}$ represents the vector with time
component $\lambda$ and transverse components $\bf b$.
In the matrix element, T represents time order and $\bar {\rm T}$
anti-time order.
The trace is over color indices in the representation of parton $c$.
$d(c)$ is the dimension of this representation\footnote{The eikonal 
cross section defined here is normalized to
{\protect $\delta(\xi)\delta^2({\bf k})$} at zeroth order.
The average over the colors of the physical incoming partons
will be absorbed into a separate overall factor.}.
Because the velocities $\beta$ and $\beta'$ of the incoming lines are
lightlike, this cross section has collinear singularities, and must be
regulated.  Infrared divergences, however,
cancel in the sum over final states \cite{cssfact}.

To organize collinear singularities, we introduce eikonal
parton distributions, which approximate the radiation
at fixed energy and transverse momentum from an energetic, lightlike parton,
\ba
{\cal R}_a^{({\rm eik})}(w,Q,{\bf k},\epsilon)
&=&
\frac{Q}{2\,d(a)}
\, \int {dy_0\over 2\pi}\; {d^2{\bf b}\over (2\pi)^2}\ {\rm
e}^{-iwQy_0/2+i{\bf b}\cdot {\bf k}}
\nonumber\\
&\ &  \hspace{-10mm} \times\; {\rm Tr}\
\langle 0|\Phi_\beta^{(a)}{}^\dagger\left(0,-\infty;0\right)\;
\Phi_\beta^{(a)}(0,-\infty;y_0 \hat n+{\bf b})|0\rangle\, ,
\label{jindef}
\ea
computed in the $\hat{n}
\cdot A=0$ gauge, just as ${\cal R}_{f/f}$, Eq.\ (\ref{Rdef}).
Similarly, by analogy to Eq.\ (\ref{phidef}), we
can construct an eikonal distribution
at fixed light-cone momentum fraction,
\ba
\phi_a^{({\rm eik})}(\xi,\mu,\epsilon)
&=&
\frac{Q}{\sqrt{2}\,d(a)}\, \int {d\lambda\over 2\pi}\ {\rm e}^{-i\xi
Q\lambda/\sqrt{2}}
\nonumber\\
&\ &  \hspace{5mm} \times\; {\rm Tr}\
\langle0|\Phi_\beta^{(a)}{}^\dagger(0,-\infty;0)\;
\Phi_\beta^{(a)}(0,-\infty;\lambda v)|0\rangle\, ,
\label{phieikdef}
\ea
which as usual requires renormalization of its ultraviolet divergences, and
regularization for its collinear divergences.  As in Eq.\ (\ref{phidef}),
we omit the ordered exponential in the opposite-moving light cone
direction, between $0$ and $\lambda v$.  Also like the $\overline{\rm MS}$
distribution, $\phi_{f/f}$, $\phi_a^{({\rm eik})}$ is a pure
counterterm, and is independent of
the momentum scale $Q$ and of the direction of $\beta$.  It is also
flavor-independent
among quarks and antiquarks, differing, of course, for gluons.
Note that $\xi$ in $\phi_a^{({\rm eik})}$ plays the role of $1-x$
in $\phi_{f/f}$. That is, we fix the light-cone component
of the emitted radiation, since the eikonal line does not
have a definite initial-state momentum.

Purely virtual diagrams in both
${\cal R}^{({\rm eik})}$
and $\phi^{({\rm eik})}$ enter as overall factors, which can be used
to normalize
these functions.
We choose to define the virtual contributions by the requirements that
\begin{equation}
\int_0^1 dw\; \int d^2{\bf k}\; 
{\cal R}_a^{({\rm eik})}(w,Q,{\bf k},\epsilon) =
\int_0^1 d\xi\; \phi_a^{({\rm eik})}(\xi,\mu,\epsilon)
=1\, .
\label{eikdistnorm}
\end{equation}
These conditions ensure that both functions are sums
of plus distributions in terms of the variables $\xi$ or $w$, integrated
over the interval from zero to unity.  This choice does not affect
the $N$-dependence
of the functions at all, but ensures that factorization does not
introduce spurious collinear singularities.
This condition also enables us to define an evolution equation
for the eikonal light-cone distribution of the form of
Eq.\ (\ref{evolmsbar}), with solution
\be
\tilde \phi_{f}^{\rm (eik)}(N,\mu,\epsilon)
=
\exp\left[ \int_0^{\mu^2}{d\mu'{}^2\over \mu'{}^2}\; \gamma_{ff}^{\rm
(eik)} \left( N,\as(\mu') \right)  \right]\, ,
\label{eikmsbarexp}
\ee
which differs from (\ref{msbarexp}) only in the eikonal approximation
to the anomalous dimension.
The eikonal anomalous dimensions, $\gamma_{ff}^{\rm (eik)}$, are found from
the plus distributions of the
splitting functions, when written as in Eq.\ (\ref{splitAB}), subject
to the normalization
condition (\ref{eikdistnorm}).
To leading power in $N$,
the moments of the eikonal distribution in $D$ dimensions are given
by
\be
\tilde \phi_{f}^{\rm (eik)}(N,\mu,\epsilon)
=
\exp\left[ - \ln \bar N\ \int_0^{\mu^2}{d\mu'{}^2\over \mu'{}^2}\;
A_f \left(\as(\mu') \right)  \right]\, ,
\label{eikmsbarexp2}
\ee
where we define
\be
\bar N \equiv N{\rm e}^{\gamma_E}\, .
\label{barNdef}
\ee
  The discussion on the
dimensional regularization
and $\MS$ definition of $\phi_{a/a}(N,\mu,\epsilon)$ given after Eq.\
(\ref{msbarexp})
applies as well to its eikonal analog in Eq.\ (\ref{eikmsbarexp}).

Essentially the same arguments  (see Appendix A) for the joint factorization of
the partonic cross section are valid for the eikonal
cross section, $\sigma^{\rm (eik)}_{a\bar b}$.
We may therefore factorize $\sigma^{\rm (eik)}_{a\bar b}$,
in terms of energy and transverse momentum distributions, as in Eq.\
(\ref{qtthfact}),
\ba
\sigma^{\rm (eik)}_{c d}(\xi, Q,{\bf k},\epsilon)
&=& \nonumber \\
&\ & \hspace{-25mm}
\int dw_c d^2{\bf k}_c\,{\cal R}_{c}^{({\rm eik})}(w_c,Q,{\bf k}_c,\epsilon)\;
\int dw_d d^2{\bf k}_d\, {\cal R}_{d}^{({\rm eik})}(w_d,Q,{\bf k}_d,\epsilon)
\nonumber \\
&\ & \hspace{-25mm}  \times
\int dw_s d^2{\bf k}_s\ U_{c d}(w_s,Q,{\bf k}_s)\
\delta(\xi-w_c-w_d-w_s)\;
\; \delta^2 \left({\bf k}+ {\bf k}_c+{\bf k}_d+{\bf k}_s\right)\, .
\nonumber\\
\label{sigeikUfact}
\ea
Alternately, we may factorize the eikonal
cross section in terms of eikonal light-cone distributions,
\ba
\sigma^{\rm (eik)}_{cd}(\xi, Q,{\bf k},\epsilon)
&=&
\int dw_c\,\phi_{c}^{({\rm eik})}(w_c,\mu,\epsilon)\;
\int dw_d\, \phi_{ d}^{({\rm eik})}(w_d,\mu,\epsilon)
\nonumber\\
&\ & \quad \times \int d\xi'\; \delta(\xi-w_c-w_d-\xi')\; \hat
\sigma^{\rm (eik)}_{cd}(\xi', Q,{\bf k})\, ,
\label{sigeikphifact}
\ea
where partons $c$ and $d$ are implicitly in a color singlet state.
This is the eikonal approximation to
Eq.\ (\ref{dyqtcofact}) for incoming partons $c$ and $\bar d$,
identifying $w_i\sim 1-x_i$, $\xi\sim 1-Q^2/S$ and $\xi'\sim 1-Q^2/\hat s$.
In Eq.\ (\ref{sigeikUfact}) and (\ref{sigeikphifact}), respectively,
the  ${\cal R}$'s and $\phi$'s absorb collinear singularities associated
with soft gluons.
The remaining functions $U_{c d}$ and $\hat \sigma_{c d}^{\rm (eik)}$
are then infrared safe.

All of the refactorizations in Eqs.\ (\ref{cssfact}),
(\ref{thfact}) and (\ref{qtthfact}) involve functions that are
gauge-dependent.  We have already noted the
gauge-dependence of the functions ${\cal R}$ and ${\cal R}^{\rm (eik)}$
above.  The soft function $U_{ab}$ also inherits gauge
dependence through ${\cal R}^{\rm (eik)}$.  To the extent that
the factorization formulas are valid, however, all
gauge dependence is guaranteed to cancel in the
cross sections.  In Appendix A we study the theoretical basis
of these refactorizations; for the purposes of
the following discussion, we accept their validity.

\subsection{Transforms and the soft function}

As usual, the refactorized cross sections are displayed
most conveniently in terms of their appropriate transforms,
in this case, Laplace and Fourier.  Transforms
that we will need below are:
\ba
\bar {\cal R}_{f/f}(N,{\bf b}Q,\epsilon)
&=&
\int_0^\infty dx\; {\rm e}^{-N(1-x)}\, \int {d^2{\bf k}}\, {\rm
e}^{-i{\bf b}\cdot {\bf k}}\
{\cal R}_{f/f}(x,Q,{\bf k},\epsilon)
\nonumber\\
\bar {\cal R}_{a}^{({\rm eik})}(N,{\bf b}Q,\epsilon)
&=&
\int_0^\infty dw\; {\rm e}^{-Nw}\, \int {d^2{\bf k}}\, {\rm
e}^{-i{\bf b}\cdot {\bf k}}\
{\cal R}_a^{({\rm eik})}(w,Q,{\bf k},\epsilon)
\nonumber\\
\bar U_{c\bar d}(N,{\bf b}Q) &=&
\int_0^\infty dw\; {\rm e}^{-Nw}\, \int {d^2{\bf k}}\, {\rm
e}^{-i{\bf b}\cdot {\bf k}}\
U_{c\bar d}(w_s,Q,{\bf k}) \nonumber\\
\bar \sigma^{\rm (eik)}_{c\bar d}(N,{\bf b}Q,\epsilon)
&=&
\int_0^\infty d\xi\; {\rm e}^{-N\xi}\, \int {d^2{\bf k}}\, {\rm
e}^{-i{\bf b}\cdot {\bf k}}\
\sigma^{\rm (eik)}_{c\bar d}(\xi,Q,{\bf k},\epsilon)\, .
\label{doubletrans}
\ea
These transforms simplify the double convolutions of the
partonic cross section, Eq.\ (\ref{qtthfact}).  On the other hand,
the moments of the hard scattering
functions of Eq.\ (\ref{dyqtcofact}) are determined from
the partonic cross sections via Eq.\
(\ref{normalmoment}).   In this way, we find
\ba
\hat  \sigma_{a  b\to V}\left(N,Q_T^2/Q^2,Q^2/\mu^2,\mu,\alpha_s(\mu)\right)
&=&
     \sigma_{a  b\to V}^{(\rm{H)}}(Q^2)\;
{1 \over
\tilde\phi_{a/a}(N,\mu)\;  \tilde\phi_{b/b}(N,\mu)}
\nonumber \\
&\ & \hspace{-60mm} \times
\int{d^2 {\bf b} \over (2\pi)^2}\; {\rm e}^{i{\bf b}\cdot {\bf Q}_T}\;
\bar {\cal R}_{a/a}(N,{\bf b}Q,\epsilon)\;
\bar {\cal R}_{ b/ b}(N,{\bf b}Q,\epsilon)\;  \bar U_{ab}(N,{\bf b}Q)
\, ,
\label{qttransform}
\ea
where we define a combination of short-distance function and Born
cross section as
\be
\sigma_{ab\to V}^{({\rm H)}}(Q^2) 
\equiv
\pi\
\sigma_{ab\to V}^{(0)}(Q^2)\; h_{ab}^{\rm
(j)}(\as(\mu))\, .
\label{sigHsigB}
\ee
The factor $\pi$ relates the azimuthally symmetric $d\sigma/dQ^2d^2{\bf Q}_T$
to $d\sigma/dQ^2dQ_T^2$, while the Born cross section 
$\sigma_{ab\to V}^{(0)}$ absorbs the color average for the initial-state
partons, referred to above.  In Eq.\ (\ref{qttransform}), we have
approximated $N+1$ by $N$ in the arguments of the light-cone
distributions, as is acceptable to leading power in $N$.  Again,
we suppress the renormalization scale $\mu=Q$; the explicit $\mu$
here has the interpretation of a factorization scale.

Another useful form of Eq.\ (\ref{qttransform}) is
\ba
\hat  \sigma_{a  b\to V}\left(N,Q_T^2/Q^2,Q^2/\mu^2,\mu,\alpha_s(\mu)\right)
&\ &
\nonumber \\
&\ & \hspace{-45mm}  =
\hat \sigma_{a b\to V}^{\rm (H)}(Q^2)\;
\int {d^2 {\bf b} \over (2\pi)^2}\; {\rm e}^{i{\bf b}\cdot {\bf
Q}_T}\; \bar c_{a/a}(N,{\bf b},Q,\mu)\;
\bar c_{b/b}(N,{\bf b},Q,\mu) \, ,
\label{qtcform}
\ea
where the functions $c$ and their eikonal analogs are defined by
\ba
\tilde c_{f/f}(N,{\bf b},Q,\mu)
&=&
{\bar {\cal R}_{f/f}(N,{\bf b}Q,\epsilon)\, \left[\, 
\bar U_{f\bar f}(N,{\bf b}Q)\, \right]^{1/2}
\over \tilde\phi_{f/f}(N,\mu,\epsilon)}\, ,
\nonumber\\
\tilde c_{f}^{\rm (eik)}(N,{\bf b},Q,\mu)
&=&
{\bar {\cal R}_{f}^{\rm (eik)}(N,{\bf b}Q,\epsilon)\, \left[\, \bar 
U_{f\bar f}(N,{\bf b}Q)\, \right]^{1/2}
\over \tilde\phi_{f}^{\rm (eik)}(N,\mu,\epsilon)}\, .
\label{Cfact}
\ea
The $c$'s will appear as building blocks in direct photon and other
cross sections with color flow into the final state 
as part of the hard scattering.

The eikonal cross section has many of the same properties
as its partonic counterpart.
Moments of the eikonal hard-scattering function at fixed $\bf k$
are found from (\ref{sigeikphifact}):
\be
\hat \sigma^{\rm (eik)}_{c d}(N,Q,{\bf k},\mu)
=
{ \tilde \sigma^{\rm (eik)}_{c d}(N,Q,{\bf k},\epsilon)
\over
\tilde\phi_{c}^{\rm (eik)}(N,\mu,\epsilon)\, \tilde\phi_{d}^{\rm
(eik)}(N,\mu,\epsilon)}\, ,
\label{hatsigeikratio}
\ee
where to avoid unnecessary clutter in our notation, we identify the
transforms of the functions $\hat\sigma$ only through its
arguments.  Then, using the eikonal transforms in Eq.\ (\ref{doubletrans})
in the eikonal joint convolution (\ref{sigeikUfact}), we have
\ba
\hat \sigma^{\rm (eik)}_{c d}(N,Q,{\bf k},\mu)
&=&
{1 \over
\tilde\phi_{c}^{({\rm eik})}(N,\mu,\epsilon)\;  \tilde\phi_{d}^{({\rm
eik})}(N,\mu,\epsilon)}
\nonumber \\
&\ & \hspace{-45mm} \times
\int{d^2 {\bf b} \over (2\pi)^2}\; {\rm e}^{i{\bf b}\cdot {\bf k}}\;
\bar {\cal R}_c^{({\rm eik})}(N,{\bf b}Q,\epsilon)\;
\bar {\cal R}_d^{({\rm eik})}(N,{\bf b}Q,\epsilon)\;  
\bar U_{c d}(N,{\bf b}Q)\, ,
\label{eikqttransform}
\ea
with the same coherent function $U_{c\bar d}$.

For completeness, and for reference below, we observe that the 
eikonal function $U_{ij}$ at measured energy and transverse momenta 
may be defined by its transforms, through
\begin{equation}
\bar U_{c d}(N,{\bf b}Q) =
{\bar \sigma^{\rm (eik)}_{c d}(N,{\bf b}Q,\epsilon)
\over
\bar {\cal R}_{c}^{({\rm eik})}(N,{\bf b}Q,\epsilon)\
\bar {\cal R}_{ d}^{({\rm eik})}(N,{\bf b}Q,\epsilon)
}\, ,
\label{Umoment}
\end{equation}
where collinear singularities cancel in the ratio.
The other forms of the soft function in Eqs.\ (\ref{cssfact}) and
(\ref{thfact})
differ only in the components
of the total final-state momentum that are fixed. Note that this expression
is essentially a rewriting of the refactorization for the eikonal
cross section, Eq.~(\ref{eikqttransform}).

The behavior of the hard scattering function at large moment $N$
and impact parameter ${\bf b}$ may be studied either in terms
of $U$ and the distributions ${\cal R}$, or, as we see in the
next section, by relating the partonic and eikonal
functions given in Eqs.\ (\ref{qttransform}) and (\ref{eikqttransform}).
In the remainder of the paper, we apply this formalism
to derive our jointly resummed cross sections.

\section{Joint Resummation for Electroweak Annihilation}

As pointed out in Refs.\ \cite{CLS,Gatheral,cttwcls}, 
color-singlet cross sections with symmetric phase space exponentiate at
high moments that force the phase space to an ``elastic"
limit, where only soft gluon radiation is allowed.
This is the case for doubly-transformed
cross sections, and for the singly-transformed
cross section in threshold resummation \cite{dyresumgs,dyresumct}.
The elastic limit is naturally associated with the eikonal approximation.
As we now show, the full leading-power
$N$- and $b$-dependence of electroweak annihilation
cross sections can be deduced for 
quark-antiquark annihilation and gluon fusion
directly from eikonal cross sections.

\subsection{Partonic and eikonal cross sections}

Near threshold,
that is, to leading power in $N$, all real-gluon emission
in the partonic hard-scattering function, Eq.\ (\ref{qttransform})
may be treated in eikonal approximation, Eq.\ (\ref{eikqttransform}).  Since
the function $U$ is the same in the partonic cross
section and its eikonal approximation, the difference,
for fixed ${\bf b}$, resides
entirely in the parton distributions, and we have
\be
{\bar {\cal R}_{a/a}(N,{\bf b}Q,\epsilon)\;
\bar {\cal R}_{b/b}(N,{\bf b}Q,\epsilon)
\over
\tilde\phi_{a/a}(N,\mu,\epsilon)\;  \tilde\phi_{b/b}(N,\mu,\epsilon)}
=
V_{ab}(Q,\mu)\
{\bar {\cal R}_{a}^{({\rm eik})}(N,{\bf b}Q,\epsilon)\
\bar {\cal R}_{b}^{({\rm eik})}(N,{\bf b}Q,\epsilon)
\over
\tilde\phi_{a}^{\rm (eik)}(N,\mu,\epsilon)\;  \tilde\phi_{b}^{\rm
(eik)}(N,\mu,\epsilon)}\, ,
\label{Vvirt}
\ee
where $V_{ab}(Q,\mu)$ is an overall factor, entirely from virtual corrections,
which are of the same graphical form in ${\cal R}$ and $\phi$.
The function $V_{ab}(Q,\mu)$ is therefore independent of ${\bf b}$ and $N$
to order $N^0$;
its $Q$ and $\mu$ dependence may be determined as follows.

As shown explicitly in Eq.\ (\ref{msbarexp}) and (\ref{eikmsbarexp}) above,
light cone distributions and their eikonal approximations  in the
$\MS$ scheme are fully
determined by the splitting functions.
In moment space,  the
leading power in $N$ comes entirely from the
transforms of $[1/(1-z)]_+$ and $\delta(1-z)$ contributions to the
splitting functions of Eq.\ (\ref{splitAB}).
The former, which can only arise from the combination of real-gluon and virtual
corrections, are fully represented in the eikonal distributions,
$\phi^{\rm (eik)}$.  As a consequence, to leading power,
the ratios $\phi_a^{\rm (eik)}(N,\mu)/\phi_{a/a}(N,\mu)$ only
receive contributions
from the left-over $\delta(1-z)$ terms in the splitting functions,
\be
{\tilde \phi_{a}^{\rm (eik)}(N,\mu_F,\epsilon)\, \tilde \phi_{b}^{\rm
(eik)}(N,\mu_F,\epsilon) \over
\tilde \phi_{a/a}(N,\mu_F,\epsilon)\, \tilde \phi_{b/b}(N,\mu_F,\epsilon)}
=
\exp\; \left\{ -\int_0^{\mu_F^2} {d\mu'{}^2\over \mu'{}^{2+2\epsilon}}\;
{\left[B_a(\as(\mu'))+B_b(\as(\mu'))\right]}
\right\}\, ,
\label{phiratio}
\ee
with $\mu_F$ the factorization scale, with explicit dimensional
regularization, and with
the functions $B_a$ given by Eq.\ (\ref{explicitB}).
The collinear divergences in this expression
cancel in the ratio in Eq.\ (\ref{Vvirt}).  The ratio of the
${\cal R}$-functions must thus
take the same form at leading power in $N$, but with an upper limit
on the $\mu'$-integral given by the renormalization scale in ${\cal R}$,
which, as above, we choose to be $Q$,
\be
{\bar {\cal R}_a^{({\rm eik})}(N,{\bf b}Q,\epsilon)\
\bar {\cal R}_b^{({\rm eik})}(N,{\bf b}Q,\epsilon) \over
\bar {\cal R}_{a/a}(N,{\bf b}Q,\epsilon)\;
\bar {\cal R}_{b/b}(N,{\bf b}Q,\epsilon)}
=
\exp\; \left\{ -\int_0^{Q^2} {d\mu'{}^2\over \mu'{}^{2+2\epsilon}}\;
{\left[B_a(\as(\mu'))+B_b(\as(\mu'))\right]}
\right\}\, .
\label{Rratio}
\ee
As a result, the uniquely determined
form of $V_{ab}(Q,\mu)$ is
\be
 V_{ab}(Q,\mu_F)
= \exp \; \left\{ \int_{\mu_F^2}^{Q^2} {d\mu'{}^2\over \mu'{}^{2}}\;
{\left[B_a(\as(\mu'))+B_b(\as(\mu'))\right]}
\right\}\, .
\label{Vexp}
\ee
For electroweak annihilation, of course,
$B_a=B_b$, but the
result is quite general.

We are now ready to combine Eqs.\ (\ref{qttransform}) for the partonic
hard-scattering function, Eq.\ (\ref{eikqttransform}) for its eikonal
approximation, and Eq.\ (\ref{Vexp}) for the ratio $V$, to
derive an expression for the refactorized electroweak
cross section that we will study below.
The result is:
\ba
\hat  \sigma_{ab\to V}\left(N,Q_T^2/Q^2,Q^2/\mu_F^2,\mu_F,\alpha_s(\mu)\right)
&\ &
\nonumber \\
&\ & \hspace{-60mm} = \sigma_{ab\to V}^{(\rm{H)}}(Q^2)\;
\exp \; \left\{ \int_{\mu_F^2}^{Q^2} {d\mu'{}^2\over \mu'{}^{2}}\;
{\left[B_a(\as(\mu'))+B_b(\as(\mu'))\right]}
\right\}
\nonumber \\
&\ & \hspace{-55mm} \times
\int{d^2 {\bf b} \over (2\pi)^2}\; {\rm e}^{i{\bf b}\cdot {\bf Q}_T}\;
\hat \sigma^{\rm (eik)}_{ab}(N,{\bf b},Q,\mu)
\, .
\label{qttransform2}
\ea
Equivalently, the {\em hadronic} cross section, given
as an inverse transform from moment space, takes the
form
\ba
{d\sigma_{AB\to V}\over dQ^2\, dQ^2_T}
&=&
\sum_{ab}\;
 \sigma_{ab\to V}^{({\rm H)}}(Q^2)\;
\exp \; \left\{ \int_{\mu_F^2}^{Q^2} {d\mu'{}^2\over \mu'{}^{2}}\;
\left[B_a(\as(\mu'))+B_b(\as(\mu'))\right]\right\}\; 
\nonumber \\
&&\times \int_{\cal C} {dN \over 2 \pi i}\;
\tilde{\phi}_{a/A}(N+1,\mu_F) \tilde{\phi}_{b/B}(N+1,\mu_F)\; \tau^{-N}
\nonumber \\
&&\times
\int {d^2 {\bf b} \over (2\pi)^2} {\rm e}^{i {\bf b} \cdot {\bf Q}_T}\
\hat \sigma^{\rm (eik)}_{c\bar d}(N,{\bf b},Q,\mu) \, ,
\label{physsig}
\ea
with $\mu_F$ the factorization scale and $\tau=Q^2/S$.
Corrections are order $1/N$ in the hard scattering.  The sums over 
$a$ and $b$ include quarks, antiquarks and gluons.
Thus, for the case of electroweak annihilation, a direct
examination of the eikonal cross section will determine
the large-moment and impact parameter behavior of the
partonic hard-scattering functions.
Here and below, ${\cal C}$ is a contour to the
right of $N$-plane singularities in the various transform functions.

Eq.\ (\ref{physsig}) reduces the computation
of the cross section in $\MS$ scheme to 
the determination of the eikonal cross sections.
Alternately, one may treat the factorized functions $\psi$
and $U$ separately, applying renormalization-group arguments
\cite{dyresumgs,cscss}.  The result must be the same.
In the remainder of this section, however, we shall 
examine the eikonal cross section directly. 

\subsection{Exponentiation}

We employ a number
of general features of eikonal cross sections, where the 
graphical analysis introduced in Refs.~\cite{Gatheral,GS81,FT} 
is particularly helpful. Moments of cross sections in the eikonal
approximation exponentiate at the level of integrands, with exponents
given by the moments of a set of graphical functions
that Gatheral \cite{Gatheral} termed ``webs".
Webs can be generated uniquely from cut diagrams
in eikonal cross sections order-by-order.
They are defined both in terms of graphical topology (irreducibility
under cuts of the eikonal lines) and color structure.
The lowest-order web is simply a single gluon exchanged between 
the lines. Beyond lowest order, each web is itself a  cut diagram, 
and can be integrated over the momentum, $k$, that it contributes 
to the final state. A very useful additional feature of webs is that 
at fixed $k$ they have no overall ultraviolet divergences.

Quite generally, then, the joint moment and impact parameter dependence
of the eikonal cross section may be expressed as
\ba
\bar\sigma_{ab}^{\rm (eik)}(N,{\bf b}Q,\epsilon)
&=& \exp\, \Bigg\{\, 2\, \int {d^{4-2\epsilon}k\over
\Omega_{1-2\epsilon}}\; \theta\left({Q\over \sqrt 2} - k^+\right)\;
\theta\left({Q\over \sqrt 2} - k^-\right)
\nonumber\\
&\ & \hspace{-20mm} \times
w_{ab} \left(k^2,{k\cdot \beta k\cdot \beta'\over
\beta\cdot\beta'},\mu^2,\alpha_s(\mu),\epsilon\right)\,
\left( {\rm e}^{-N(k\cdot \hat{n}/Q)-i{\bf b}\cdot 
{\bf k}}\;-1\right) \bigg\}\, ,
\label{web1}
\ea
where $w_{ab}$ represents the web at fixed total momentum $k^\mu$,
where $\hat{n}^{\mu} = \delta_{\mu 0}$, 
and where for convenience we choose the factor 
$\Omega_{1-2\epsilon}$ to be 
$2\pi^{1-\epsilon}/\Gamma(1-\epsilon)\sim 2 \pi 
(\pi {\rm e}^{\gamma_E})^{-\epsilon}$, equal to
the dimensionally-continued transverse angular integral.
In this form, the single-gluon emission
contribution to the web is normalized to be
\be
w^{(1){\rm (real)}}_{a\bar a}(k)
=
{2C_a \alpha_s\over\pi}\; \left(4\pi\mu^2{\rm
e}^{-\gamma_E}\right)^\epsilon\; \frac{1}{k_T^2}\; \delta_+(k^2)\, .
\ee
That the overall coefficient is twice the
one-loop term in the function $A_a(\alpha_s)$ in Eq.\ (\ref{explicitA}) is 
not, of course, a coincidence, and we shall see below how
this relation arises.  The $\epsilon$-dependent factor 
matches the $\MS$ collinear subtraction.
In Eq.\ (\ref{web1}), we explicitly limit
the phase space for gluon emission by the
plus and minus momenta of the annihilating partons
at partonic threshold.  As in the case of the eikonal
distributions ${\cal R}^{\rm (eik)}$ and $\phi^{\rm (eik)}$
in Eq.\ (\ref{eikdistnorm}),
dependence on the choice of upper limits is exponentially
suppressed in $N$ for real
gluon emission, but must be specified to set the
scale of virtual corrections.
We determine purely virtual corrections in the eikonal cross section by
demanding that it  be normalized
to unity at ${\bf b}=0$, $N=0$.  This choice ensures that the
perturbative expansion of
$\sigma^{\rm (eik)}$ is a sum of plus distributions.

The $k$-dependence of $w_{ab}(k,\beta,\beta')$ in Eq.\ (\ref{web1})
is strongly constrained by the invariance of the web under rescalings of
the light-like eikonal velocities, $\beta^\mu$ and $\beta'{}^\mu$.
We then observe, based on the lack of overall UV divergences in the webs,
that they obey\footnote{ We note that to derive this relation, the webs should
be renormalized appropriately in terms of their external eikonal lines.  These
renormalization factors vanish in Feynman gauge, and we shall ignore
them below.  The overall result for the eikonal cross section,  of
course, is gauge independent.}
\be
\mu{d\over d\mu}\; w_{ab} \left(k^2,{k\cdot \beta k\cdot \beta'\over
\beta\cdot\beta'},\mu^2,\alpha_s(\mu),\epsilon\right)
=0\, .
\label{webrgi}
\ee
We know even more about webs, because the purely virtual web is the 
logarithm of
the lightlike eikonal form factor, discussed extensively
in treatments of
the Sudakov form  factor in QCD \cite{KR,jccsud}.   From these investigations,
and by comparison to resummations for the Drell-Yan cross section
\cite{dyresumgs,dyresumct},
we learn that the webs may have at most a single overall infrared  divergence,
coupled with a single overall collinear divergence.
Additional logarithmic
singularities can arise only through the renormalization of subdiagrams.
As a result, at fixed values of $k_T$, relative to the axis
determined by $\vec \beta$ and $\vec \beta'$, the integral of the web over 
$k^2$ is finite.
The web integrals are ultraviolet divergent for $k_T\to \infty$ and are
collinear and infrared singular at $k_T=0$ once $k^2$ is integrated.

\subsection{The exponent}

The exponentiated eikonal cross section contains a considerable
amount of information, which follows from 
the properties of the web described above.
We use the azimuthal symmetry of the web
$w_{ab}$ in Eq.\ (\ref{web1}) to organize the
transverse and light-cone integrals of $k$ into the form
\ba
\bar\sigma_{ab}^{\rm (eik)}(N,{\bf b}Q,\epsilon)
&=&
\exp\, \Bigg[\,
2\, \int {d^{2-2\epsilon}k_T\over \Omega_{1-2\epsilon}}\;
\int_0^{Q^2-k_T^2} dk^2\;
w_{ab} \left(k^2,k_T^2+k^2,\mu^2,\alpha_s(\mu),\epsilon\right)
\nonumber\\
&\ & \hspace{-15mm} \times
\left(\int_{(k_T^2+k^2)/\sqrt{2}Q}^{Q/\sqrt{2}} {dk^+\over 2k^+}\;
{\rm e}^{-N\sqrt{2}\left({k^+\over 2Q}+{k_T^2+k^2\over
2Qk^+}\right)-i{\bf b}\cdot {\bf k}_T} - \ln\sqrt{Q^2\over
k_T^2+k^2}\; \right) \Bigg]\, ,
\label{firstA}
\ea
where the light-cone variables refer to the frames for which
$\beta_{T}=\beta'_{T}=0$.
For large $N$, the $k^+$ integral is well-approximated (up to
exponentially-suppressed
contributions) by a Bessel function, and we find
\ba
\bar\sigma_{ab}^{\rm (eik)}(N,{\bf b}Q,\epsilon)
&=&
\exp\, \Bigg\{\,
2\, \int {d^{2-2\epsilon}k_T\over \Omega_{1-2\epsilon}}\;\;
\int_0^{Q^2-k_T^2} dk^2\;
w_{ab} \left(k^2,k_T^2+k^2,\mu^2,\alpha_s(\mu),\epsilon\right)
\nonumber\\
&\ & \hspace{-15mm} \times
\left[\, {\rm e}^{-i{\bf b}\cdot {\bf k}_T}\;
K_0\left(2N\sqrt{k_T^2+k^2\over Q^2}\right) - \ln \sqrt{Q^2\over
k_T^2+k^2}\, \right]
+{\cal O}\left({\rm e}^{-N}\right)\ \Bigg\}\, .
\label{nextA}
\ea
In view of our comments above about the infrared sensitivity of the web,
we are particularly interested in the limit that $k_T^2+k^2$ vanishes.
    From the behavior of $K_0(x)$ for small values of its argument,
\be
K_0(x)\sim -\ln(x{\rm e}^{\gamma_E}/2)\, ,
\label{K0behave}
\ee
we see that the momentum dependence of $K_0$ 
cancels the logarithm in the collinear limit, leaving a factor $\ln(N{\rm
e}^{\gamma_E})$.  This remainder generates a collinear logarithmic 
singularity at $k_T^2=0$, which is canceled by the moments of eikonal  
distributions, as we now show.

To construct the eikonal hard-scattering functions, $\hat\sigma^{\rm (eik)}$,
given by Eq.\ (\ref{hatsigeikratio}), we substitute
Eq.\ (\ref{nextA}) for the eikonal cross section,
and the explicit expression (\ref{eikmsbarexp}) for the
eikonal distributions, into the Fourier transform of 
$\hat\sigma^{\rm (eik)}$ to derive
\ba
\hat\sigma_{ab}^{\rm (eik)}(N,{\bf b},Q,\mu)
&=&
\exp\, \Bigg[\,
2\, \int {d^{2-2\epsilon}k_T\over \Omega_{1-2\epsilon}}\;
\Bigg\{\int_0^{Q^2-k_T^2} dk^2\;
w_{ab} \left(k^2,k_T^2+k^2,\mu^2,\alpha_s(\mu),\epsilon\right)
\nonumber \\
& & \hspace{-30mm} \times
\left[\, {\rm e}^{-i{\bf b}\cdot {\bf k}_T}\;
K_0\left(2N\sqrt{k_T^2+k^2\over Q^2} \right) - \ln\sqrt{Q^2\over
k_T^2+k^2}\, \right]
+ {1\over (k_T^2)^{1-\epsilon}}\ln \bar N\, \sum_{d=a,b}
A_d\left(\as(k_T)\right) \Bigg\}\Bigg]\, ,
\nonumber\\
\label{nextA2}
\ea
where we have relabeled the variable $\mu'$ in Eq.\ (\ref{eikmsbarexp})
as $k_T$, and where $\bar N$ is defined in Eq.\ (\ref{barNdef}).
Eq.\ (\ref{nextA2}) has a lot in common with standard resummations
in logarithms of $N$ and $b$, although it still includes an extra integral over
$k^2$.  It may be further simplified,
however, using properties of the webs.

We continue by rearranging Eq.\ (\ref{nextA2})
into a form that isolates double logarithmic behavior,
\ba
\hat\sigma_{ab}^{\rm (eik)}(N,{\bf b},Q,\mu)
&=&
\exp\, \Bigg[\,
2\, \int {d^{2-2\epsilon}k_T\over \Omega_{1-2\epsilon}}\;
\Bigg\{\int_0^{Q^2-k_T^2} dk^2\;
w_{ab} \left(k^2,k_T^2+k^2,\mu^2,\alpha_s(\mu),\epsilon\right)
\nonumber \\
& & \hspace{-15mm} \times
\left[\, {\rm e}^{-i{\bf b}\cdot {\bf k}_T}\; K_0\left({2Nk_T\over Q}
\right) + \ln\left({k_T\over Q}\right)\, \right]
+ {1\over (k_T^2)^{1-\epsilon}}\ln \bar N\, \sum_{d=a,b}
A_d\left(\as(k_T)\right) \Bigg\}\Bigg]
\nonumber \\
& & \hspace{-20mm} \times
\exp\, \Bigg\{\, 2\, \int {d^{2-2\epsilon}k_T\over
\Omega_{1-2\epsilon}}\; \int_0^{Q^2-k_T^2} dk^2\;
w_{ab} \left(k^2,k_T^2+k^2,\mu^2,\alpha_s(\mu),\epsilon\right)
\nonumber \\
& & \hspace{-15mm}
\times \left[\; {\rm e}^{-i{\bf b}\cdot {\bf k}_T}\,
\left\{K_0\left(2N\sqrt{k_T^2+k^2\over Q^2}\right)
- K_0\left({2Nk_T \over Q }\right) \right\}
+\ln \left( \sqrt{k_T^2+k^2}  \over k_T\right) \; \right]\Bigg\}\, .
\label{thirdA}
\ea
Let us deal with the two exponentials on the right-hand side of this
relation in turn.
The first exponential in Eq.\ (\ref{thirdA}) begins at the leading
logarithm (LL). We have seen that the webs contain no internal collinear
or infrared divergences.  Also, because the web requires no
overall ultraviolet subtraction for $k_T$ fixed, the
$k^2$ integrals of $w_{ab}$ in Eq.\ (\ref{nextA2}) converge
on a scale set by $k_T$, independent of $N$, $b$ or $Q$.
At the same time,
the factorizability of the eikonal cross section requires the
cancellation of all singularities at $k_T=0$ in Eq.\ (\ref{thirdA}).
We may thus formally expand the integral of the web over $k^2$
in inverse powers of $Q^2$, with a leading coefficient that
behaves as $1/k_T^2$ for $k_T\to 0$, which must
match the collinear singularity of the subtraction:
\be
\int_0^{Q^2-k_T^2} dk^2\;
w_{ab} \left(k^2,k_T^2+k^2,\mu^2,\alpha_s(\mu),\epsilon\right)
     =
{A_a\left(\as(k_T)\right)+A_b\left(\as(k_T)\right) \over
(k_T^2)^{1-\epsilon}}
+ {\cal A}_{ab}\left(\alpha_s(k_T),k_T,Q\right)\, ,
\label{wintexpA}
\ee
where the function ${\cal A}_{ab}$ behaves as $(k_T^0/Q^2)$ for
$k_T\to 0$.  In this expression we have used the renormalization-scale
invariance of the webs, Eq.\ (\ref{webrgi}), to set the scale of
the coupling at $k_T$, which is the only remaining kinematic variable.
Given this result, the $k_T$ integral in the first exponential of 
(\ref{thirdA}) is seen to be finite, and we may remove the dimensional 
regularization on this factor. Leading threshold logarithms in the 
perturbative expansion of the exponent,
for example $\as^n\ln^{n+1}N$, are generated by the explicit
logarithms of $k_T$ and $N$,
which multiply  the $1/k_T^2$ behavior isolated in Eq.\ (\ref{wintexpA}).

In the second exponential
on the right-hand side of Eq.\ (\ref{thirdA}),
the term in square brackets behaves smoothly for $k_T^2\sim  k^2\to 0$,
as well as for $k_T\to 0$ with $k^2$ fixed.
As a result, this term has a finite $\epsilon\to 0$ limit,
and begins with next-to-leading (NLL) logarithms,
for example,  $\as^n\ln^n N$.  With $\MS$ eikonal distributions,
however, even NLL logarithms are absent, because at leading order the web 
$w_{ab}$ is proportional to $\delta(k^2)$ (i.e., one-gluon exchange), which 
vanishes in this factor.

Using Eq.\ (\ref{wintexpA}) in (\ref{thirdA}), and setting the number of
dimensions to four,
we derive an expression for the resummed cross section in transform space,
\be
\hat\sigma_{ab}^{\rm (eik)}(N,{\bf b},Q,\mu)
=
\exp\left[D_{ab}^{\rm (eik)}(N,b)\right]\
\exp\left[ E_{ab}^{\rm (eik)}(N,b,Q,\mu)\, \right]\, ,
\label{eiksigmahatE}
\ee
where the leading $N$ and $b$-dependence (LL and NLL)
is entirely contained in the exponent
\ba
E_{ab}^{\rm (eik)}(N,b,Q,\mu_F)  &\ & \nonumber\\
&\ & \hspace{-20mm} =
\int_0^{Q^2} {d k_T^2\over k_T^2}\; \left\{\ \sum_{i=a,b}
A_i\left(\as(k_T)\right)\; \left[\, J_0 \left( b k_T \right) \; 
K_0\left({2Nk_T\over Q} \right) + \ln\left({\bar N k_T\over
Q}\right)\, \right]\,
\right\}
\nonumber\\
&\ & \hspace{-10mm}
- \ln \bar N\ \int_{\mu_F^2}^{Q^2} {d k_T^2\over k_T^2}\sum_{i=a,b}
A_i\left(\as(k_T)\right)\, .
\label{Eelaborate}
\ea
The second term on the right accounts for the difference between the physical
scale $Q$ and the factorization scale $\mu_F$.

The factor $\exp [D_{ab}]$ in Eq.\ (\ref{eiksigmahatE}) contains corrections
in the form of an infrared safe
expansion in $\as$ plus NNLL and nonleading powers in $N$ and $b$,
\ba
D_{ab}^{\rm (eik)}(N,b)
&=&
\int_0^{Q^2} dk_T^2\; {\cal A}_{ab}\left(\alpha_s(k_T),k_T,Q\right)
\left[\, {\rm e}^{-i{\bf b}\cdot {\bf k}_T}\; K_0\left({2Nk_T\over Q}
\right) + \ln\left({k_T\over Q}\right)\, \right]
\nonumber \\
& & \hspace{-15mm}
+\int_0^{Q^2} d k_T^2 \;  \int_0^{Q^2-k_T^2} dk^2\;
w_{ab} \left(k^2,k_T^2+k^2,\mu^2,\alpha_s(\mu),\epsilon\right)
\nonumber \\
& & \hspace{-15mm}
\times \left[\; {\rm e}^{-i{\bf b}\cdot {\bf k}_T}\,
\left\{K_0\left(2N\sqrt{k_T^2+k^2\over Q^2}\right)
- K_0\left({2Nk_T \over Q }\right) \right\}
+\ln \left( \sqrt{k_T^2+k^2}  \over k_T\right) \; \right]\, .
\label{Delaborate}
\ea
These, rather elaborate, expressions are accurate to all logarithms
in $N$ and $b$,
and implicitly contain as well the structure of power 
corrections \cite{powerCS,irrold},
which we hope
to study in future work.  Here, we only note that the
expansion of the function $K_0(x)$ for small $x$ contains, up to a single 
logarithm, only even powers in  $x$.  This simple observation is enough
to ensure that the threshold resummation of
perturbation theory to any order
implies the presence of even powers of $Q^{-1}$ (and $b$) only 
\cite{powerCS,nphh}.
In this paper, we shall regard the above results as
the starting point for a joint NLL resummation in $N$ and $b$
for electroweak annihilation, and for high-$p_T$
photon and hadron cross sections.
As noted above, the entire NLL result is associated with $E_{ab}$,
although $D_{ab}$ may contribute beginning at NNLL.

\subsection{Hadronic cross sections}

The explicit form of the jointly resummed cross section is now found
by inserting Eq.\ (\ref{eiksigmahatE}) in (\ref{physsig}).  The
factorization scale
dependence (denoted $\mu_F$) may be exhibited explicitly
by combining the second term in Eq.\ (\ref{Eelaborate}) with the
corresponding term in (\ref{physsig}):
\ba
{d\sigma_{AB\to V}\over dQ^2\, dQ^2_T}
&=&
\sum_{ab}\;
\hat \sigma_{ab\to V}^{({\rm H)}}(Q^2) \; \int_{\cal C} 
{dN \over 2 \pi i}\;
\exp \; \left\{ \int_{\mu_F^2}^{Q^2} {d\mu'{}^2\over \mu'{}^{2}}\; \sum_{i=a,b}
\gamma_{ii}(N,\as(\mu'))\right\}\;
\nonumber \\
&&\times
\tilde{\phi}_{a/A}(N+1,\mu_F) \tilde{\phi}_{b/B}(N+1,\mu_F)\; \tau^{-N}
\nonumber \\
&&\times
\int {d^2 {\bf b} \over (2\pi)^2} {\rm e}^{i {\bf b} \cdot {\bf Q}_T}\
\exp\left[ E_{ab}^{\rm (eik)}(N,b,Q,Q)\, \right]
\exp\left[D_{ab}^{\rm (eik)}(N,b)\right]\, ,
\label{physsigresum}
\ea
where $\gamma_{ii}$ is the full $N^0$ ($\ln N$ and constant) term in the
$N$th moment of the diagonal splitting function for parton $i$.
Notice that we have set $\mu=Q$ in $E^{\rm (eik)}$.
In the resummed cross section
(\ref{physsigresum}), dependence on the factorization scale $\mu_F$
is suppressed compared to NLO 
\cite{scalereduce,thrphen,qqbaresumBCMN,scale,KidOw,KidOw1}, because the
eikonal anomalous dimensions compensate the evolution of
parton distributions to leading power in $N$, at all orders in $\alpha_s$.

Eq.\ (\ref{physsigresum}) is our most general form in
this paper for the Drell-Yan cross section, which we
will approximate to NLL in Sec.~5.  Applications to other
processes are, as we shall see, conveniently carried out
in terms of the coefficient function $\bar c_{i/i}$, Eq.\ (\ref{Cfact}).
By comparing Eqs.\ (\ref{qttransform2}) and (\ref{qtcform}), and once
again using the explicit $\mu_F$ dependence of the resummed exponent,
we derive an analogous expression for
the product of coefficient functions $\bar c_{i/i}$:
\ba
\bar c_{a/a}(N,{\bf b},Q,\mu)\;
\bar c_{b/b}(N,{\bf b},Q,\mu)
&=&
\exp \; \left\{ \int_{\mu_F^2}^{Q^2} {d\mu'{}^2\over \mu'{}^{2}}\; \sum_{i=a,b}
\gamma_{ii}(N,\as(\mu'))\right\}
\nonumber\\
&\ & \times
\exp\left[ E_{ab}^{\rm (eik)}(N,b,Q,Q)+D_{ab}^{\rm (eik)}(N,b)\right]\, .
\label{CCE}
\ea
This result will be useful for direct photon and other
hard-scattering processes with factoring initial-state
interactions, but with final-state color flow.

\section{Single-particle Inclusive Cross Section}

In the following, we apply the methods outlined above to
$p_T^3 d \sigma_{AB\to {\rm \gamma}+X}(x_T) / dp_T$,  the prompt
photon inclusive cross section at measured $p_T$.
Unlike electroweak annihilation, however, we will not take the
limit $p_T\to 0$.
As a result, we have no explicit logarithms of $p_T$ to resum in the
hard-scattering
cross section.  Threshold resummation has been carried out for this
process in \cite{1pIresumLOS,1pIresumCMN},
and its consequences studied phenomenologically in \cite{thrphen,KidOw},
but there has been no generally accepted method to incorporate
the kind of recoil corrections that are so important in describing
the low-$Q_T$ limit of electroweak annihilation.  These effects must, 
however, be present in the
hard-scattering functions for this process at some level, even
if they cancel almost completely.  Our goal here is to develop
a framework in which we can identify them systematically.
This is a prerequisite to any reliable estimates of their influence.

Models of ``intrinsic" transverse
momentum seem to suggest that recoil effects of magnitude similar 
to those in electroweak annihilation may be important
\cite{kteff1,kteff2}. Recent experimental results 
appear difficult to explain without them \cite{dgamexp}.   
The inputs to these analyses are primarily
nonperturbative, in the form of Gaussian distributions in partonic transverse
momenta, whose widths may be compared directly to nonperturbative parameters
in electroweak annihilation \cite{LiLai}.
At the same time, the interpretation of the theory and experiment
is not without controversy \cite{kmrkt,afgkpw}.  For these reasons as well,
it is of interest
to reanalyze the prompt photon cross section in the light of the joint
resummation \cite{Liunified} procedure introduced 
for electroweak annihilation above, in the formalism of collinear
factorization.  

\subsection{Hard-scattering functions for inclusive prompt photons}

The prompt photon cross section may be written in factorized form as
\be
{p_T^3 d \sigma_{AB\to \gamma+X}(x_T) \over dp_T }
=
\sum_{ab}
\int dx_a\; \phi_{a/A}(x_a,\mu)\,
\int dx_b\; \phi_{b/B}(x_b,\mu)\;
p_T^3{d\hat \sigma_{ab\to \gamma}\left(\hat
x_T^2,p_T/\mu,\alpha_s(\mu)\right)
\over dp_T}\; ,
\label{dgamptcofact}
\ee
where we define $x_T$ and $\hat x_T$ by
\be
x_T^2 \equiv { 4p_T^2 \over  S},
\quad \quad \hat x_T^2 \equiv { 4p_T^2 \over \hat s}\, ,
\label{hatxtdef}
\ee
with $\hat s=x_ax_bS$.
To leading power in $N$, moments of $p_T^3d\hat\sigma/dp_T$ are
\ba
p_T^3{d\hat \sigma_{ab\to \gamma}\left(N,p_T/\mu,\alpha_s(\mu)\right)
\over dp_T}
&=&
\int_0^1 d\hat x_T^2\; (\hat x_T^2)^{N-1}\;
p_T^3\; {d \hat \sigma_{ab\to \gamma}\left(\hat
x_T^2,p_T/\mu,\alpha_s(\mu)\right)\over dp_T}
\nonumber\\
&\ & \hspace{-20mm} =
{1\over \tilde\phi_{a/a}(N+1,\mu)\;  \tilde\phi_{b/b}(N+1,\mu)}\;
\int_0^1 d x_T^2\; (x_T^2)^{N-1} p_T^3{d\sigma_{ab\to \gamma}\over dp_T}\, .
\label{xtmoment}
\ea
Following our discussion of electroweak annihilation, we will
use a combination of  threshold and $Q_T$ resummation to estimate
higher-order corrections in the partonic hard-scattering function,
$p_T^3\hat\sigma_{ab\to \gamma}/dp_T$,  always working
to leading power in the moment variable $N$.

At fixed
photon  rapidity $\hat\eta$ in the partonic center-of-mass system, 
singularities arise when the partonic center-of-mass energy
$\sqrt{\hat s}$, reaches the threshold value of the final state
necessary to produce the photon at that $p_T$ and $\hat\eta$.
Any such final state is kinematically equivalent to the photon recoiling
against
a massless jet.  The minimum invariant mass at
measured $p_T$ is
\be
\hat s_{\rm min}= 4p_T^2\cosh^2\hat \eta = {4p_T^2 \over \hat x_T^2}\, .
\label{hatshatcosh}
\ee
This kinematic relation changes, however, when
we take into account the recoil
of the photon-jet pair transverse momentum, denoted ${\bf Q}_T$ below,
against perturbative initial-state radiation.
In effect, when the incoming partons get a kick in
the direction of the observed photon through initial-state
radiation, the minimum invariant mass necessary to produce
the photon at measured $p_T$ decreases.
These effects are present in higher orders in $\hat\sigma$,
but are not reflected directly in logs of $1-\hat x_T^2$.

We can only begin to take recoil into account systematically, however, when
we have defined what we mean by initial-state
radiation, and what by the short-distance scattering.  This is precisely
an issue of refactorization, discussed above in
Sec.\ 2.  There we showed that refactorization into generalized initial-state
parton distributions, soft radiation and a short-distance
process is quite natural to
leading power in threshold singularities, $(1-z)^{-1}$,
for electroweak annihilation.   As we shall see, leading power at
threshold for
prompt photons allows refactorization
in an analogous manner, including now a new function for the final-state
jet, and a generalization of the function for soft radiation.
The essential point, however, is that at partonic threshold
it is possible to identify a $2\to 2$ short-distance scattering,
involving only lines off-shell by ${\cal O}(p_T^2)$,
that underlies the production of the high-$p_T$ photon.
Our aim is to treat this short-distance scattering in the same manner
as the short-distance cross section in electroweak annihilation,
in terms of its invariant mass squared, $\tilde s\equiv Q^2$,
and its transverse momentum ${\bf Q}_T$ (defined in the {\it hadronic}
center-of-mass system).  When ${\bf Q}_T$ is in the direction
of the observed photon, $\tilde s$ may be less than $\hat s$,
the partonic invariant mass squared in Eq.\ (\ref{dgamptcofact}).

To take recoil into account in joint resummation, we identify the
transverse momentum of the photon
in the center-of-mass system of the final-state photon-jet pair as
one-half the relative transverse momentum of the photon and recoiling
jet at partonic threshold,
\be
{\bf p}'_T={\bf p}_T-{1\over 2}\, {\bf Q}_T\, .
\label{pprimedef}
\ee
In terms of the kinematics of the short-distance 
$2\to 2$ subprocess, we can define a natural ``scaling" variable
for this refactorized  scattering, analogous to $x_T$ and $\hat
x_T$ in  Eq.\
(\ref{hatxtdef}),
\be
\tilde x_T^2 \equiv { 4|{\bf p}'_T|^2 \over Q^2} = {1\over
\cosh^2\tilde\eta}\, ,
\label{tildexdef}
\ee
with $\tilde\eta$ the rapidity in the refactorized
hard-scattering center-of-mass system.  
The variables $x_T^2$ and $\tilde x_T^2$ are related by
\be
\tilde x_T^2=x_T^2\, \left({ S \over Q^2}\; {|{\bf p}'_T|^2\over |{\bf
p}_T|^2}\right)\, ,
\label{xttoxttilde}
\ee
a relation that we will use below in the analysis of moments.

We will estimate the partonic cross section as the integral
over ${\bf Q}_T$ and $Q^2=\tilde{s}$ of the doubly-resummed cross section
at measured ${\bf Q}_T$ and $Q^2$, limiting the ${\bf Q}_T$ integral
to a cut-off scale $\bar \mu$.  That is, we will write the cross
section as
\be
p_T^3\; { d \sigma_{ab\to \gamma} \over dp_T}
=
p_T^3\; { d \sigma^{({\rm resum})}_{ab\to \gamma} \over dp_T}
+\Delta(p_T,\bar{\mu})\, ,
\label{formal1}
\ee
with
\be
p_T^3{ d \sigma_{ab\to \gamma}^{({\rm resum})} \over dp_T}
=
\int dQ^2 d^2{\bf Q}_T\; p_T^3 {d \sigma_{ab\to \gamma}^{({\rm resum})} \over
dQ^2\,
d^2{\bf Q}_T\, dp_T}\; \theta\left(\bar{\mu}-|{\bf Q}_T|\right)\, .
\label{formal2}
\ee
We have introduced a cut-off $\bar\mu$ in the recoil transverse 
momentum ${\bf Q}_T$,
which we include to avoid going outside the range where the
approximations for joint resummation fail, that is,
where  the recoil transverse momentum becomes competitive
with the hard scattering.  The need for a 
matching condition for the resummed to fixed-order expressions
at high recoil is familiar from $Q_T$-resummation in
electroweak annihilation.   Nonsingular finite-order terms, corrected
for the matching \cite{cscss,yuancollab} are included in 
$\Delta(p_T,\bar{\mu})$. The implementation of such a matching procedure 
remains to be carried out in the new joint resummation formalism.  We shall,
however, exhibit the theta function in ${\bf Q}_T$ in each
of our expressions below, as a reminder of its importance.

In our analysis below we will determine the jointly resummed cross section
at fixed $Q^2$ and ${\bf Q}_T$.
To develop a jointly resummed cross section, we begin, as for 
electroweak annihilation, with
a study of perturbation theory near partonic threshold.

\subsection{Leading regions in single-photon production}

The derivation of a refactorization formula
for the single-photon inclusive cross section
is quite similar to the analysis that leads
to Eq.\ (\ref{qtthfact}) for electroweak
annihilation.
In the photon cross section, however,
final-state interactions play an important role.
To analyze their contributions, we need an
analysis of the leading regions in the momentum
space of cut diagrams that produce logarithmic
corrections to the cross section.  The analysis
is quite similar to that carried out 
for threshold resummation in heavy quark \cite{qqbaresumKS,qqbaresumBCMN}
and jet production \cite{jetresum}, but now taking into account transverse
momenta  near threshold.
In this analysis, we shall neglect, for the time being,
fragmentation components in the prompt photon cross section,
which are relatively
modest in a significant kinematic region \cite{thrphen}.

\begin{figure}[t]
\begin{center}
\hspace*{-7mm}
\epsfig{file=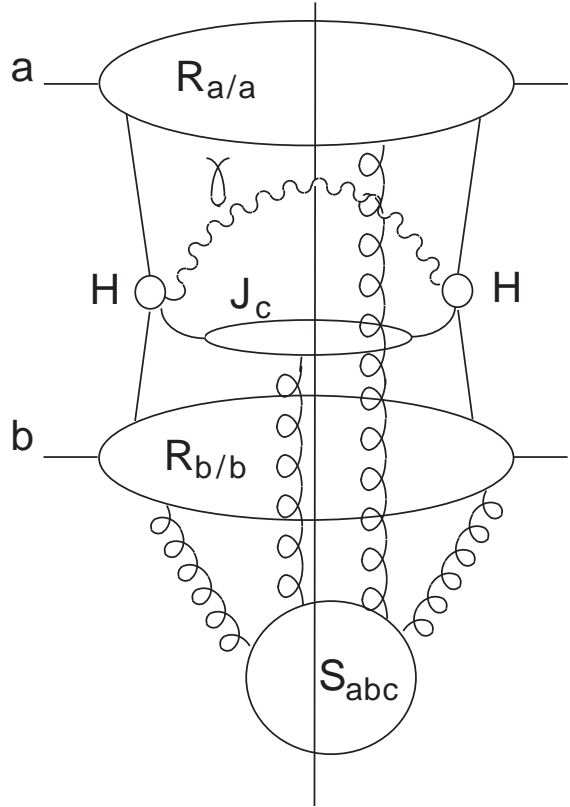,width=8cm}
\end{center}
\caption{Leading region for prompt photon production
near partonic threshold.  The vertical line represents
the final state, including the photon, $\gamma$, and
the recoil jet, $J_c$.}
\label{dgamleadingfig}
\end{figure}

The relevant leading regions for the $2\to 2$ partonic subprocess
$a+b\to c + \gamma$ are
illustrated by Fig.\ \ref{dgamleadingfig}, which may be compared
to Fig.\ \ref{EWAleadingfig} for electroweak annihilation.
In addition to the jets ${\cal R}_{a/a}$ and  ${\cal R}_{b/b}$ associated
with the incoming partons, and the short-distance
subdiagrams $H$, there is also a subdiagram $J_c$ that account for partons
collinear to the outgoing parton $c$, and
a new soft subdiagram $S_{abc}$, which accounts for soft radiation
from the final as well as initial hard partons.   These leading regions
are of the general class discussed in Ref.\ \cite{cssfact}, and
identified for hard-scattering
cross sections in Ref.\ \cite{LibbySt}, on the basis
of analyticity properties and power
counting bounds.   In the general
case, there may be an arbitrary number of noncollinear jets
in the final state.  Leading power in the threshold variable, $1-\hat{x}_T^2$,
limits their number to a single jet recoiling against the hard photon.

The role of the final state jet $J_c$ in threshold resummation is
well-understood, and has been treated in Refs.\
\cite{1pIresumLOS,1pIresumCMN,thrphen}.
The outgoing jet may interact with soft radiation,
through subdiagram $S_{abc}$ in Fig.\ \ref{dgamleadingfig}.
We may think of the soft function as associated with
coherent soft radiation, describing emission and absorption
by sources with specified four-velocities and color charges.
The role of soft gluon functions encountered in
threshold resummation has been extensively studied elsewhere
\cite{colevol}. We will give a formal definition
of the fully factorized soft function in subsection \ref{Sdefsec}

There is a potential complication for the joint resummation
of the single-photon cross section in the influence
of soft radiation on
the recoil transverse momentum ${\bf Q}_T$ of the photon-$J_c$ pair.
By analogy to electroweak annihilation,
we seek to resum logarithms of $1-\hat{x}_T^2$ as well as
logarithms of the total transverse momentum
of the partons involved in the underlying $2\to 2$ hard
scattering, ${\bf Q}_T$.  This transverse momentum
is defined relative to the (initial-state) {\it hadronic} center-of-mass
system, evaluated at fixed observed photon transverse momentum, ${\bf p}_T$,
in that frame.  Referring to Fig.\ \ref{dgamleadingfig},
we must ask, however, whether ${\bf Q}_T$ is the same on
both sides of the cut, i.e., for the amplitude and
for its complex conjugate.   An imbalance in the relative
transverse momenta of the two hard-scatterings would
make it necessary to introduce a more complicated
convolution than in Eq.\ (\ref{qtthfact}).
Such an imbalance, however, can arise only from the transverse momentum
that flows through
the soft function $S_{abc}$ between the incoming jet subdiagrams 
${\cal R}_{f/f}$ and the outgoing recoil jet $J_c$.  In the absence of
such a flow, the transverse momentum of the final-state jet
is a dependent quantity, and does not appear directly in
the transverse momentum logarithms. More importantly,
a single ${\bf Q}_T$ describes the relative transverse momenta
of the hard scattering on both sides of the cut.

\begin{figure}[t]
\begin{center}
\hspace*{-7mm}
\epsfig{file=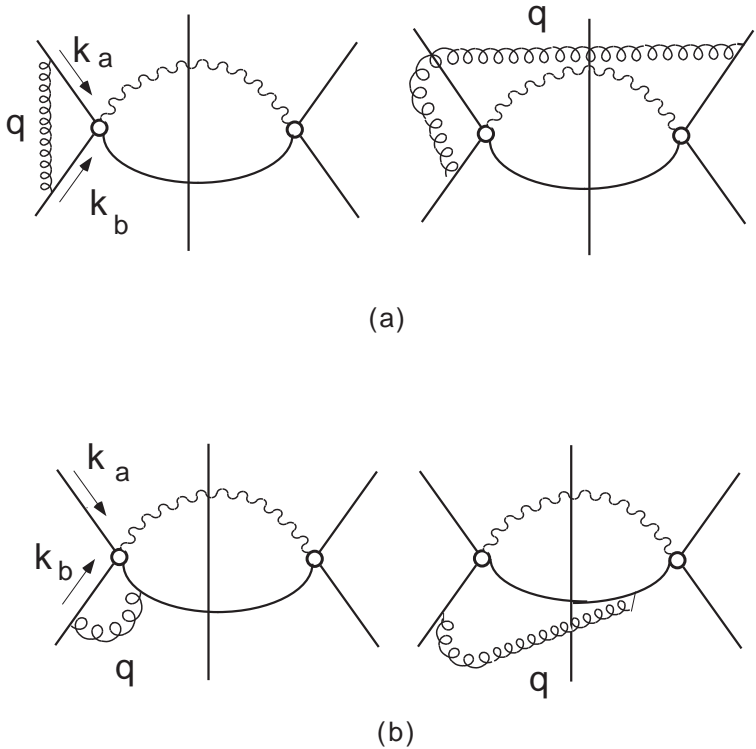,width=10cm}
\end{center}
\caption{Diagrams that illustrate the cancellation of
infrared divergences: (a) initial-state
interactions, which require different diagrams, with different
parton momenta ${\bf k}_a$ and ${\bf k}_b$. (b) initial-final-state 
interference, for which the cancellation proceeds through
cuts of a single diagram, at fixed ${\bf k}_a$ and ${\bf k}_b$.}
\label{ISFScontrastfig}
\end{figure}

We must therefore study the flow of
transverse momentum through the soft subdiagram.
We begin with diagrams that have initial-state
interactions only, in which soft gluons, as
in electroweak annihilation, couple only
to the initial-state jets, ${\cal R}_{a/a}$ and 
${\cal R}_{b/b}$ in Fig.\ \ref{dgamleadingfig}.
The transverse momentum that they carry into the final
state must come from the hard scattering, through the initial-state partons,
$a$ and $b$, equally in both the amplitude and the complex conjugate
amplitude, that is, with the same ${\bf Q}_T$ on the two sides of
the cut diagram of Fig.\ \ref{dgamleadingfig}.   At the same time, infrared
divergences associated with coherent soft gluons cancel as
we sum over {\it different} cut diagrams, with {\it different}
connections of soft gluons to the jet diagrams.
In the diagrams necessary to cancel infrared divergences,
therefore, the initial-state parton
transverse  momenta ${\bf k}_a$ and ${\bf k}_b$ will each vary.
Examples are shown in Fig.\ \ref{ISFScontrastfig}a.
This imperfect match means that the cancellation
proceeds through plus distributions in ${\bf k}_a^2$ and
${\bf k}_b^2$, and can produce logarithms
of ${\bf b}$ in the impact parameter space conjugate to ${\bf Q}_T$.
These purely initial-state interactions can be treated
exactly as in Sec.\ 2 above.

Consider next soft connections to the
final-state jet, including interference between
initial- and final-state interactions, as in Fig.\ \ref{ISFScontrastfig}b,
where the soft function $S_{abc}$
now has contributions in which soft radiation is emitted by
an initial state line, and absorbed by the final
state jet.
As usual, we must sum over final
states to cancel the  infrared divergences
associated with this soft radiation.
There is, however, a crucial difference between
the cancellation of initial-state and initial-final
coherent soft radiation.
In the case of initial-state radiation, as just described,
we must sum over different diagrams to effect the cancellation.
Final-state, and initial-final-state interference
divergences, however, cancel
in the sum over cuts of {\it individual} uncut diagrams,
as  discussed in context of factorization proofs in \cite{LibbySt}.
Li and Lai also observed that the  $k_T$-dependence
due to final state interactions cancels in the
context of prompt photon production \cite{LiLai}.
Hence, unlike singularities that arise from purely initial-state interactions,
the cancellation of final-state singularities
can be effected at {\it fixed} transverse momentum
for all lines in a cut diagram, in particular, for fixed pair
transverse momentum ${\bf Q}_T$ of the initial-state partons
$a$ and $b$, and hence for the short-distance process in
both the amplitude and its complex conjugate.  In contrast to
initial-state interactions, the transverse-momentum dependence
of the final-state soft radiation cancels algebraically, rather
than through plus distributions.

It is worth
noting that the above result requires that we resum logarithms
of the transverse momentum {\it at the short-distance scattering}, rather
than of the observed photon-jet pair in the final state.  In the
latter case, the transverse momentum at the short-distance scattering depends
on which of the soft gluons attached to the final-state jet
are virtual and which are real, and the cancellation of final state
infrared divergences
reverts from algebra \cite{LibbySt} to  plus distributions, and may produce logarithms
in impact parameter space.
     When the recoil jet is observed independently, therefore,
a somewhat different analysis is necessary, which we shall not
carry out here.

To summarize our considerations so far, only
the transverse momenta of coherent soft radiation
associated  with initial-state hard partons
must be taken into account in joint resummation.
In contrast, final-state and coherent soft radiation linking the
outgoing jet with one or both of the incoming jets produces
no logarithms in the pair transverse momentum.

The cancellation between the final
states in Fig.\ \ref{ISFScontrastfig}b still requires an
integral over energy \cite{cssrv,cssfact}.  For this reason,
both initial- and final-state interactions
contribute logarithms to threshold  resummation.
We must incorporate the distinction between
initial-state and initial-final interference
logarithms into the
refactorized convolution that generalizes
Eq.\ (\ref{qtthfact}) for electroweak annihilation to the case of
prompt photons. In the next subsection, we show that
it is possible to do this by separating purely initial-state 
soft radiation from initial-final interference, at least to the 
level of next-to-leading logarithms in both threshold and
pair transverse momenta.

\subsection{The soft function}
\label{Sdefsec}

To derive an analog of the electroweak annihilation refactorization
formula, Eq.\ (\ref{qtthfact}), for direct photon production, it is
necessary to identify a function that summarizes
final-state soft gluon radiation.  In particular, we want to separate
those effects associated with initial-state radiation, which
are sensitive to transverse momenta, from those from the final state,
which are not.  In this subsection, we will construct such a function.

The soft function will
be built from nonabelian phase operators, following the discussion of Sec.\ 2.3
above.  We begin by generalizing Eq.\ (\ref{Wdef}),
which describes the annihilation of a pair of
phase operators to a product that
corresponds to the color flow
\cite{qqbaresumKS,qqbaresumBCMN,colevol} at the hard scattering, for the 
partonic process, $a+b\to \gamma+c$:
\ba
\left[\, {\cal W}_{abc}(X)\, \right]_{e_c;e_b,e_a}
&=&
\left[\, \Phi^{(c)}_{\beta_c}{}(\infty,0;X)\, \right]_{e_c,d_c}\;
\left(\, c_{abc}\, \right)_{d_c;d_b,d_a}
\nonumber\\
&\ & \hspace{5mm} \times
\left[\, \Phi^{(b)}_{\beta_b}(0,-\infty;X)\, \right]_{d_b,e_b}\;
\left[\, \Phi^{(a)}_{\beta_a}(0,-\infty;X)\, \right]_{d_a,e_a}\, .
\label{wabcdef}
\ea
As in Sec.\ 2.3, we go on construct an eikonal cross section
in the form of Eq.\ (\ref{sigeikdef}), 
at fixed soft-gluon energy, parameterized as
$w_s Q$, and fixed transverse momentum ${\bf k}$ \cite{qqbaresumKS,jetresum},
\ba
\sigma^{\rm (eik)}_{abc}(w_s,Q,{\bf k},\mu,\epsilon)
&=&
Q\,  \int
{d\lambda \over 2\pi} \; {d^2{\bf b}\over (2\pi)^2}\ {\rm
e}^{-i\lambda w_s Q+i{\bf b}\cdot {\bf k}}\;\nonumber\\
&\ & \times {1\over C_A C_F}\, {\rm Tr}\
\langle 0|\,  {\rm \bar T}\, \left[{\cal W}_{abc}^\dagger(0)\right]\,
{\rm T}\left[{\cal W}_{abc}(\lambda \hat n+{\bf b})\right]\, |0\rangle\, ,
\label{pgsigeikdef}
\ea
where the trace is over the external color indices ($e_i$) of the
operators.
In this expression, $\hat n$ is the unit vector in the time direction, and
we leave the renormalization scale $\mu$ free.
The relevant transforms  with respect to $w_s$
and ${\bf k}$ are
Laplace and Fourier, respectively,
\be
\bar\sigma^{{\rm (eik)}}_{abc}
\left( {N\mu\over Q},{\bf b}Q,\epsilon\right)
=
\int_0^\infty d w_s\; {\rm e}^{-Nw_s}\, \int {d^2{\bf k}}\, {\rm
e}^{-i{\bf b}\cdot {\bf k}}\
\sigma^{\rm
(eik)}_{abc}(w_s,Q,{\bf k},\mu,\epsilon)\, .
\label{dgameikxsec}
\ee
The soft radiation function for the $a+b\to\gamma+c$
subprocess in prompt photon
production is now constructed from $\bar\sigma^{{\rm (eik)}}(N,{\bf b})$,
by analogy to the soft function $U$ for electroweak annihilation,
Eq.\ (\ref{Umoment}).

We find the soft function
by dividing the transformed eikonal cross section (\ref{dgameikxsec})
by functions that
eliminate double counting with the external jets near partonic
threshold: both incoming, and, in
this case, outgoing.  For the incoming lines, these functions are the
$\bar {\cal R}_a^{\rm (eik)}(N,{\bf b})$, defined in 
Eq.\ (\ref{doubletrans}). Similarly, for the outgoing jet, we identify 
a new partonic function, which will appear in the refactorization
formula, along with its eikonal partner.
As we have seen, infrared divergences associated with final
state interactions cancel at fixed recoil for the
hard scattering.  We may therefore treat the outgoing jet inclusively
in its transverse momentum.  The relevant functions
are then the same as those encountered in pure threshold resummation
\cite{jetresum}.  For example, for a quark jet they may be defined as
two-point functions,
\ba
\tilde J_c^{\rm (eik)}(N\mu/Q) &=& \int dw_c\; {\rm e}^{-Nw_c}\;
\int  d\sigma\; {\rm e}^{-i(w_cQ/\sqrt{2})\sigma}
    \nonumber\\
&\ & \hspace{10mm} \times\ {1\over N_C}
\ {\rm Tr} \langle 0|\Phi^{(q)}_{\beta_c}{}^\dagger(0)\
   \Phi^{(q)}_{\beta_c}{}(\sigma\beta_c)|0\rangle\ \; ,
\nonumber\\
\tilde J_c(N\mu/Q,Q/\mu) &=&
\int dw_c\; {\rm e}^{-Nw_c}\
\int d\lambda d\sigma\; {d^2{\bf y}\over
(2\pi)^2}{\rm e}^{-i(Q/\sqrt{2})\lambda-i(w_cQ/\sqrt{2})\sigma}
    \nonumber\\
&\ & \hspace{5mm} \times\ {\pi Q\over \sqrt{2} N_C}\;
{\rm Tr} \langle 0|q_c(0)\; \gamma\cdot
\bar
\beta_c\; \bar q_c(\lambda\bar\beta_c+\sigma\beta_c+{\bf y})|0\rangle
\nonumber\\
&\equiv&
\int dw_c\; {\rm e}^{-Nw_c}\; J_c(w_c,Q) \; ,
\label{outjetdefs}
\ea
for the eikonal and partonic jet, respectively.  In the partonic jet function,
$\beta_c$ is the light-like velocity vector in the direction of the jet,
and $\bar\beta_c$ is the velocity vector opposite to $\beta_c$ in
the overall partonic center-of-mass.  The matrix elements are evaluated in
$n\cdot A=0$ gauge. The traces refer to color and Dirac indices.
Outgoing gluon jets may be
defined analogously, with operators similar to those used for
fragmentation functions \cite{matdefs}.  In momentum space, $w_cQ^2$ is
the squared invariant mass of the outgoing jet.

The above reasoning leads to the following generalization of the soft
function to prompt photon processes:
\be
\bar S^{(ab\to \gamma c)}\left( {Q\over N\mu},{\bf b}Q,\alpha_s(\mu),
n\right) =
{\bar \sigma^{{\rm (eik)}}_{abc}\left( {Q\over
N\mu},{\bf b},\alpha_s(\mu),\epsilon\right)
    \over
\bar {\cal R}_{a}^{\rm (eik)}(N\mu/Q,{\bf b}Q,\epsilon)\,
\bar {\cal R}_{b}^{\rm (eik)}(N\mu/Q,{\bf b}Q,\epsilon)\,
\tilde J_{c}^{\rm (eik)}(N\mu/Q)}\, .
\label{Sdef}
\ee
As in the case of electroweak annihilation, Eq.~(\ref{Umoment}),
this expression is a rewriting of the refactorization of the 
eikonal cross section into incoming and outgoing jets and soft
radiation. Again, the ${\cal R}_{a,b}^{\rm (eik)}$ 
remove initial-state eikonal
radiation from $\hat \sigma_{abc}^{\rm (eik)}$, and
$\tilde J_{c}^{\rm (eik)}(N)$ removes dependence on the outgoing jet.
In this form, however, the soft function inherits the gauge dependence
implicit in Eq.~(\ref{Sdef}) of the incoming and outgoing jet functions. 
We can eliminate this dependence, and simplify the
overall formalism, by following an observation made in Refs.\
\cite{qqbaresumKS,colevol} and
employed in \cite{1pIresumLOS}.  We define a variant
soft function in transform space by dividing $\bar S$
by a factor $\bar U_{i\bar i}^{1/2}(N\mu/Q,{\bf b}Q)$ for each of the
incoming jets $a$ and $b$, and by $\bar U_{c\bar c}^{1/2}(N,{\bf 0})$
for the outgoing jet,
\ba
\bar S'{}^{(ab\to \gamma c)}\left( {Q\over N\mu},{\bf
b}Q,\alpha_s(\mu)\right) &=&
\nonumber\\
&\ & \hspace{-40mm}
{ \bar S^{(ab\to \gamma c)}\left( {Q\over N\mu},{\bf
b}Q,\alpha_s(\mu),n \right)
\over
\bar U_{a\bar a}^{1/2}(N\mu/Q,{\bf b}Q)\,
\bar U_{b\bar b}^{1/2}(N\mu/Q,{\bf b}Q)\,
\bar U_{c\bar c}^{1/2}(N\mu/Q,{\bf 0})}\, .
\label{Sprimedef}
\ea
By shifting factors of $\bar U^{1/2}$ from the soft radiation
function to the jets in transform space, we produce slightly
modified jet functions, of the convolution form
\ba
{\cal R}_{a/a}'(x,Q,{\bf k},\epsilon) &=&
    \int dw dy\; \delta(w +y-x) \int d{\bf k}_r d{\bf k}_u\
\delta^2({\bf k}_r+{\bf k}_u + {\bf k})
\nonumber\\
&\ & \hspace{30mm} \times\ {\cal R}_{a/a}(w,Q,{\bf k}_r,\epsilon)
U_{a\bar a}^{1/2}(y,Q,{\bf k}_u)
\nonumber\\
&=& \int_x^1 d\xi\ \phi_{a/a}(\xi,\mu)\, 
\tilde c_{a/a}\left({x\over\xi},{\bf k},Q,\mu\right)
+{\cal O}(1/N)\, ,
\nonumber\\
J'_c(w_c,Q)
&=& \int dw' dw_u \delta(w'+w_u-w_c)\ J_c(w',Q)\; 
\int d{\bf k}_u\; U_{c\bar c}^{1/2}(w_u,{\bf k}_u)\, ,
\label{Rprimedef}
\ea
where in the second equality for ${\cal R}_{a/a}'$, 
$\bar c_{a/a}$ is the double inverse transform
of the infrared safe function $\bar c_{a/a}$ defined in Eq.\ 
(\ref{Cfact}) above.

\subsection{Refactorization, recoil and the resummed cross section}

The refactorization formula for prompt photon production 
generalizes the corresponding expression for electroweak annihilation,
Eq.\ (\ref{qtthfact}), by including the outgoing jet function $J_c'$ 
in Eq.\ (\ref{Rprimedef}), and the modified soft
radiation function $S'$, Eq.\ (\ref{Sprimedef}).  In transform
space the refactorization is in terms of products; in momentum
space in terms of convolutions,
\ba
     p_T^3\; {d \sigma_{ab\to c\gamma}^{({\rm resum})} \over dQ^2\,
d^2{\bf Q}_T\, dp_T}
&=&
\int dx_a d^2{\bf k}_a\, {\cal R}'_{a/a}(x_a,{\bf k}_a,Q)\;
\int dx_b d^2{\bf k}_b\, {\cal R}'_{b/b}(x_b,{\bf k}_b,Q)\nonumber \\
&\ & \hspace{-10mm} \times
\; \int d w_c\; J'_c(w_c,Q)\;
\int dw_s \, \int d^2{\bf k}_s\ S'{}^{(ab\to \gamma
c)}\left(w_s,Q,{\bf k}_s,\as(\mu) \right)
\nonumber \\
&\ & \hspace{-10mm} \times
\;
{1\over S}\ \delta(1-Q^2/S-(1-x_a)-(1-x_b)-w_s-w_c)\; 
\delta^2 \left({\bf Q}_T+ {\bf k}_a+{\bf k}_b\right)
\nonumber\\
&\ & \hspace{-10mm} \times
\; C_\delta^{ab\to \gamma c}(\as(\mu),\tilde x_T^2)\; 
p_T^3\; {dp_T'\over dp_T}\; {d\hat \sigma_{ab\to c\gamma}^{(0)}
(\tilde x_T^2) \over dp'_T}\, .
\label{qt1pIfact}
\ea
Eq.\ (\ref{qt1pIfact}) generates, order-by-order in
perturbation theory, the same singularities at ${\bf Q}_T=0$
as the full cross section, to leading power in $1-Q^2/S$.
The factor $dp_T'/dp_T$ compensates for the difference in
phase space between fixing $p_T$ and  $p_T{}'$.
The function $C_\delta^{(ab\to \gamma c)}=1+{\cal O}(\alpha_s)$,
times the Born cross section, is the perturbative
short-distance function, which in the case of prompt photon
production (as opposed to electroweak annihilation) 
depends on $\tilde x_T^2$.~\footnote{From Eq.~(\ref{xttoxttilde}), 
$\tilde x_T^2$ is determined by $p_T$, $Q_T$ and $Q$.}  
This means that, beyond the lowest order,
the short-distance scattering function need not have the same angular
dependence as the Born cross section.
The short-distance function contains 
only corrections from (virtual) lines that are off-shell
by at least ${\cal O}(p_T)$.  It contains no real-gluon
emission, since all radiation has, to leading power in 
$1-Q^2/S$, been absorbed into the long-distance functions
${\cal R}'$, $J_c'$ and $S'$.
The computation of these short-distance functions is equivalent
to the  matching conditions of effective  field theories.

With $S'$ constructed as above, the soft transverse
momentum ${\bf k}_s$ is associated entirely with final-state
interactions, and is not included in the recoil momentum of
the hard subprocess.  We may therefore integrate over ${\bf k}_s$
and redefine
\be
S'{}^{(ab\to \gamma c)}\left(w_s,Q,\as(\mu) \right)
\equiv
\int d^2{\bf k}_s\ S'{}^{(ab\to \gamma c)}\left(w_s,Q,{\bf
k}_s,\as(\mu) \right)\, ,
\ee
where on the left we retain the same notation for the function, 
but omit the transverse
argument.  Our refactorization formula then simplifies to
\ba
     p_T^3\; {d \sigma_{ab\to c\gamma}^{({\rm resum})} \over dQ^2\,
d^2{\bf Q}_T\, dp_T}
&=&
\int dx_a d^2{\bf k}_a\, {\cal R}'_{a/a}(x_a,{\bf k}_a,Q)\;
\int dx_b d^2{\bf k}_b\, {\cal R}'_{b/b}(x_b,{\bf k}_b,Q)\nonumber \\
&\ & \hspace{-10mm} \times
\;  \int d w_c\; J'_c(w_c,Q)\;
\int dw_s S'{}^{(ab\to \gamma c)}\left(w_s,Q,\as(\mu)\right)
\nonumber \\
&\ & \hspace{-10mm} \times
\;
{1\over S}\ \delta(1-Q^2/S-(1-x_a)-(1-x_b)-w_s-w_c)\nonumber\\
&\ & \hspace{-10mm} \times
\; \delta^2 \left({\bf Q}_T+ {\bf k}_a+{\bf k}_b\right)
\nonumber\\
&\ & \hspace{-10mm} \times
\; C_\delta^{(ab\to \gamma c)}(\as(\mu),\tilde x_T^2)\;
p_T^3\; {dp_T'\over dp_T}\;  {d\hat \sigma_{ab\to c\gamma}^{(0)}
(\tilde x_T^2) \over    dp'_T}\, .
\label{qt1pIfact2}
\ea
To simplify the notation further, we also introduce 
a new function ${\cal F}$, which combines contributions from the final-state
jet and soft final-state radiation,
\ba
{\cal F}_{abc}(w_f)
=
\int dw_sdw_c\,  S'{}^{(ab\to \gamma c)}\left(w_s,Q,\as(\mu) \right)\;
J_c'(w_c,Q)\ \delta\left(w_f-w_s-w_c\right)\, .
\label{calFdef}
\ea
As described above, this convolution does not involve the recoil
transverse momentum.
In moment space, Eq.\ (\ref{calFdef}) becomes a product
\be
\tilde {\cal F}_{abc}(N)= \tilde  S'{}^{(ab\to \gamma c)}
\left({Q\over N\mu}\right)\, \tilde J_c'(N) + {\cal O} (1/N) \, .
\label{calFmt}
\ee
We will discuss the explicit $N$-dependence from soft gluon radiation
and the final-state jet below.

Eq.\ (\ref{qt1pIfact2}) defines recoil in the prompt photon cross section
in much the same way that Eq.\ (\ref{qtthfact}) defines it for
electroweak annihilation.  The short distance functions,
$C_\delta\; d\sigma^{(0)} (\tilde{x}_T^2) /dp_T$ include only lines
off-shell by ${\cal O}(p_T'{}^2)$, and are evaluated at zero relative 
transverse momenta for initial-state partons.  Expansions in the 
transverse momenta of the incoming partons are to be absorbed into
higher orders in $C_\delta$. In this convolution form, however,
the kinematics of the hard scattering influences the  cancellation of 
singularities at vanishing ${\bf k}_a$ and ${\bf k}_b$.  
This procedure has a straightforward interpretation 
order-by-order.  At fixed order, all contributions that
are singular at threshold for fixed $Q^2$ may be put in the form of 
Eq.\ (\ref{qt1pIfact2}). To evaluate the cross section at any fixed order, 
we would integrate each such contribution over $Q^2$ 
and ${\bf Q}_T$, as in Eq.\ (\ref{formal2}), with no
further approximations.  The result would contain finite
corrections resulting from the kinematic linkage of the
hard scattering with the cancellation of singularities in
transverse momentum.  We would then sum to all orders.
In the resummed formalism, we simply approximate the short-distance
function to ${\cal O}(\alpha_s)$,
sum the singularities to all orders
at leading or next-to-leading logarithmic accuracy,  and do the 
$Q^2$ and ${\bf Q}_T$ integrals last.
In the following, we shall derive the consequences
of this reorganization of perturbation theory. 

We are now ready to return to Eq.\ (\ref{formal2}), and derive the 
jointly-resummed partonic cross
section as an integral of the differential resummed cross section 
(\ref{qt1pIfact2}) 
over the hard-scattering scale $Q$ and its relative transverse
momentum, ${\bf Q}_T$.   
Changing variables from $Q^2$
to $\tilde x_T^2$, we derive 
\ba
{p_T^3 d \sigma^{({\rm resum})}_{ab\to \gamma} \over dp_T}
&=& \frac{p_T^4}{8 \pi S^2}\ \sum_{ij}\; \int_0^1 d\xi_i\; d\xi_j\;
{\phi}_{i/a}(\xi_i,\mu) {\phi}_{j/b}(\xi_j,\mu)\nonumber\\
&& \hspace{-25mm} \times\;
\int d^2{\bf Q}_T\; \theta\left(\bar{\mu}-|{\bf Q}_T|\right)\;
\int_0^1 d\tilde x^2_T\;
{|M_{ij}(\tilde x^2_T)|^2\over \sqrt{1-\tilde{x}_T^2}}\;
C_\delta^{(ij\to \gamma k)}(\as(\mu),\tilde
x_T^2)
\nonumber\\
&& \hspace{-25mm} \times\;
\int d^2 {\bf k}_i\; d^2 {\bf k}_j\; 
\delta^2\left({\bf Q}_T+{\bf k}_{i}+{\bf k}_{j}\right)\,
     \int_0^{\xi_i} dx_i\, \int_0^{\xi_j} dx_j\, \int_0^1 dx_f\,
\delta\left(\tilde x_T^2 - \left({{p}_T'{}^2\over p_T^2}\right)\,
{x_T^2\over x_ix_j(1-x_f)}\right)
\nonumber\\
&& \hspace{-25mm} \times\;
\bar{c}_{i/i} \left( {x_i\over \xi_i},{\bf k}_{i},
\frac{2 p'_T{}}{\tilde x_T},\mu \right)
\bar{c}_{j/j} \left( {x_j\over \xi_j},{\bf k}_{j},
\frac{2 p'_T{}}{\tilde x_T},\mu\right)
{\cal F}_{ijk}(x_f,\tilde x_T^2)\, \left({1\over x_ix_j(1-x_f)}\right)^2\, ,
\label{part1pIresumconvol}
\ea
where we have used Eq.\ (\ref{Rprimedef}) to isolate the partonic
hard-scattering function in $\MS$ scheme, and
Eq.\ (\ref{calFdef}) to summarize the contribution of soft
and jet radiation in the final state.
In the argument of the delta function, we have
used Eq.\ (\ref{xttoxttilde}) to reexpress the ratio $Q^2/S$ as
\be
{Q^2\over S}
=
\left({p_T'{}^2\over p_T^2}\right){x_T^2 \over \tilde x_T^2}\, .
\label{momentidentity}
\ee
In Eq.\ (\ref{part1pIresumconvol}), we have also reexpressed the 
Born cross sections in terms of
the $2\to 2$ invariant amplitudes $M(\tilde x_T^2)$, and have used 
the approximate
relation $p_T'(dp_T'/dp_T) = p_T$, valid up to a nonsingular term of 
${\cal O}({\bf Q}_T)$, which we neglect.

It is a relatively small step from Eq.\ (\ref{part1pIresumconvol}) to
a jointly resummed cross section for hadrons, by
replacing partonic by hadronic $\MS$ distributions,
\ba
{p_T^3 d \sigma^{({\rm resum})}_{AB\to \gamma} \over dp_T}
&=& 
\frac{p_T^4}{8 \pi S^2}\ \sum_{ij}\; \int_0^1 d\xi_i\; d\xi_j\;
{\phi}_{i/A}(\xi_i,\mu) {\phi}_{j/B}(\xi_j,\mu)\nonumber\\
&& \hspace{-25mm} \times\; \int {d^2{\bf Q}_T\over (2\pi)^2}\; 
\theta\left(\bar{\mu}-|{\bf Q}_T|\right)\; \int_0^1 d\tilde x^2_T\;
{|M_{ij}(\tilde x^2_T)|^2\over \sqrt{1-\tilde{x}_T^2}}\;
C_\delta^{(ij\to \gamma k)}(\as(\mu),\tilde
x_T^2)
\nonumber\\
&& \hspace{-25mm} \times\;
\int d^2 {\bf b}\;{\rm e}^{i {\bf b} \cdot {\bf Q}_T}\; \left({ S\over
2{p}'_T{}^2}\right) \;
     \int_0^{\xi_i} dx_i\, \int_0^{\xi_j} dx_j\, \int_0^1 dx_f\,
\delta\left(\tilde x_T^2 - \left({{p}_T'{}^2\over p_T^2}\right)\,
{x_T^2\over x_ix_j(1-x_f)}\right)
\nonumber\\
&& \hspace{-25mm} \times\;
\bar{c}_{i/i} \left( {x_i\over \xi_i},{\bf b},
\frac{2 p'_T{}}{\tilde x_T},\mu \right)
\bar{c}_{j/j} \left( {x_j\over \xi_j},{\bf b},
\frac{2 p'_T{}}{\tilde x_T},\mu\right)
{\cal F}_{ijk}(x_f,\tilde x_T^2)\, \left({1\over x_ix_j(1-x_f)}\right)^2\, ,
\label{1pIresumconvol}
\ea
where we have also replaced the convolution in transverse
momenta by a Fourier transform,
so that the functions $\bar c$ are now in impact parameter space.
  
Eq.\ (\ref{1pIresumconvol}) factorizes under $x_T^2$
moments at large $N$, up to $1/N$ corrections.
Thus, following essentially the same steps that led to Eq.\ (\ref{physsig})
for  electroweak annihilation, 
we find the physical cross section
as an inverse transform \cite{LSV},
\ba
{p_T^3 d \sigma^{({\rm resum})}_{AB\to \gamma} \over dp_T}
&=& \frac{p_T^4}{8 \pi S^2}\ \sum_{ij}\  \int_{\cal C} {dN \over 2 \pi i}\;
\tilde{\phi}_{i/A}(N,\mu) \tilde{\phi}_{j/B}(N,\mu)
\nonumber \\
&& \hspace{-25mm} \times
\int_0^1 d\tilde x^2_T\; \left(\tilde x^2_T \right)^N
{|M_{ij}(\tilde x^2_T)|^2\over \sqrt{1-\tilde{x}_T^2}}\;
C_\delta^{(ij\to \gamma k)}(\as(\mu),\tilde
x_T^2)
\nonumber\\
&& \hspace{-25mm} \times
\int {d^2 {\bf Q}_T \over (2\pi)^2}\;
\theta\left(\bar{\mu}-|{\bf Q}_T|\right)
\left( \frac{S}{4 |{\bf p}_T - {\bf Q}_T/2|^2} \right)^{N+1}
\nonumber \\
&\ & \hspace{-25mm} \times\;
\int d^2 {\bf b}\; {\rm e}^{i {\bf b} \cdot {\bf Q}_T} \;
\bar{c}_{i/i} \left( N,{\bf b},\frac{4 p'_T{}^2}{\tilde x^2_T},\mu \right)
\bar{c}_{j/j} \left( N,{\bf b},\frac{4 p'_T{}^2}{\tilde x^2_T},\mu\right)\
\tilde {\cal F}_{ijk}(N,\tilde x_T^2)\, .
\label{1pIresumeq}
\ea
In the $\bar c$'s, the impact parameter ${\bf b}$ is conjugate to 
${\bf Q}_T$ at fixed values of $Q^2={4 p'_T{}^2}/{\tilde x^2_T}$.  
In this form, however, each $\bar c_{i/i}$ also depends implicitly
on ${\bf Q}_T$, through $p'_T{}^2$.
Explicit forms for
the functions $\bar c_{i/i}$ have been given in Eq.\ (\ref{CCE}) above.
In the next subsection, we will find the $N$-dependence of
the  final-state function $\cal F$.

\subsection{Resummation for the final state}

Constructed as in Eq.\ (\ref{Sdef}), $\tilde S'{}^{(ab\to \gamma c)}$
satisfies the same
renormalization group equation as the soft functions
encountered in heavy quark \cite{qqbaresumKS} and
jet \cite{jetresum} threshold resummations,
\be
\mu{d\over d\mu}\, \ln \tilde S'{}^{(ab\to \gamma c)}\left({Q\over
N\mu},\as(\mu) \right)
=
-2{\rm Re}\, \Gamma^{(ab\to \gamma c)}_{S'}(\as(\mu))\; ,
\label{Sabcrg}
\ee
where the anomalous dimension $\Gamma^{(ab\to \gamma c)}_{S'}$
is a function of the velocities $\beta_i$ associated
with the phase lines corresponding to partons $a$, $b$ and $c$.
For the special  case of prompt photon production, $\Gamma_{S'}$
is a number, rather than a matrix, because the short-distance
cross section has only one color structure 
\cite{1pIresumLOS,1pIresumCMN}.
The solution to (\ref{Sabcrg}) is therefore a simple exponential,
\ba
\tilde S'{}^{(ab\to \gamma c)}\left({Q\over
N\mu},\as(\mu) \right)
&=&
\tilde S'{}^{(ab\to \gamma c)}\left(1,\as(Q/N) \right)
\nonumber\\
&\ & \hspace{5mm} \times
\exp\; \left[ \int_\mu^{Q/N} {d\mu'\over\mu'}
2{\rm Re}\, \Gamma^{(ab\to \gamma 
c)}_{S'}\left(\alpha_s(\mu')\right)\right]\, .
\label{Sprimesoln}
\ea
The logarithm of the soft function,
constructed in this fashion, has at most
a single logarithm of $N$ per loop, so that by
calculating $\Gamma^{(ab\to \gamma c)}_{S'}$ to one
loop, we determine $S'{}^{(ab\to \gamma c)}$
at the level of its leading logarithms, that is,
    $\as^L\ln^L N$ in the exponent, while remaining at NLL in the overall
cross section.
\begin{figure}[t]
\begin{center}
\hspace*{-7mm}
\epsfig{file=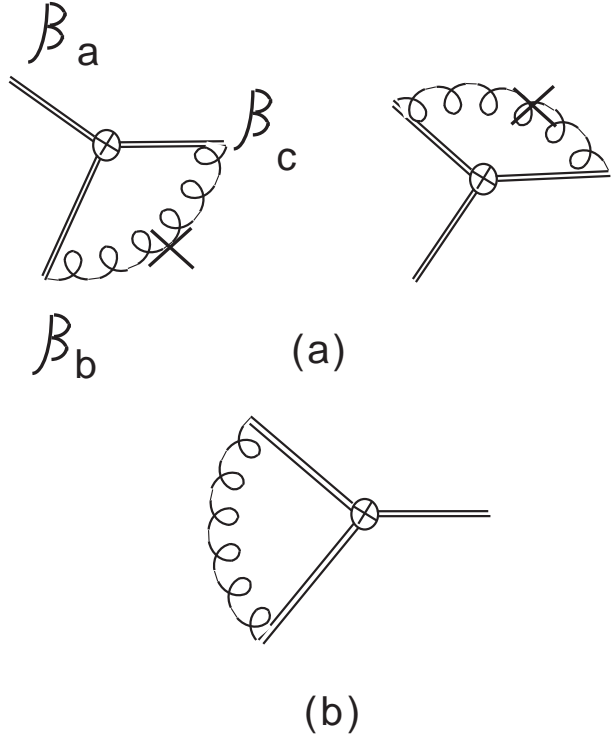,width=8cm}
\end{center}
\caption{(a) Diagrams which contribute to the soft anomalous
dimension.  The crosses on the gluon lines denote the modified
propagator given in the text.  The remainder of the $A^0=0$
gauge propagator cancels the real part of (b).}
\label{ADfig}
\end{figure}
$\Gamma^{(ab\to \gamma c)}_{S'}$ is calculated
from the $\MS$ counterterms for the three diagrams shown in
Fig.\ \ref{ADfig}, in which a single gluon is exchanged between
pairs of eikonals.  The details of these calculations,
which are to be carried out  in an axial $n\cdot A=0$
gauge, are described in \cite{colevol}. Many terms cancel, and  
in Fig.\ \ref{ADfig}a, the
additional crosses on the gluon lines denote a slightly
modified $A^0=0$ gauge propagator:
\be
{1\over k^2}\, \left(\, -g^{\mu\nu} + {\hat n^\mu k^\nu+k^\mu \hat 
n^\nu \over \hat n\cdot k}
-{1\over 2} {k^\mu k^\nu \over (\hat n\cdot k)^2}\right)\, .
\label{modprop}
\ee
The factor $1/2$ in this expression is specific to the lowest-order
calculation.  It takes into account the effect of subtracting
the factor $U_{a\bar a}U_{b\bar b}$ implicit in the definition
of $c_a c_b$, Eq.\ (\ref{Cfact}).   The ``missing"
terms in Eq.\ (\ref{modprop}) completely cancel
the real part of the diagram of Fig.\ \ref{ADfig}b.  The resulting
anomalous dimension is then
\ba
{\rm Re}\, \Gamma^{(q\bar q\to \gamma g)}_{S'} &=& {\alpha_s\over
2\pi}\, C_A\, \ln \left({tu\over s^2}\right)
\nonumber\\
{\rm Re}\, \Gamma^{(qg\to \gamma g)}_{S'} &=& {\alpha_s\over 2\pi}\,
\left[\; 2C_F\, \ln \left( {-u\over s} \right)
+ C_A\, \ln\left({t\over u}\right)\; \right]
\, ,
\label{ADexplicit}
\ea
where $s$, $t$ and $u$ are the invariants for the partonic
$2\to 2$ subprocess. These anomalous dimensions are
independent of the gauge vector $n^\mu$ at one loop
\cite{colevol}. The functions $S'{}^{(ab\to \gamma c)}$, defined as the 
solutions to Eq.\ (\ref{Sabcrg}),
with the usual boundary condition $S'{}^{(ab\to \gamma c)}(N=0)=1$,
are free of initial-state
radiation that would be sensitive to recoil at the logarithmic level.
They continue to contribute to the threshold phase
space through the energy $w_sQ$ \cite{qqbaresumKS,jetresum}, in a
manner described below.

Explicit expressions for $\tilde S'(N)$ are found from Eqs.\ (\ref{Sprimesoln})
and (\ref{ADexplicit}) above.  The transform of the recoil jet
function may be found from Ref.\ \cite{angular}, and is given in
its most general form by
\ba
\tilde J'_c(N) &=& \exp[\, E'_c(N)\, ]\nonumber\\
E'_{c}(N)
&=&
\int^1_0 dz \frac{z^{N-1}-1}{1-z}\;
\left \{\int^{(1-z)}_{(1-z)^2} \frac{d\lambda}{\lambda}
A_{c}\left[\alpha_s(\sqrt{\lambda} Q)\right] \right.
\nonumber\\ &&
\hspace{20mm} \left.
{}+B'_{c}\left[\alpha_s(\sqrt{1-z} Q) \right]
+B''_{c}\left[\alpha_s((1-z) Q) \right]\right\}\, ,
\label{Eprexp}
\ea
where the $A_c$ are given in Eq.\ (\ref{explicitA}).
In the specific normalization chosen for the
recoil jet, $J'_c$, in Eq.\ (\ref{Rprimedef}) above,
$B_c'{}'$ vanishes, while
\be
B'_{q}=\frac{\alpha_s}{\pi} \left(-\frac{3}{4}\right) C_F \, ,
\quad
B'_{g}=\frac{\alpha_s}{\pi} \left(- {\beta_0\over 4}\right) \, .
\label{Bpr}
\ee
These results, along with Eq.\ (\ref{CCE}) for
the functions $\tilde c_{i/i}$, specify the
explicit NLL resummed cross section as an inverse
Mellin moment in Eq.\ (\ref{1pIresumeq}).  We will give
the relevant expressions in Sec.\ 5.  We close this
section with a discussion of the additional considerations
necessary to include fragmentation effects in the formalism.

\subsection{Including fragmentation}

\begin{figure}
\begin{center}
\hspace*{-7mm}
\epsfig{file=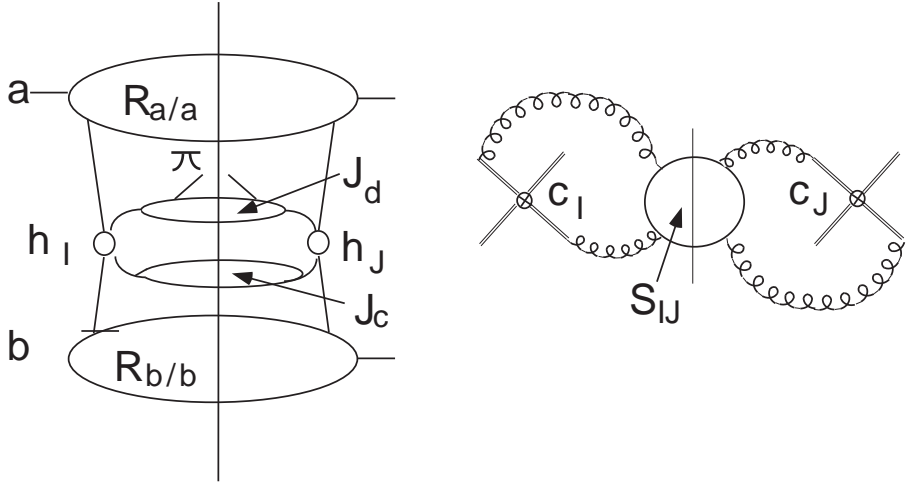,width=12cm}
\end{center}
\caption{Leading region for a single-particle inclusive cross section
in cut diagram form. $c_I$ and $c_J$ represent color tensors~\cite{colevol}.} 
\label{2to2fig}
\end{figure}

For single-particle inclusive hadron production, and even for
high-$p_T$ photon production,
we must supplement the considerations above to include final-state
fragmentation.
This is relatively easy, since the dynamics of fragmentation factorizes from
initial-state, hard virtual, and soft coherent radiation as well as from the
final-state corrections associated with the recoiling jet,
as illustrated in Fig.\ \ref{2to2fig}.  The analysis of
this section applies to pure threshold resummation 
(as discussed in  \cite{thrphen}) as well
as to joint resummation.

The underlying short-distance scattering subprocess is again a $2\to 2$
reaction, but now with two outgoing colored particles, one of which 
fragments into the observed hadron, which we may take to
be a pion (or photon):
\be
a(x_ap_a) +b(x_bp_b) \to c + d(P/z) \to \pi(P) +X\, ,
\label{2topi}
\ee
where we have exhibited the partons' momenta.
The outgoing jet, $J_d$, fragments into the observed hadron ($\pi$).
The short-distance scattering involves more than one color flow,
the set depending on the flavors of the partons in Eq.\ (\ref{2topi}).
At any order in perturbation theory, the color flow in
the amplitude and complex conjugate need not be the same.
Referring to Fig.\ \ref{2to2fig},
and adopting the notation of electroweak annihilation,
we represent the (dimensionless) short-distance function as
$p_T^3\; {d\hat \sigma_{JI}^{\rm (f)}(\tilde x_T^2) /
    dp'_T}=h^*_J\; h_I$,
where f denotes the $2\to 2$ partonic scattering of Eq.\ (\ref{2topi}).
The soft function is built in
the same way as for prompt photon production, treating
the color tensors $c_{I,J}$ for each color exchange at the hard
scattering as an effective local vertex linking 
phase operators in the flavors of the partons $a \dots d$.
The soft function is written as $S_{JI}^{\rm  (f)}$.
The same arguments regarding the
cancellation of recoil effects in the soft function
apply to single-hadron inclusive as to prompt photon
production.  Also, the normalization of the soft function
follows Eq.\ (\ref{Sprimedef}) above, for
each of the matrix elements of $S^{\rm (f)}_{JI}$,
   but now with factors $U^{1/2}$ for both outgoing jets.
The resulting anomalous dimensions for
most relevant flavor flows have been computed
in Refs.\ \cite{qqbaresumKS,colevol}.

Near threshold, in the notation of Eqs.\ (\ref{qt1pIfact}) and
(\ref{momentidentity}), the phase space is slightly modified,  
relative to the prompt photon case, because the fragmenting jet has
invariant mass $q_d^2\ge 0$, and because the
outgoing parton at the hard scattering carries momentum $P/z\ge P$,
which shifts the scaling variable $\tilde x_T$ by a factor $z$.
This changes the transverse momentum in the
hard-scattering center-of-mass system: ${\bf p}_T'\to {\bf
p}'{}'_T(z)\equiv{\bf p}_T/z-{\bf Q}_T/2$, which is
related to the short-distance 
scattering and overall invariant masses squared, $Q^2$ and $S$, by
\be
{Q^2\over S}
=
\left({|{\bf p}'{}'_T|^2\over |{\bf p}_T|^2}\right){x_T^2 \over
\tilde x_T^2} \sim \left({|{\bf p}'_T|^2\over |{\bf
p}_T|^2}\right){x_T^2 \over  z^2 \tilde x_T^2}\, .
\label{momentidentity2}
\ee
In the second relation we have expanded in $1-z$, and
have neglected corrections of the form $(1-z){\bf Q}_T$.
In place of the argument
of the first delta function in Eq.\ (\ref{qt1pIfact}) for
resummed prompt photon production,
we find the following kinematic constraint near
partonic threshold for single-particle inclusive cross
sections with fragmentation:
\ba
1 &=&  {Q^2\over S} +(1-x_a)+(1-x_b)+w_s + q_c^2/S + q_d^2/S
    \nonumber\\
&=& {|{\bf p}'_T|^2\over |{\bf p}_T|^2}{x_T^2 \over
\tilde x_T^2}
+2(1-z)+(1-x_a)+(1-x_b)+w_s + q_c^2/S + q_d^2/S\, ,
\label{pi0ps}
\ea
with $x_T^2$ as in Eq.\ (\ref{hatxtdef})
and $\tilde x_T^2$ as in (\ref{tildexdef}).

The refactorization that
generalizes Eq.\ (\ref{qt1pIfact}) 
includes an additional integral over $z$,
linked through the restriction (\ref{pi0ps}), and a 
function that describes the fragmentation of parton $d$ into
hadron $\pi(P)$. We parameterize the momentum of parton $d$ as
\be
q_d^\mu ={P^\mu\over z} + \left({z\; q_d^2 \over 2P\cdot 
\bar\beta_d}\right)\, \bar\beta_d^\mu
\equiv \frac{P\cdot \bar\beta_d}{z} \beta_d^\mu + 
\frac{z w_d Q^2}{2P\cdot \bar\beta} \, \bar\beta_d^\mu
\label{qdmtm}
\ee
where $\bar\beta_d^\mu$ is the opposite-directed unit light-cone
vector relative to
momentum $P^\mu=(P\cdot\bar\beta_d) \beta_d^\mu$,
which is itself taken to be lightlike. The second form introduces
a dimensionless variable $w_d\equiv q^2_d/S$.

The fragmentation dynamics of the
outgoing jet can be described
by a function that is quite similar to a standard fragmentation function
\cite{matdefs}, and to the inclusive jet functions (\ref{outjetdefs})
above.  As above, we illustrate the case of a quark $q_f$,
computed in $n\cdot A=0$ gauge,
\ba
J_{f/f}(w_d,P,z,\epsilon) &=&
{Q\over 8N_C}\sum_X\; \int {d\lambda d\sigma\over 2\pi}\
{\rm e}^{-i(P\cdot\bar\beta_d/z)\lambda-iq_d\cdot\beta_d\sigma}
\nonumber\\
&\ & \hspace{10mm} \times\ {\rm  Tr}\;
\langle 0|\, q_f(0)\, |f(P),X\rangle
\, \rlap /{\bar\beta}_d\, \langle f(P),X| \bar 
q_f(\lambda\beta_d+\sigma\bar\beta_d)\,
|0\rangle
\nonumber\\
&=&
\int_z^1 {d\xi\over \xi}\ e_{f/f}\left( {z\over \xi},{w_d},P,\mu_F\right)\;
D_{f/f}(\xi,\mu_F,\epsilon)\, ,
\label{Jdef}
\ea
with $\mu_F$ a factorization scale.
In the first, defining, relation, the trace is over both
color and  Dirac indices.  A sum over spins is understood.  In the
second equality,
$D_{f/f}$ is a fragmentation
function, which we may define in 
   $\MS$ factorization scheme.  As usual, at partonic
threshold, the infrared-safe coefficient function
$e_{d/d}$ may be taken diagonal in flavor, up to
corrections that vanish as $1/N$ in moment space.

It is convenient to define a final state threshold function
by analogy to Eq.\ (\ref{calFdef}), as a convolution of
the soft function with the recoiling jet,
\ba
{\cal F}_{JI}^{\rm (f)}\left(w_f\right)
\equiv \frac{1}{S} \int dw_s\, d q_c^2 \,
\delta\left(w_f-w_s-q_c^2/S\right)
\
S_{JI}'{}^{\rm (f)}\left(w_s,Q,\mu \right)\; J'_c(q_c^2)\, .
\label{calFJIef}
\ea
Laplace moments reduce the convolution to a product,
\ba
\tilde {\cal F}_{JI}^{\rm (f)}(N)
&=& \int_0^\infty dw_f\; {\rm e}^{-Nw_f}\ {\cal F}_{JI}^{\rm (f)}(w_f)
\nonumber\\
&=& \tilde S_{JI}'{}^{\rm (f)}\left({Q\over
N\mu},\as(\mu) \right)\; \tilde J'_c(N)\, .
\ea
It is now straightforward to generalize each of Eqs.\ (\ref{qt1pIfact2}),
(\ref{1pIresumeq})  and (\ref{1pIresumconvol}), corresponding to
refactorization at
partonic threshold, and to the jointly-resummed hadronic cross section
written as an inverse moment, and
in convolution form,  respectively.  We give below the first
two of these generalized expressions explicitly.

The refactorized partonic single-particle inclusive cross section
is given by
\ba
     p_T^3\; {d \sigma_{\rm f}^{({\rm resum})} \over dQ^2\,
d^2{\bf Q}_T\, dp_T}
&=&
\int dx_a d^2{\bf k}_a\, {\cal R}'_{a/a}(x_a,{\bf k}_a,Q)\;
\int dx_b d^2{\bf k}_b\, {\cal R}'_{b/b}(x_b,{\bf k}_b,Q)\nonumber \\
&\ & \hspace{10mm} \times\; 
\delta^2 \left({\bf Q}_T+ {\bf k}_a+{\bf k}_b\right)
\, \int dz \, \int  d w_d \nonumber \\
&\ & \hspace{10mm} \times
\;  \sum_{IJ}\
\int d w_f\; {\cal F}_{JI}^{\rm (f)}\left(w_f\right)\;
p_T^3\; {dp_T'\over dp_T}\;
{d\hat \sigma_{JI}^{\rm (f)}(\tilde x_T^2) \over
    dp'_T}
\nonumber \\
&\ & \hspace{10mm} \times
\; \int_z^1 {d\xi\over \xi}\ e_{d/d}\left( {z\over 
\xi},{w_d},P,\mu_F \right)\;
D_{d/d}(\xi,\mu_F,\epsilon)\;
\nonumber \\
&\ & \hspace{10mm} \times
\;
{1\over S}\ \delta\left(1-{Q^2/S}-(1-x_a)-(1-x_b)-w_f-w_d\right)\;
\, ,
\label{qt1pIhadfact2}
\ea
where $\tilde x_T$ is specified by Eq.\ (\ref{momentidentity2}).
As indicated in Eq.\ (\ref{formal2}), the
resummed cross section is found
by  integrating over $Q^2$ and ${\bf Q}_T$.
The form convenient for $x_T^2$-moments is found by changing variables
from $Q^2$ to $\tilde x_T^2$.  
We compute the hard-scattering cross section in
moment space, by dividing by perturbative distributions 
for the incoming partons,
and by a perturbative fragmentation function for the parton
that fragments into the observed particle.

The resummed hadronic cross section is found by
replacing the perturbative distributions
and fragmentation function by their hadronic analogs in moment space,
and inverting the
transform with respect to $x_T^2$.
The result is
\ba
{p_T^3 d \sigma^{({\rm resum})}_{AB\to \gamma} \over dp_T}
&=& \frac{p_T^4}{8 \pi S^2}\ \sum_{\rm f}\  \int_{\cal C} {dN \over 2 \pi i}\;
\tilde{\phi}_{a/A}(N,\mu) \tilde{\phi}_{b/B}(N,\mu)\ \tilde D_{\pi/d}(2N+3,\mu)
\nonumber \\
&& \hspace{-25mm} \times
\int_0^1 d\tilde x^2_T\; \left(\tilde x^2_T \right)^N
{|M^{(\rm f)}_{IJ}(\tilde x^2_T,\alpha_s(\mu))|^2\over 
\sqrt{1-\tilde{x}_T^2}}\;
C_\delta^{\rm (f)}(\as(\mu),\tilde
x_T^2) \nonumber \\ 
&& \hspace{-25mm} \times
\int {d^2 {\bf Q}_T \over (2\pi)^2}\;
\Theta\left(\bar{\mu}-|{\bf Q}_T|\right)
\left( \frac{S}{4 |{\bf p}_T - {\bf Q}_T/2|^2} \right)^{N+1} \\
&\ & \hspace{-25mm} \times
\int d^2 {\bf b} \; {\rm e}^{i {\bf b} \cdot {\bf Q}_T} \;
\bar{c}_{a/a} \left( N,{\bf b},\frac{4 p_T^2}{\tilde x^2_T},\mu \right)
\bar {c}_{b/b} \left( N,{\bf b},\frac{4 p_T^2}{\tilde x^2_T},\mu\right)\
\tilde {\cal F}_{JI}^{\rm (f)}(N,\tilde x_T^2)\;
\tilde e_{d/d}\left( 2N+3,N,P,\mu_F \right)
\, , \nonumber
\label{1pIresumeq2}
\ea
where we define
\be
\tilde e_{d/d}\left( M,N,P,\mu_F \right)
\equiv
\int dy\, dw_d\ {\rm e}^{-M(1-y)-Nw_d}\
e_{d/d}\left( y,w_d,P,\mu_F \right)\, .
\label{emoment}
\ee
In Eq.\ (\ref{1pIresumeq2}), $M^{\rm (f)}_{IJ}$ denotes the projection of the
amplitude squared at lowest
nonvanishing order on each of the color flows in the amplitude
and complex conjugate, and $C_\delta^{\rm (f)}$ is the corresponding
expansion in $\alpha_s$ of the complete hard-scattering function.
A similar form was found in Ref.\ \cite{thrphen} in the context of
threshold resummation for the fragmentation component of
high-$p_T$ photon production.  The function $\tilde{e}_{d/d}$
in Eq.\ (\ref{emoment}) is a double moment,
with respect to the jet invariant mass squared, $w_dQ^2$ and the
scaling variable, $y$.   Its double logarithmic behavior
\cite{thrphen}, however, is determined
entirely by the latter, because in the momentum configurations that give
rise to double logarithms, we have $1-y \gg w_d$, corresponding to the
collinear emission of soft gluons.  Finally, we note once
again the dependence of the resummed cross section on a
cutoff scale $\bar\mu$.  At this scale, the resummed cross section
must be matched to a finite-order, or partially resummed
cross section.  The investigation of the best
implementation of this procedure remains
for future work.

\section{Exponents at NLL}

In this section we apply joint resummation to
electroweak annihilation and prompt photon
production, and exhibit the relevant exponents to
next-to-leading logarithm. For prompt photons, these expressions 
were the basis of the phenomenological estimates
given in Ref.\ \cite{LSV}.  We reserve for
future work  the corresponding results for
single-particle annihilation including
fragmentation, since they will require 
slightly more elaborate calculations involving color mixing
at the hard scattering \cite{qqbaresumKS,colevol}.

\subsection{Electroweak annihilation}

Starting from Eq.\ (\ref{physsigresum}),
we can identify an explicit expression for the resummed electroweak
annihilation cross section that is
accurate to NLL in both $N$ and $b$ \cite{LSV}.
We recall first  that the exponent $D_{ab}$
in (\ref{physsigresum}) contributes only
at NNLL.  For the NLL exponent $E_{ab}$,
we have used for guidance the NLL
approximation to threshold resummation introduced in Ref.\ \cite{thrphen}.
Our expression for the
perturbative exponent $E_{ab}^{\rm PT}$ in Eq.\ (\ref{physsigresum}) is 
\ba
E_{ab}^{\rm PT}(N,b,Q,Q)
&=&
\int_{(b/c_1+\bar N/Q)^{-1}}^{Q}
{d\mu'\over \mu'}\;
\left[\, A_a(\as(\mu'))+ A_b(\as(\mu'))\, \right]\, \; 2\ln{\bar N \mu'\over Q}
\, ,
\label{Elog}
\ea
where as above, we define $\bar N=N{\rm e}^{\gamma_E}$.
Eq.\ (\ref{Elog}) approaches the
normal forms of $Q_T$-resummation (in $b$-space) as
$b\to \infty$ at fixed $N$, and of threshold resummation
for $N\to \infty$ at fixed $b$.  We have introduced an
explicit dimensionless scale $c_1$ into the $b$-dependence
of the lower limit in the first integral.
In Eq.\ (\ref{Elog}), the factorization scale,
set to $\mu_F=Q$ in the upper limit
of the integral, replaces the  conventional upper limit $c_2 Q$
in the $Q_T$-resummation formalism \cite{ktresumold,cscss}.

Following the format of \cite{thrphen},
we find the following closed expression for $E_{a\bar a}^{\rm PT}$,
accurate to NLL in both $N$ and $b$,
\begin{equation}
E_{a\bar a}^{\rm PT} (N,b,Q,\mu_F) = \frac{2}{\alpha_s (\mu)}
h_a^{(0)} (\lambda,\beta) +
2h_a^{(1)} (\lambda,\beta,Q,\mu,\mu_F)   \;  ,
\end{equation}
where for $a=q$ or $g$ we define
\begin{eqnarray}
h_a^{(0)} (\lambda,\beta) &=& \frac{A_a^{(1)}}{2\pi b_0^2}
\left[ 2 \beta + (1 - 2 \lambda) \ln(1-2 \beta) \right]\, ,
\label{hsubadef0}
\end{eqnarray}
and
\begin{eqnarray}
h_a^{(1)} (\lambda,\beta,Q,\mu,\mu_F) &=&
\frac{A_a^{(1)} b_1}{2\pi b_0^3} \left[ \frac{1}{2} \ln^2 (1-2 \beta) +
\frac{1 - 2 \lambda}{1-2\beta} (2 \beta + \ln(1-2 \beta) ) \right] \nonumber \\
&+& \frac{1}{2\pi b_0} \left( - \frac{A_a^{(2)}}{\pi b_0} +
A_a^{(1)} \ln \left( \frac{Q^2}{\mu^2} \right) \right)
\left[ 2 \beta \frac{1 - 2 \lambda}{1-2\beta}+ \ln(1-2 \beta) \right]
\nonumber \\
&-&
\frac{A_a^{(1)}}{\pi b_0} \lambda \ln \left( \frac{Q^2}{\mu_F^2}
\right) \; ,
\label{hsubadef}
\end{eqnarray}
in terms of moment variables,
\begin{eqnarray}
\lambda &=& b_0 \alpha_s (\mu)  \ln \left (\bar N  \right) \; ,
\nonumber \\
\beta &=& b_0 \alpha_s (\mu)
\ln \left( \bar N  + b Q/c_1 \right) \, .
\label{varsdef}
\ea
The coefficients $A_a^{(1)}$ and $A_a^{(2)}$ in Eqs.\ (\ref{hsubadef0})
and (\ref{hsubadef}) have been given in Eq.~(\ref{explicitA}).
The last term on the right-hand side of Eq.\ (\ref{hsubadef}) is
the contribution at NLL of the first exponent of Eq.\ (\ref{physsigresum}),
including the anomalous dimensions.
The beta  function coefficients in these expressions are given by
\ba
b_0 &=& \frac{11 C_A - 4 T_R N_F}{12 \pi}\, , \nonumber \\
b_1 &=& \frac{17 C_A^2-10 C_A T_R N_F-6 C_F T_R N_F}{24 \pi^2}\, .
\end{eqnarray}
We recall that Eq.\ (\ref{physsigresum}) was derived choosing
the renormalization scale as $Q$, the mass of the produced boson.
In the expressions immediately above, $\mu_F$ is the factorization scale, while
the argument $\mu$ denotes the renormalization scale dependence
that arises when the scale of $\alpha_s$ is shifted from $Q$
to $\mu$.

Using these results in Eq.\ (\ref{physsigresum}), we derive
doubly-resummed expressions for the production
of W, Z and Higgs
at measured ${\bf Q}_T$,
through electroweak annihilation,
accurate to NLL in both transforms.  Compared to existing $Q_T$-resummation
formalisms  \cite{yuancollab,evst}, we anticipate modest changes 
associated with
the additional threshold resummation, especially for the W and Z,
when their mass is far below collider energies, as at the
Tevatron. Nevertheless,
we expect a decreased sensitivity to the choice of factorization
scale \cite{scalereduce,thrphen,qqbaresumBCMN,scale,KidOw,KidOw1}.
At the same time, it may be of interest to study the $b$-space
integral in (\ref{physsigresum}) in the ``minimal-principal value"
prescription introduced in Ref.\ \cite{LSV}.
We shall not pursue these phenomenological
implications here, however.

In addition to a perturbative exponent, we expect that nonperturbative
contributions, of the sort familiar from $Q_T$-resummation, will
be phenomenologically important whenever $Q_T\ll Q$ 
\cite{ktresumold,cscss,yuancollab,evst}.
As noted above, the form of power corrections associated with the
running coupling may be inferred from Eqs.\ 
(\ref{eiksigmahatE})-(\ref{Delaborate}),
with the result that only even powers in $Q^{-1}$ and $b$ are
required \cite{powerCS,nphh}.

\subsection{Prompt photons}

We now turn to the prompt photon cross section, and show 
how the NLL resummed cross section, already studied in a preliminary
fashion in Ref.\ \cite{LSV}, is derived.  We start with
Eq.\ (\ref{1pIresumeq}), which expresses the resummed cross section
in terms of doubly-transformed initial-state coefficient
functions $\bar c$, given in Eq.\ (\ref{CCE}), and
final-state functions ${\cal F}$, specified by
Eqs.\ (\ref{calFmt}), (\ref{Sprimesoln}) and (\ref{Eprexp}).
All logarithms of $b$ and $N$ exponentiate, and we find the form
\ba
{p_T^3 d \sigma^{({\rm resum})}_{AB\to \gamma} \over dp_T}
&=& \sum_{ij} \frac{p_T^4}{8 \pi S^2} \int_{\cal C} {dN \over 2 \pi i}\;
\tilde{\phi}_{i/A}(N,\mu_F) \tilde{\phi}_{j/B}(N,\mu_F)
\nonumber\\
&\ & \hspace{5mm} \times
\; \int_0^1 d\tilde x^2_T \left(\tilde x^2_T \right)^N\;
{|M_{ij}(\tilde x^2_T)|^2\over \sqrt{1-\tilde{x}_T^2}}\,
\;
C_\delta^{(ij\to \gamma k)}(\as(\mu),\tilde
x_T^2)
\nonumber \\
&& \hspace{10mm} \times
\int {d^2 {\bf Q}_T \over (2\pi)^2}\;
\Theta\left(\bar{\mu}-Q_T\right)
\left( \frac{S}{4 {\bf p}_T'{}^2} \right)^{N+1}\nonumber\\
&\ & \hspace{15mm} \times
\int d^2 {\bf b} \,
{\rm e}^{i {\bf b} \cdot {\bf Q}_T} \,
\exp\left[E_{ij\to \gamma k}\left( N,b,\frac{4 p_T^2}{\tilde
x^2_T},\mu_F \right)\right]\, .
\label{1pIresumE}
\ea
The final factor in this expression, the inverse transform
of the exponential, was described as a ``profile" 
function in Ref.\ \cite{LSV}, and was denoted $P_{ij}(N,{\bf Q}_T,Q,\mu)$.

The exponential moment dependence at NLL 
is given explicitly by
\begin{eqnarray}
E_{ij\rightarrow \gamma k}^{\rm PT} (N,b,Q,\mu_F) &=&
H_i (N,b,Q,\mu_F)  + H_j (N,b,Q,\mu_F) + F_k (N,Q) + G_{ijk} (N)\, .
\label{Eij}
\ea
Here the contributions from the initial state jets in $\bar c_{i/i}$ are 
computed from Eq.\ (\ref{Elog}), and are contained in
the functions
\ba
H_a (N,b,Q,\mu_F) &=& \frac{1}{\alpha_s (\mu)} h^{(0)}_a (\lambda,\beta) +
h^{(1)}_a (\lambda,\beta,Q,\mu,\mu_F) \, ,
\ea
with the $h^{(i)}_a$ given above in Eq.\ (\ref{hsubadef}).
Contributions
in $\tilde {\cal F}_{ijk}$ 
 from the final state jets
are the LL and NLL functions
\ba
F_a (N,Q) &\equiv& \frac{1}{\alpha_s (\mu)} f^{(0)}_a (\lambda) +
f^{(1)}_a (\lambda,Q,\mu)\, ,
\ea
while those  in $\tilde {\cal F}_{ijk}$ from the  soft functions,
are NLL only,
\ba
G_{abc} (N) &\equiv& g^{(1)}_{abc} (\lambda)\, .
\end{eqnarray}
The new functions $f_a^{(i)}$ are found from Eqs.\ (\ref{Eprexp}) 
and (\ref{Bpr}), while the $g_{abc}^{(1)}$ are computed following
Eqs.\ (\ref{Sprimesoln}) and (\ref{ADexplicit}), with the results: 
\begin{eqnarray}
f^{(0)}_a (\lambda) &=& 2 h^{(0)}_a (\lambda/2,\lambda/2) - h^{(0)}_a
(\lambda,\lambda) \; ,\\
f^{(1)}_a (\lambda,Q,\mu) &=&
2 h^{(1)}_a (\lambda/2,\lambda/2,Q,\mu,Q) -
h^{(1)}_a (\lambda,\lambda,Q,\mu,Q) \nonumber \\
&+& \frac{A_a^{(1)} \ln 2}{\pi b_0}
\left( \ln(1-2 \lambda) - \ln(1-\lambda) \right) 
- \frac{B_a^{(1)}}{\pi b_0}  \ln(1-\lambda) \;  , \\
g^{(1)}_{q\bar{q}g} (\lambda) &=& \frac{C_A}{2\pi b_0} 
\ln(1-2 \lambda) \ln \left( \frac{t u}{s^2} \right) \; ,\\
g^{(1)}_{qgq} (\lambda) &=& \frac{1}{\pi b_0} \ln(1-2 \lambda)
\left[ C_F \ln \left( \frac{-u}{s} \right) +\frac{1}{2}
C_A \ln \left( \frac{t}{u} \right)  \right]\, ,
\end{eqnarray}
where the coefficients $A^{(1)}_a$ and $B^{(1)}_a$ have been defined in 
Eqs.~(\ref{explicitA}) and (\ref{explicitB}), respectively.
These formulas were employed in Ref.\ \cite{LSV} to
test the sensitivity of the inclusive prompt photon cross section
to joint resummation\footnote{We note a slight difference in
notation from Ref.\ {\protect \cite{LSV}}, where the
functions {\protect $C_\delta$} were not exhibited separately.}.

Again, to the perturbative expressions, nonperturbative corrections
must in general be added.  For prompt photons, we do not yet have
all-orders expressions of the full generality of Eqs.\ 
(\ref{eiksigmahatE})-(\ref{Delaborate})
for electroweak annihilation, so we cannot yet conclude that all 
nonperturbative
corrections are of even powers in $b$ and $N$.  We consider this the most
likely scenario, however, and in Ref.\ \cite{LSV}, the significant
role of a modest nonperturbative term proportional to $b^2$ in
the exponent was highlighted.  Clearly, this issue bears further study.

\subsection{Recoil and enhancement}

The jointly-resummed electroweak annihilation cross section,
Eq.\ (\ref{physsigresum}), differs from 
$b$-space formalisms for $Q_T$-resummation \cite{ktresumold,yuancollab} in the
inclusion of threshold logarithms at
leading power in the Mellin moment variable $N$.  
At NLL, these corrections are included through the modified lower 
limit in the exponent, Eq.\ (\ref{Elog}), which stabilizes the 
exponent to an integration range $Q/N < \mu' <Q$ at $b=0$.  
This limit gives exactly the threshold resummed exponent 
for the $Q_T$-integrated cross section  \cite{dyresumgs,dyresumct}.  

In view of the above, we see that, in the $\mu'$ integral 
in the exponent, Eq.\ (\ref{Elog}), the effects of 
$Q_T$-resummation and of threshold resummation are additive.  
This feature may be understood from the distinct origins
of threshold and transverse momentum logarithms.  
The logarithms of $N$ in threshold resummation, which enhance the
cross section, come primarily from subtracting negative corrections 
to the perturbative parton distributions in the construction of the 
hard-scattering function in the $\MS$ (or other) scheme
\cite{dyresumgs}.
These subtractions produce enhancements because virtual 
corrections suppress the denominator of Eq.\ (\ref{normalmoment})
at the edge of phase space, i.e., at large $N$.  The logarithms of $b$, 
however, are
associated with real-gluon emission, in the numerator
of Eq.\ (\ref{normalmoment}), the partonic cross section.  
 They are negative in the exponent, because
they cancel the divergent lower limit of the virtual $\mu'$ integral
in Eq.\ (\ref{Elog}).  Notice that the integrand changes sign
at $\mu'=Q/\bar N$, which is the dividing line between 
threshold and transverse momentum dependence.  Very roughly speaking,
it is possible to resum threshold and transverse momentum
logarithms simultaneously because they come from ``different"
gluons: gluons from the parton distributions for the former and 
from the unsubtracted partonic cross section for the latter.
In essence, when recoil is taken into account, the Sudakov
suppression associated with large values of $b$ in (\ref{Elog})
does not cancel the threshold enhancement, but redistributes it
over a range of partonic invariant masses and momentum transfers at
the hard scattering.

In the jointly-resummed prompt photon cross section, Eq.\ (\ref{1pIresumE}),
the explicit enhancement of the integrand is 
associated with the inclusion of recoil,
relative to the threshold-resummed cross section.
The recoil allows $\tilde{x}_T^2 < \hat{x}_T^2$ for $p_T{}'<p_T$,
even while the full partonic invariant $\hat{s}=\xi_1 \xi_2 S$ is 
bounded from below by $4p_T^2$. In moment form, Eq.~(\ref{1pIresumeq}),
this enhancement appears through the factor $(S/4p_T'{}^2)^N\ge 
(S/4p_T^2)^N$. Eq.\ (\ref{1pIresumE}) reduces to the inclusive 
threshold-resummed cross section in Ref.\ \cite{thrphen}
if we neglect recoil, that is, if we set $b$ to zero in
the functions $c_{i/i}$, or at NLL, the exponent
$E_{ij}$ in Eq.\ (\ref{Eij}). Then,
the exponent reverts to its form in threshold resummation, and 
the $b$ integral gives a delta function, which eliminates
the ${\bf Q}_T$ integral, and freezes ${\bf p}_T'={\bf p}_T$.

\section{Conclusions}

We have described a joint resummation procedure
for threshold and  transverse momentum singularities.
The resummation organizes a well-defined set of corrections
to single-particle and electroweak annihilation cross
sections at measured transverse momentum, to all orders
in perturbation theory.   Although the arguments for
specific applications are somewhat involved, the basic
observation is relatively simple: for those contributions that
are singular at partonic threshold,
the transverse momentum, $Q_T$, of the 
short-distance scattering can be identified meaningfully.  
The factorization
properties of perturbation theory near partonic
threshold allow us to control logarithms simultaneously
in $Q_T$ and the relevant threshold variable, $1-z$,
for electroweak annihilation, or $1-\hat x_T^2$ for
single-particle inclusive (1PI) cross sections.
For electroweak annihilation cross sections, $Q_T$
may be identified with the observed final-state boson. 
For high-$p_T$ 1PI cross sections,
$Q_T$ must be integrated to derive the hard-scattering
function in the formalism of collinear
factorization.  The integral over singular
distributions in $Q_T$ leaves behind finite remainders that modify,
and may enhance,
the predictions of threshold resummation. 

Joint resummation 
extends our control over a
class of effects that can have an important phenomenological
influence.  At the same time, the resummed expressions 
afford new insights into nonperturbative power corrections
in hard-scattering cross sections, which may be competitive with
resummed  perturbation theory, or even overshadow it.  A preliminary
study points to the importance of both high orders in
perturbation theory and of nonperturbative effects in prompt
photon production at fixed target energies \cite{LSV}.

For electroweak annihilation at measured $Q_T\ll Q$, the gap between existing 
formalisms that resum in transverse momentum, and the joint
resummation described here, should be relatively easy to close.
Such an application would produce, we conjecture, a decrease in sensitivity
to the factorization scale.   

Work remains to make  our joint resummation formalism a tool for 
phenomenological predictions in 1PI cross sections.  From a numerical
point of view, the resummed expressions involve extra
integrals compared even to threshold resummation, so 
purely computational considerations make it more challenging
to implement. Perhaps more significantly, it will
be necessary to develop an appropriate matching formalism for
large recoil.   Nevertheless, we believe that the joint
resummation formalism sheds valuable light on
the reliability of perturbative calculations in hadronic
scattering, and on the influence of nonperturbative effects. 

\subsection*{Acknowledgements}  We have benefited from
conversations with many of our colleagues on
the issues discussed in this paper.  We would like to
thank in particular  S.\ Catani, J.C.\ Collins, 
M.\ Fontannaz, J.\ Huston, N.\ Kidonakis,
G.\ Korchemsky, M.\ Mangano, P.\ Nason,
J.\ Owens, W.K.\ Tung, and M.\ Zielinski for helpful comments
that have influenced this work.

The work of G.S.\ and W.V.\ was supported in part by the 
National Science Foundation,
grant PHY9722101. G.S. acknowledges the hospitality of
Brookhaven National Laboratory.  
The work of E.L.\ is part of the research program of the
Foundation for Fundamental Research of Matter (FOM) and
the National Organization for Scientific Research (NWO).
W.V. is grateful to RIKEN, Brookhaven National Laboratory and the U.S.
Department of Energy (contract number DE-AC02-98CH10886) for
providing the facilities essential for the completion of this work.

\begin{appendix}

\section{Status of the Refactorizations}

Specific arguments for the refactorized
cross section, Eq.\ (\ref{thfact}) for threshold resummation
were given in Ref.\ \cite{jetresum}.
The $Q_T$-refactored expression, Eq.\ (\ref{cssfact}) was analyzed
in some detail at one
loop in Ref.\ \cite{dyqtfact}, although explicit
arguments for its validity to all orders have not,
to our knowledge, appeared in the literature.  Eq.\ (\ref{qtthfact})
is new to this paper.
In fact, we believe that all three of these refactorized
cross sections, Eqs.\ (\ref{cssfact}), (\ref{thfact}) and (\ref{qtthfact})
are on a theoretical footing similar to that of collinear factorization,
Eq.\ (\ref{dyqtcofact}), but, of course,
with different corrections as indicated above.
We  summarize below the ingredients of general factorization
proofs for these relations, modeled after the arguments for
collinear factorization given in Refs.\ \cite{cssrv,cssfact} and summarized
recently in Ref.\ \cite{novio}.  The following reasoning applies explicitly
to electroweak annihilation cross sections; the extension to single-particle
inclusive cross sections is straightforward \cite{LibbySt}.

\subsection{Leading regions and cut diagrams}

     Each of the functions in the refactorized expressions for 
electroweak annihilation,
Eqs.\ (\ref{cssfact}), (\ref{thfact}) and (\ref{qtthfact}), corresponds to
one of the subdiagrams in Fig.\ \ref{EWAleadingfig}, which represents a general
leading region, in the terminology of Ref.\ \cite{cssrv}.
A leading region in momentum space is one that gives rise to a contribution
to the cross section that is leading power in the hard momentum $Q$.
In general, leading regions can be classified by a set of on-shell virtual 
lines, whose vanishing denominators produce logarithmic corrections.
In the following, an on-shell momentum is one whose invariant mass is
much less than $Q$, and a soft momentum means one all of whose components
are much less than $Q$ in the specified leading region \cite{cssrv}.

Fig.\ \ref{EWAleadingfig} is in cut diagram notation, in which
$C$ represents a particular final state, with a graphical
contribution to the amplitude on the left of $C$, and a contribution
to the complex conjugate amplitude on the right of $C$.  In
this notation,
the  cross section
is a sum over all cuts $C$ of forward-scattering diagrams $G$,
consistent with the specified final state.  In this case,
all relevant final states contain an electroweak vector
boson $V$, with momentum $Q^\mu$.
In Fig.\ \ref{EWAleadingfig}, the subdiagrams $J_{A}$, $J_{B}$, $U$ 
and $H$ include, respectively, on-shell lines with momenta
collinear to $p_A$ (to be absorbed into ${\cal P}_{a/a}$,
$\psi_{a/a}$ or ${\cal R}_{a/a}$), lines with on-shell momenta parallel to
$p_B$ (to ${\cal P}_{b/b}$, $\psi_{b/b}$ or ${\cal R}_{b/b}$),
lines with soft momenta (to $U$),
and lines off-shell by order $Q$ (to $H$).

Following the notation of Ref.\ \cite{cssrv}, we represent
the contribution from region $L$ of graph $G$ to the cross section as
\ba
G_L &=&
\sum_C \int {dk_A^+\over 2\pi}\, {dk_B^-\over 2\pi}\;
H^{(C)}\left(k_A^+,k_B^-\right)\
\prod_\ell \int {d^4q_\ell \over (2\pi)^4}\; \prod_j \int {d^4 \bar
q_j \over (2\pi)^4}\
J_A^{(C)}\left(k_A^+,q_\ell^\alpha\right)^{\{\mu_1\dots\mu_n\}}\nonumber\\
&\ & \quad \times U^{(C)}\left(q_\ell^\alpha,\bar
q_j^\beta\right)_{\{\mu_1\dots\nu_1\dots\}}
\  J_B^{(C)}\left(k_B^-,\bar q_j^\beta\right)^{\{\nu_1\dots\nu_n\}}\, .
\label{GL}
\ea
In this expression, we have suppressed the flavor labels ($a$ and $b$ in
Eq.\ (\ref{dyqtcofact})
and so on), and the corresponding  Lorentz indices, that link the jets with the
hard scattering.   Eq.\ (\ref{GL}) is a representation of the most
general leading-power
contribution to the cross section $d\sigma/d^4Q$, considered as an integral
in the space of loop and final-state momenta.   We shall argue that
for each such leading region $L$, $G_L$ may be rewritten as in Eq.
(\ref{cssfact}), (\ref{thfact}) or (\ref{qtthfact}), up to the
corrections indicated in those expressions.   For this discussion, it
is convenient to introduce the lightlike vectors $v^\mu$ and $u^\mu$,
in the directions of the incoming momenta $p_a$ and $p_b$,
according to
\be
v^\mu={p_a^\mu \over p_a^+}=\delta_{\mu +}\, , \quad u^\mu
={p_b^\mu\over p_b^-}=\delta_{\mu -}\, .
\ee
Already at the first step in our factorization argument, Eq.\
(\ref{GL}), we have made
the approximation that the
short-distance function $H(k_A^+,k_B^-)$ depends on
the large light-cone components of the quarks that annihilate to
produce the electroweak vector boson.  In particular, we
identify $(k_A^+/p_a^+)=x_a$ and $(k_B^-/p_b^-)=x_b$ in Eq.\
(\ref{dyqtcofact}).
We know of no other approximation that can be extended beyond lowest order to
formulate the short-distance function in a consistent fashion, as a
collinear-finite quantity \cite{cssrv}.
Thus, we neglect transverse momenta in the calculation of $H$.
This implies that in our factorized cross section, as in Eq.\
(\ref{cssfact}), the measured transverse momentum,
$Q_T$ does not appear in the short-distance function directly, but
only through its kinematic linkage (``recoil'')
with the total final-state transverse momenta
of the jets $J$, and of the soft subdiagram $U$.  
Notice that the short distance
function $H$ in Eq.\ (\ref{GL}) need not be identical to the
partonic hard scattering function $\hat \sigma$, which will
in general absorb infrared-safe corrections from $U$.

\subsubsection{The soft approximation and the eikonal function}

The desired factorized expression for each case discussed above
reduces to an identity, if we
make the following substitutions for the jet functions $J^{(C)}_A$
and $J^{(C)}_B$,
\ba
J_A^{(C)}\left(k_A^+,q_\ell^\alpha\right)^{\{\mu_1\dots\mu_n\}}
&\to&
J_A^{(C)}\left(k_A^+,(q_\ell\cdot v)
u^\alpha\right)^{\{\xi_1\dots\xi_n\}}u_{\xi_1}\dots u_{\xi_n}
v^{\mu_1}\dots v^{\mu_n}
\nonumber\\
J_B^{(C)}\left(k_B^-,\bar q_j^\beta\right)^{\{\nu_1\dots\nu_n\}}
&\to&
J_B^{(C)}\left(k_B^-, (\bar q_j\cdot u)
v^\beta\right)^{\{\lambda_1\dots\lambda_n\}}v_{\lambda_a}\dots
v_{\lambda_n}
u^{\nu_1}\dots u^{\nu_n}\, .
\label{softapprox}
\ea
These substitutions are the ``soft approximation" \cite{cssrv}.  In
the soft approximation, the
momenta and polarizations of all soft gluons
connected to $J_A$ are approximated by their minus components, that is,
the components
moving opposite to the direction of lines in the jet.
Similarly, for $J_B$, only plus components are kept.
Notice that although the transverse components of soft momenta are
neglected in the jet functions, no approximation is made in the soft
function itself.

\begin{figure}
\begin{center}
\hspace*{-7mm}
\epsfig{file=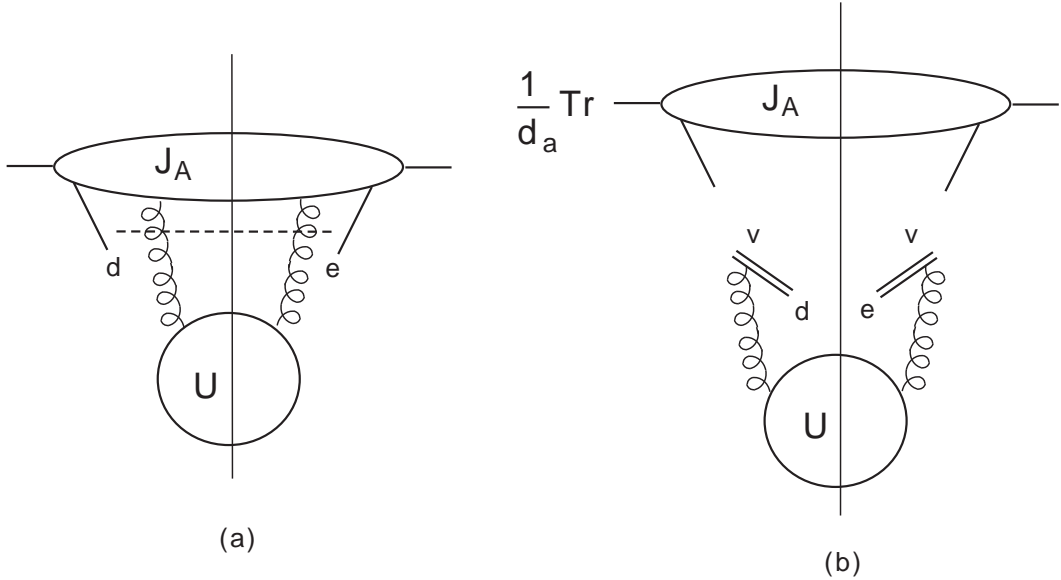,width=14cm}
\end{center}
\caption{(a) Gluons attached to jet $J_A$ in the
soft approximation, indicated by  the dashed line.
Color indices $d$ and $e$ are to be contracted with
the hard scattering.  (b) Result of the Ward identity
discussed in the text, in which the jet is replaced
by an eikonal (double) line with velocity vector, $v^\mu$.
Note the color trace on the remaining jet functions,
normalized by $d_a$, the dimension of the color
representation of parton $a$.}
\label{WIfig}
\end{figure}

Once the soft approximations (\ref{softapprox})  have been carried out,
the decomposition of $G_L$, Eq.\ (\ref{GL}), into the appropriate
convolutions in  transverse momentum and light-cone or energy
fraction, requires only that we sum over all connections of gluons
from the soft diagrams $U$ to the corresponding jet subdiagrams.
The graphical Ward identities of the theory ensure this result,
illustrated by Fig.\ \ref{WIfig} \cite{cssfact}.  After the sum over diagrams,
the coupling of the soft gluons to jets is replaced by
their coupling to eikonal lines, as in Eq.\ (\ref{Phidef}), in
the directions of the jet momenta, which serve as sources for
the gluons in the functions $U$.  The Ward identities,
which are essentially algebraic, also do not require an integral
over soft-gluon momenta, which may be treated as fixed.  The use of
Ward identities is thus consistent with the restrictions on final-state
momenta necessary to define cross sections at observed transverse momenta or
near threshold.

\subsubsection{Justification of the soft approximation}

We have now seen that a refactorization
appropriate to each of our theorems is assured if the soft approximation
can be justified.  What must
therefore be verified is the accuracy of the soft approximation,
as it is used in each factorized or refactorized cross section.  As
discussed, for example, in Ref.\ \cite{cssrv}, the soft approximation
fails, on a graph-by-graph basis, for $J_A^{(C)}$ whenever 
$q_\ell^-\ll q_{\ell,T}$ for any
one of the soft lines that connect $J_A^{(C)}$ to $U$, and
in $J_B^{(C)}$ whenever $\bar q_j^{\, +}\ll \bar q_{j,T}$ for one of the lines
that connect $J_B^{(C)}$ with $U$.
The central step in the proof of collinear factorization \cite{cssfact}
is to show that all regions where the soft approximation fails
cancel in the sum over final states.  Equivalently, we must show
that the $q_\ell^-$ integrals linked to $J^{(C)}_A$, considered as
contour integrals
in the complex plane, are not pinched between coalescing
singularities with $q_\ell^-\ll q_{\ell,T}$.   For a fixed final state,
however, it is easy to identify such pinches, between poles
in the upper half-plane from the Feynman denominators of ``spectator"
lines, which carry large plus momenta into the final state, and poles
in the lower half-plane
from the denominators of ``active" lines, which carry plus momenta
into the electroweak annihilation \cite{CoSt,BBL}. The former may be 
thought of as characteristic of final-state
interactions, and the latter of initial-state interactions.
After the sum over final states,
however, final-state interactions cancel, and
all remaining singularities in $q_\ell^-$ from the function $J_A$
are in the lower half-plane.  As a result, after the sum,
the $q_\ell^-$ integrals are no longer pinched in
the dangerous region, and the soft approximation may be carried out.

For collinear factorization, Eq.\ (\ref{dyqtcofact}), the cancellation
of final state interactions is verified once the internal {\it minus} momentum
integrals internal to jet $J_A$, as well as the
internal {\it plus} momentum integrals of the remainder of the diagram,
are carried out.  Details can be found in Ref.\ \cite{cssfact}; here we shall
only need the result that no other integrals are necessary.
We must verify that corresponding arguments
apply to the $Q_T$, threshold and joint refactorized cross sections.

\subsubsection{$Q_T$ refactorization}

An essential feature of the cancellation of final-state interactions
in the collinear-factorized cross section is that it requires
integrals over light-cone momenta only  \cite{cssfact}.  The
cancellation is therefore consistent with
{\it fixed} transverse momenta.
The new feature of the $Q_T$-refactorized
Eq.\ (\ref{cssfact}), relative to the collinear-factorized
cross section, Eq.\ (\ref{dyqtcofact}), is simply that
in the former the total transverse momentum
of the soft and jet functions is frozen at $-Q_T$.   The arguments for
factorization as in Eq.\ (\ref{dyqtcofact}) are therefore adequate for
refactorization as in Eq.\ (\ref{cssfact}), where $Q_T$ is
the total transverse momentum carried by quanta that
are absorbed by the short distance function $|H|^2$ in Eq.\ (\ref{GL}).

Eq.\ (\ref{cssfact}) refers to a cross section in which
the observed final-state momentum is that of a single vector boson only, with
all real QCD radiation incorporated into the jets or soft function.
In general, the short distance function, $|H|^2$, like
the jet and soft functions, includes corrections
associated with QCD radiation into the final state.  The
structure of the leading regions in Eq.\ (\ref{GL}), however,
ensures that these corrections are  not singular when the
total transverse momentum of the extra, ``hard" QCD radiation
vanishes.  By construction, singular behavior is entirely in the
jets and soft function.  As a result, such regions,
although of leading power in $Q^2$,
are not singular as a power at $Q_T=0$, when $Q_T$ is the
{\em vector boson} momentum \cite{cscss,dyqtfact}.  These nonsingular
contributions, which
begin at next-to-leading order, are absorbed into the correction
$Y_{\rm kt}$ in the $Q_T$-refactorized cross section, Eq.\ (\ref{cssfact}).

\subsubsection{Threshold refactorization}

The situation for threshold resummation  is a bit more
subtle.  In this case, we want to fix the energy of
radiation from the jets and soft function in the
hard-scattering center-of-mass frame.
This clearly puts restrictions on the light-cone momentum
integrals needed to ensure the cancellation of final
state interactions.  Near threshold, however, {\it all} radiation
into the final state,
including the radiation within the jets, is soft
compared to $Q$, because near threshold
the total energy of final state radiation is of order $(1-\tau)Q$,
with $\tau=Q^2/S$.  All radiation,
including ``spectator" gluons in the jets may thus be absorbed
into the soft function $U$ in Eq.\ (\ref{GL}).
Corrections to this result are less singular by
a power at $\tau=1$ than the, leading, $1/(1-\tau)$ behavior in
perturbation theory.
The jet subdiagrams that remain after this factorization are purely virtual,
with only ``active" lines, whose energies flow into the hard
scattering \cite{dyresumgs}.
In this case, as pointed out in Sec.\ 3.2 of Ref.\ \cite{jetresum},
the same mechanism for the cancellation
of final-state interactions in collinear factorization,
Eq.\ (\ref{dyqtcofact}),
ensures that the soft
function near threshold is free of overlapping soft-collinear logarithms,
even when the light-cone integrals are restricted.  Collinear singularities
are present in the soft subdiagram, which is now
the eikonal cross section discussed above, and they 
factorize in the usual way,
as in Eq.\ (\ref{Umoment}) above (integrated over $k_T$).

We can now reorganize the cross section in moment space to
derive Eq.\ (\ref{thfact}).
We remove factorizing, purely collinear singularities
from $U$ by multiplying and dividing  the eikonal cross section by
eikonal jet functions $\psi_f^{({\rm eik})}$, corresponding to the incoming
partons, and defined by analogy to Eq.\ (\ref{jindef}), in the
appropriate (Laplace)
transform space.  The ratio of the soft function to eikonal jets
in transform space, as in Eq.\ (\ref{Umoment}),  is free of collinear
singularities altogether.  The eikonal
jets combine with the virtual remainders from the original
factorization to form the functions $\psi$ at fixed energy,
near threshold, as in Eq.\ (\ref{thfact}).

We expect there to be singular but integrable remainders
in the soft function, after the sum over final state interactions.
These remainders are precisely the logarithms associated with the 
soft function, $U$ \cite{jetresum}.
Corrections resulting from the factorization of final-state radiation
into $U$, which are ${\cal O}\left([1-\tau]^0\right)$, are absorbed
into $Y_{\rm th}$ in Eq.\ (\ref{thfact}),
and begin at one loop.

\subsubsection{Joint refactorization}

The step from the
threshold-refactorized expression (\ref{thfact}), to the
jointly refactorized Eq.\ (\ref{qtthfact}) is essentially equivalent
to the step from collinear factorization to the $Q_T$-refactorized
form (\ref{cssfact}).
The extra ingredient is again to fix transverse momenta
for the soft radiation and the jets,  a condition that we have already
argued is consistent with the cancellation of final-state
interactions, and hence with the validity of the soft approximation.

Of course, as emphasized in Refs.\ \cite{cssrv,cssfact}, complete factorization
proofs should rest on arguments based upon systematic subtraction procedures.
This level of sophistication is still to be attained in hadronic scattering
cross sections.  Nevertheless, we consider the arguments outlined above
to be adequate to
justify an analysis based on the refactorization theorems in Secs.\ 2 and 4.

\section{One-loop parton distributions and refactorization}
\label{sec:one-loop-distr}

In this appendix we present for illustration,
at one-loop accuracy, the various generalized
distributions of section 2.3. To keep the presentation brief, we will 
only present results for quark distributions. We begin with the familiar 
distributions, defined at fixed `plus'-component of their momentum, and 
then turn to those at fixed energy. The latter, with transverse momentum
fixed also, are then used to verify the joint refactorization
(\ref{qtthfact}) to one loop for electroweak annihilation. 
Illustrations of joint refactorization for the one-loop prompt photon
cross section will be given elsewhere. 

\subsection{Quark distribution at fixed light-cone momentum fraction}
\label{sec:partonic-light-cone}

We give first the one-loop partonic light-cone distribution (\ref{phidef}) 
in $n\cdot A = A^+=0$, $n^2=0$ gauge. The familiar result 
is~\cite{qseml,cfp}~:
\begin{eqnarray}
\label{eq:19}
     \phi_{q/q}(x,\mu,\epsilon)
    =  \delta(1-x)- \left( \frac{1}{\epsilon} +\ln (4\pi)-
\gamma_E \right)  \frac{\alpha_s(\mu)}{2\pi}\, C_F 
\left[ \frac{1+x^2}{(1-x)_+}  +\frac{3}{2} \,\delta(1-x) \right] \; .
\end{eqnarray}
\subsection{Quark distribution at fixed light-cone fraction
and transverse momentum ${\bf k}$}
\label{sec:partonic-light-cone-1}

Next we compute ${\cal P}_{q/q}(x,{\bf k},\mu,\epsilon)$, Eq.~(\ref{Pdef}),
in $n\cdot A =0$ gauge with $n^2\neq 0$. For simplicity, we choose 
${\bf n}_T=0$. One finds
\begin{eqnarray}
\label{eq:11}
 {\cal P}_{q/q}(x,{\bf k},\mu,\epsilon)
    &=&  \delta(1-x)\,\delta^{D-2} ({\bf k}) \; \left[ 1 + 
\psi^{(1),V}_{q/q}(x,2p\cdot n,\epsilon)  \right]  \nonumber \\
&&\hspace*{-2.5cm} + \;
\frac{\alpha_s(\mu)C_F}{2\pi^2}   \, 
\left( 4\pi^2\mu^2 \right)^\epsilon \, \left\{ 
\frac{1-x}{{\bf k}^2} \left[ 1-\epsilon + \frac{2x\, \nu}{{\bf k}^2+
(1-x)^2 \nu} \right] - \frac{2 \nu(1-x)}{({\bf k}^2+(1-x)^2 \nu)^2}
\right\} \; , 
\end{eqnarray}
with $\psi^{(1),V}_{q/q}$ the same function of the gauge
vector $n^{\mu}$ given below for the one-loop virtual correction to
$\psi_{q/q}$, and where $\nu = (2p\cdot n)^2/|n^2|$. It is 
not difficult to check that
\begin{eqnarray}
  \label{eq:14}
\int  \d^{d-2}{\bf k}\,{\cal P}_{q/q}(x,{\bf k},\mu,\epsilon)&=&
\frac{\pi^{1-\epsilon}}{\Gamma (1-\epsilon)} \int_0^{\mu^2} d{\bf k}^2
\left( {\bf k}^2 \right)^{-\epsilon}
\,{\cal P}_{q/q}(x,{\bf k},\mu,\epsilon) \nonumber \\
&=&  \phi_{q/q}(x,\mu,\epsilon) \; +\; {\rm finite} \; ,
\end{eqnarray}
i.e., that the singularities in ${\cal P}_{q/q}$ match those 
of $\phi$, up to finite ($n$- and $\mu$-dependent) remainders.

\subsection{Quark distribution at fixed energy}
\label{sec:part-energy-distr}

  For the one-loop energy distribution (\ref{psidef}), the computation
is described in detail in Ref.~\cite{dyresumgs}.
We work in $n\cdot A =0,\, n^\mu=\delta^{\mu 0}$ gauge and treat the 
$1/(n\cdot k)$ terms in the gluon propagator in principal value 
prescription. The result is
\begin{eqnarray}
     \label{eq:8}
     \psi_{q/q}(x,2p_0,\epsilon)
    =  \delta(1-x)+
    \psi^{(1),R}_{q/q}(x,2p_0,\epsilon)  +   
    \psi^{(1),V}_{q/q}(x,2p_0,\epsilon)   \, ,
\end{eqnarray}
where $p_0=Q/2$, and the real and virtual contributions are given by
\begin{eqnarray}
\label{eq:21}
\psi^{(1),R}_{q/q}(x,2p_0,\epsilon) &=& \frac{\alpha_s(\mu)C_F}{2\pi}\,
\left(\frac{4\pi\mu^2}{\nu}\right)^\epsilon\,
\frac{\Gamma(2-\epsilon)}{\Gamma(2-2\epsilon)}
\left(\frac{-1}{\epsilon}\right)\,
\left[
  \frac{1+x^2-\epsilon(1-x)^2}{(1-x)^{1+2\epsilon}}\right]\,,\\
 \psi^{(1),V}_{q/q}(x,2p_0,\epsilon)   &=& 
-\frac{\alpha_s(\mu)C_F}{\pi}\,\delta(1-x) \Bigg\{
\left(\frac{4\pi\mu^2}{\nu}\right)^\epsilon\, 
\left(\frac{1}{2\epsilon^2}\right)
\left(\frac{1-\epsilon}{1-2\epsilon}\right) 
\Gamma(1-\epsilon)\Gamma(1+2\epsilon) \nonumber \\
&& \hspace*{-2cm}
\times \; {\rm Re}
\left[\frac{1}{2}\left({\rm sign}(n^2)\right)^\epsilon 
+\frac{1}{2}\left({\rm sign}(n^2)\,e^{-2\pi\, i}\right)^\epsilon  \right]
 + \frac{3}{4}\left(\frac{1}{\epsilon}+\ln(4\pi)-\gamma_E\right) \Bigg\}\,,
\end{eqnarray}
where again $\nu = (2p\cdot n)^2/|n^2|$. 
$\psi^{(1)}_{\bar{q}/\bar{q}}$ is identical.
The counterterm contribution to $\psi^{(1),V}_{q/q}$ 
results from $\MS$ fermion wavefunction 
renormalization in the $n\cdot A=0$ gauge, and the double pole in $\epsilon$ 
reflects an overlapping soft and collinear divergence. 
The expansion in $\epsilon$ of (\ref{eq:8}) reads
\begin{eqnarray}
\psi_{q/q}(x,2p_0,\epsilon)&=& \delta(1-x) + 
\left(\frac{\alpha_s C_F}{\pi}\right)
\Biggl\{-\frac{1}{2}\left(\frac{1}{\epsilon}+\ln(4\pi)-\gamma_E\right)\,
\Bigg[ \frac{1+x^2}{(1-x)_+} + \frac{3}{2} \, \delta (1-x) \Bigg]
\nonumber\\&\ & 
+2 \left[{\ln (1-x) \over 1-x}\right]_+ -\left[{1\over 1-x}\right]_+
-\ln\left({\mu^2\over \nu}\right)\left[{1\over 1-x}\right]_+ +
\frac{\pi^2}{6} \, \delta (1-x)\, \Biggr\} \; ,
\label{eq:15}
\end{eqnarray}
plus non-singular terms. 
Note at ${\cal O}(\epsilon^0)$ the appearance of the double-logarithmic 
plus-distribution $\left[\ln(1-x)/(1-x)\right]_+$, which is a remnant of the 
$1/\epsilon^2$ cancellation between real and virtual contributions.

\subsection{Quark distribution at fixed energy and transverse momentum}
\label{sec:part-energy-distr-1}

The one-loop result for the distribution 
${\cal R}_{q/q}(x,{\bf k},2p_0,\epsilon)$  in Eq.~(\ref{Rdef}) is
\begin{eqnarray}
\label{eq:17}
{\cal R}_{q/q}(x,{\bf k},2p_0,\epsilon) &=&  \delta(1-x)\,
\delta^{D-2} ({\bf k}) + {\cal R}^{(1),R}_{q/q}(x,{\bf k},2p_0,\epsilon)  +   
{\cal R}^{(1),V}_{q/q}(x,{\bf k},2p_0,\epsilon)  \; ,
\end{eqnarray}
where 
\begin{eqnarray}
\label{eq:16}
{\cal R}^{(1),R}_{q/q}(x,{\bf k},2p_0,\epsilon) &=& {\cal F}
(x,{\bf k},2p_0,\epsilon)\, P_{qq}^\epsilon(x) \,
\left[\frac{1}{{\bf k}^2} - \frac{2}{(2p_0)^2\,(1-x)^2}\right]\,,\\
{\cal R}^{(1),V}_{q/q}(x,{\bf k},2p_0,\epsilon)   &=& \delta^{D-2}
({\bf k}) \psi^{(1),V}_{q/q}(x,2p_0,\epsilon)\; .
\end{eqnarray}
In these expressions, we define
\begin{eqnarray}
     \label{eq:5}\nonumber
{\cal F}(x,{\bf k},2p_0,\epsilon) &=& \frac{\alpha_s C_F}{2\pi^2}
\left(4\pi^2\mu^2\right)^\epsilon
\left(1-\frac{4 {\bf k}^2}{(1-x)^2 (2p_0)^2}\right)^{-1/2}\,,\\
P_{qq}^\epsilon(x) &=& \frac{1+x^2-\epsilon(1-x)^2}{1-x}\,.
\end{eqnarray}
${\cal R}^{(1)}_{\bar{q}/\bar{q}}$ is identical to ${\cal R}^{(1)}_{q/q}$.
Consistency requires that
\begin{eqnarray}
\label{eq:6}
\int  \d^{d-2}{\bf k}\,{\cal R}^{(1),i}_{q/q}(x,{\bf k},2p_0,\epsilon)
&=& \frac{\pi^{1-\epsilon}}{\Gamma(1-\epsilon)}\, 
\int_0^{p_0^2(1-x)^2}d{\bf k}^2\,({\bf k}^2)^{-\epsilon}
\,{\cal R}^{(1),i}_{q/q}(x,{\bf k},2p_0,\epsilon) \nonumber\\
&= & \psi^{(1),i}_{q/q}(x,2p_0,\epsilon),\;\;\; i=R,V \; ,
\end{eqnarray}
which is straightforward to verify. Note that
in ${\cal R}$ the double-logarithmic singularities in $1-x$ and ${\bf k}$ 
are generated by the term
\begin{eqnarray}
  \label{eq:23}
 \int_0^{p_0^2(1-x)^2}d{\bf k}^2 ({\bf k}^2)^{-1-\epsilon} \frac{1}{1-x}
\left(1-\frac{{\bf k}^2}{(1-x)^2p_0^2}\right)^{-1/2} \,.
\end{eqnarray}

\subsection{Drell-Yan cross section}
\label{sec:drell-yan-cross}

The lowest order cross section for the reaction 
\begin{equation}
     \label{eq:2}
     q(p_1) + \bar{q}(p_2) \rightarrow \gamma^*(q)
\end{equation}
reads in $D=4-2\epsilon$ dimensions
\begin{equation}
     \label{eq:1}
\frac{d \sigma_{q\bar{q} }} {dQ^2 d^{D-2}{\bf Q}_T}=
\frac{4\pi \alpha^2}{3 N_C s \, Q^2}(1-\epsilon)\,
\delta \left( 1-\frac{Q^2}{s} \right)\,\delta^{D-2}({\bf Q}_T)
\equiv \sigma_0\, \delta(1-z)\,\delta^{D-2}({\bf Q}_T)\,.
\end{equation}
The one-gluon radiative correction to this cross section, resulting 
from the reaction
\begin{equation}
\label{eq:3}
     q(p_1) + \bar{q}(p_2) \rightarrow \gamma^*(q)+g(k)\,,
\end{equation}
is straightforwardly computed, using for example the expressions given 
in~\cite{aem}. In terms of the quantities defined in (\ref{eq:5}), 
the result is written compactly as
\begin{eqnarray}
     \label{eq:4}
     \frac{d^2\sigma_{q\bar{q}}^{(1),R}(z,Q^2,{\bf Q}_T)}
{dQ^2\,d^{D-2}{\bf Q}_T} &=&
2\,\sigma_0 {\cal F}(z,{\bf Q}_T,Q/\sqrt{z},\epsilon)
\;\left[\frac{1}{Q_T^2} P_{qq}^\epsilon(z) -\frac{2z}{(1-z)Q^2}\right] \; ,
\end{eqnarray}
with $P_{qq}^{\epsilon}$ given in~(\ref{eq:5}).
One can easily verify that upon integration over $d^{D-2} {\bf Q}_T$ 
this expression gives the real-emission correction to the inclusive 
Drell-Yan cross section given in Eq.~(88) of~\cite{aem}.

\subsection{Joint refactorization at one loop}
\label{sec:refact-at-one}

Let us now illustrate the refactorization of the Drell-Yan 
cross section in Eq.~(\ref{qtthfact}) at one loop. We will 
see that its singularities in ${\bf Q}_T$ and $1-z$ are accounted
for by the distributions ${\cal R}_{i/i},\, i=q,\bar{q}$, and the soft
function $U_{q\bar{q}}$. 

At one loop, the right hand side of Eq.~(\ref{qtthfact}) expands into
a sum of the one-loop expressions for its factors. 
Virtual corrections in Eq.\ (\ref{qtthfact}) are 
exactly equivalent to those of Eq.\ (\ref{thfact}) in
threshold resummation \cite{dyresumgs}.
Thus, we restrict ourselves to the real contributions, and 
check that the hard-scattering function found by expanding
(\ref{qtthfact}) to one loop:
\begin{eqnarray}
     \label{eq:9}
\sigma_0\, h^{\rm (j)(1)}_{q\bar q}
&=&
      {d \sigma^{(1),R}_{q\bar{q}} \over dQ^2 d^{D-2}{\bf Q}_T}
(z,Q^2,{\bf Q}_T) \nonumber\\
&\ & \hspace{-10mm}
-\sigma_0\, 
\left[ {\cal R}^{(1),R}_{q/q}(z,{\bf Q}_T,Q/\sqrt{z},\epsilon)+
{\cal R}^{(1),R}_{\bar{q}/\bar{q}} (z,{\bf Q}_T,Q/\sqrt{z},\epsilon) 
+ U^{(1),R}_{q\bar q}(1-z,{\bf Q}_T)\right]\; ,
\end{eqnarray}
is free of singularities at ${\bf Q}_T=0$ and (after ${\bf Q}_T$ 
integration) at $1-z=0$. The soft function $U_{q\bar q}^{(1),R}$ may be
found from Eq.\ (\ref{Umoment}).  At one loop, it is quite simple to
determine, because only the interference graphs contribute in the ratio,
and even in these diagrams only the $k^{\mu} k^{\nu}/(n\cdot k)^2$ term
in the gluon polarization tensor survives. The result is: 
\begin{equation}
\label{eq:10}
U^{(1),R}_{q \bar{q}}\left( 1-z,{\bf Q}_T\right) 
= \;{\cal F}(z,{\bf Q}_T,Q/\sqrt{z},\epsilon)
\;  {8\over Q^2(1-z)^3}\; .
\end{equation} 
Using Eqs.~(\ref{eq:17}), (\ref{eq:4}) and
(\ref{eq:10}), we then find 
\begin{equation}
    \label{eq:6C}
\sigma_0\, h^{\rm (j)(1)}_{q\bar q}
=    -\sigma_0 \,{\cal F}(z,{\bf Q}_T,Q/\sqrt{z},\epsilon)
\;\frac{8 (1+z)}{Q^2 (1-z)^2}
\; .
\end{equation}
This is the desired behavior, because,
when integrated over ${\bf Q}_T$, 
the result is nonsingular as $z\to 1$ (see the discussion after 
Eq.~(\ref{qtthfact})).

\end{appendix}

\end{document}